\documentclass[%
 reprint,
 amsmath,amssymb,
 aps,
prb,
]{revtex4-1}
\usepackage{mathrsfs}

\newcommand{\pin}{\par\noindent}
\usepackage{graphicx}% Include figure files

\usepackage{dcolumn}% Align table columns on decimal point
\usepackage{bm}% bold math
\usepackage{color}
\usepackage[colorlinks=false]{hyperref}
\hypersetup{
    colorlinks=True,
    linkcolor=blue,
    filecolor=blue,      
    urlcolor=blue,
}
\usepackage{subfiles}

\begin{document}

%\preprint{APS/123-QED}

\title{Origin of Topological Order in a Cooper Pair Insulator}
%\title{Microscopic Origin of Superconductivity-Vortex Condensate Duality}% Force line breaks with \\

\author{Siddhartha Patra}
\email{sp14ip022@iiserkol.ac.in}
% \altaffiliation[Also at ]{Physics Department, Indian Institute of Science Education and Research Kolkata.}%Lines break automatically or can be forced with \\
\author{Siddhartha Lal}
\email{slal@iiserkol.ac.in}
\affiliation{%
 %Indian Institute of Science Education and Research Kolkata \\
 %Department of Physical Sciences \\
 %Mohanpur-741246 ,India
  Department of Physical Sciences, Indian Institute of Science Education and Research-Kolkata, W.B. 741246, India
}%

%\affiliation{
%\affiliation{
%\affiliation{%

\date{\today}% It is always \today, today,

\pagebreak
\begin{abstract}
While a topologically ordered counterpart of the s-wave superconductor has been proposed in the literature on phenomenological grounds, its microscopic origin remains unknown. In meeting this goal, we employ the recently developed unitary renormalisation group (URG) method on a generalised model of electrons in two spatial dimensions with attractive interactions. We show that the effective Hamiltonian obtained at the stable low-energy fixed point of the RG flow corresponds to a gapped, insulating state of quantum matter we call the Cooper pair insulator (CPI). Detailed analyses show that the CPI ground state manifold displays several signatures of topological order, including a four-fold degeneracy when placed on the torus. Spectral flow based arguments reveal the emergent gauge-theoretic structure of the effective theory: the CPI effective Hamiltonian can be written entirely in terms of non-local Wilson loops. Further, it contains a topological $\theta$-term whose coefficient is quantised in keeping with the requirement of invariance of the ground state under large gauge transformations. This $\theta$-term is known to be equivalent to the Chern-Simons term in two spatial dimensions. Passage from the CPI to the metal by tuning the $\theta$ coefficient reveals a plateaux structure of the CPI ground state in terms of the steadily decreasing number of condensed Cooper pairs. Investigations reveal that the long-ranged many-particle entanglement content of the CPI ground state is driven by inter-helicity two-particle scattering processes. The plateau with $\theta=0$ possesses the largest bipartite entanglement entropy (EE), scaling logarithmically with subsystem size ($L$) and falling rapidly upon tuning $\theta$ towards the metal. The bipartite EE for inter-plateau transitions shows universal signatures in its variations with $L$ and $\theta$. While the EE signatures for plateaux and transitions can be distinguished at low temperatures, such distinctions are smeared out as temperature is raised. We also study the passage from the CPI to the s-wave BCS superconducting ground state under RG, and find that the RG flow promotes fluctuations in the number of condensed Cooper pairs and lowers those in the conjugate global phase of the ground state wavefunction. Consequently, we find that the distinct signatures of long-ranged entanglement in the CPI are replaced by the well-known short-ranged entanglement of the BCS state. Analysing the effects of Josephson effects reveals that while a Cooper pair tunnel coupling between two CPI systems does not lead to phase stiffening, breaking the global $U(1)$ phase rotation symmetry in one of them induces a phase coherence in the other. However, the generation of a Josephson current requires the separate breaking of the $U(1)$ symmetry in the two systems. Finally, we study the renormalisation of the entanglement in $k$-space for both the CPI and BCS ground states. The topologically ordered CPI state is shown to possess an emergent hierarchy of scales of entanglement, and that this hierarchy collapses in the BCS state. Our work offers clear evidence for the microscopic origins of topological order in this prototypical system, and lays the foundation for similar investigations in other systems of correlated electrons. 
%Discuss (i) nature of entanglement in CPI ground state and its being driven by inter-helicity two-particle scattering processes (ii) passage to BCS superconductivity (iii) Josephson effects (iv) Entanglement RG. 
%\begin{description}
%\end{description}
\end{abstract}

\pacs{Valid PACS appear here}% PACS, the Physics and Astronomy
\maketitle

\tableofcontents

%\section{Introduction}

\section{\label{sec:intro}Introduction}
\label{section:intro}
\pin
Superconductivity is undoubtedly one of the best studied example of an emergent collective phenomenon in a system of interacting electrons. While Cooper's theory~\cite{cooper1956} demonstrates that the presence of an attractive pairing interaction can lead to the formation and condensation of two-electron bound states (Cooper pair), the celebrated BCS theory~\cite{bcs} well describes the superconducting nature of this condensate of Cooper pairs in terms of a ground state wavefunction with a fluctuation in the number of Cooper pairs. Importantly, the BCS theory provides microscopic insight into various superconducting properties like the Meissner-Ochsenfeld effect, phase stiffness and the supercurrent, the transition temperature etc. The phenomenological Ginzburg Landau theory~\cite{gltheory} captures well the criticality of the superconducting transition in terms of a second order transition, involving the spontaneous breaking of the global $U(1)$ symmetry of the electronic Hamiltonian; this is known to be equivalent to the abelian Higgs field theory. The many-particle entanglement properties of the BCS wavefuction have also been established more recently
%cooper-pair number fluctuating ground state wavefunction of BCS \cite{bcs} helped to easily compute the entanglement properties 
~\cite{bcs_puspus,bcs_tullio}.
\pin
In the presence of a magnetic field \cite{SIT_mag_1,SIT_mag_2,SIT_mag_3} and/or disorder\cite{SIT_dis_1,SIT_dis_2,SIT_dis_3}, a superconducting thin film has been observed to undergo a transition to an insulating state of matter (see Ref.\cite{SIT_rev} for a review). Several experimental~\cite{SIT_insulator_exp_2,SIT_insulator_exp_3} and theoretical~\cite{CooperInsulator_theory_1,CooperInsulator_theory_2} and works offer evidence shows that this insulator~\cite{SIT_insulator_1,SIT_insulator_4} has Cooper pairs present in it. 
%Application of Disorder breaks the translational-invariance symmetry and the magnetic field breaks the time-reversal symmetry of the system. 
Recent experiments~ \cite{SIT1,SIT2,SIT3,SIT4,SIT5,SIT6,SIT7,SIT8,SIT9,SIT10,
SIT11,SIT12,SIT13,SIT14,SIT15} reveal the existence of an intervening 
%topological 
metallic phase, called the ``Bose metal" phase~\cite{BoseMetal1}, lying between the superconducting and insulating phases. Despite considerable effort, the precise nature of such a Bose metal remains unclear. 
%as standard theories do not show any intermediate metallic phase. 
Notable among various theoretical efforts on the Bose metal 
%there are many theories describing this phase.  
is the development of a gauge theory of Josephson junction arrays (JJA)~\cite{diamantini_jja,diamantini_bosemetal}, 
%has found the intermediate metallic phase. It is shown by 
%Diamantini et al. \cite{diamantini_bosemetal} that 
which show the Bose metal can be described by a topological Chern-Simons field theory emerging out of the nonlocal interaction
%winding 
between the quasiparticle and 
%-2e charge and 
vortex excitations of the superconducting system.
%degrees of freedom in the system. 
Diamantini et al.~\cite{diamantini2020} show that such a topological insulating phase possesses, at finite temperatures, a longitudinal conductance mediated by time-reversal symmetry preserved counter-propagating edge modes. 
%Similar nontrivial winding between cooper-pair and vortex gives rise to topological Chern-simons theory for gapped BCS superconductors also. 
Hansson et al. \cite{hos} further established that the abelian Higgs model equivalent of the s-wave superconductor possesses other important features of topological order, i.e., a nontrivial ground state degeneracy revealed on a multiply connected spatial manifold (such as a torus) and charge fractionalization. The degenerate ground states are labelled by topological quantum numbers corresponding to the eigenvalues of Wilson and 't-Hooft loops~\cite{hos,fradkinshenker,fradkin2013field}. More recently, 
the authors of Ref.\cite{moroz} have shown that the gapped topological bulk for s-wave pairing does not possess gapless edge states. Indeed, such a state displays a vanishing Hall conductance. Further, they also study the nature of the topological order in other spin-singlet superconductors~\cite{moroz}. The Cooper pair insulating phase has also been studied in lattice bosonic superconductors using vortex-boson duality~\cite{dasguptahalperin1981,fisherlee1989}, and its topological order investigated in the two dimensional variant~\cite{vestergren2005}.
%More recently, this has been generalized this to the case of other singlet superconductors~\cite{moroz}. The authors of Ref.\cite{moroz} also show that such a gapped topological bulk with gapless edge states lead to a vanishing Hall conductance, arising from time reversal and parity invariance of the Chern-Simons effective field theories that describe these ground states. 
%Nontrivial braiding statistics between Wilson Loop and ‘t Hooft\cite{hooft} loop gives rise to 4-fold topological degeneracy on torus and 2-fold on cylinder \cite{hos,moroz,fradkinshenkar,fradkin2013field}.  All these Chern-simons field theories are translationally invariant, time-reversal invariant and parity invariant. This topological insulating phase is our focus for this work.
%While the phenomenological field theoretic understanding of
\pin 
Insight into such a topologically ordered insulating phase (or Bose metal) from a microscopic approach, however, remains to be developed. Indeed, this is the primary goal of this work.
%, there is no microscopic theory available. In this work, 
Thus, we are going to present microscopic Hamiltonian for this novel phase of quantum matter, demonstrating it to be an insulator with topological ordered gapped ground state manifold. Importantly, we will find that such a state is emergent purely from quantum fluctuations arising from inter-particle interactions. Further, such quantum fluctuations preserve translational and time-reversal invariance, and are not necessarily
%the standard Insulator of the SIT this Insulating behavior is neither 
driven by either disorder or the coupling to an external magnetic field.
\pin
In meeting this goal, we begin with a generalized Hamiltonian for a Fermi liquid with a short-ranged repulsive density-density interaction, as well as attractive pairing term. Suitably rewritten in terms of Anderson pseudospins~\cite{apt}, this corresponds to a reduced BCS Hamiltonian with additional repulsive interactions familiar from Fermi liquid theory.
%with an onsite density-density repulsive term. Using Anderson pseudo spin transformation \cite{apt} we rewrite this Hamiltonian in the Cooper channel ($n_{k\uparrow}=n_{-k\downarrow}$) in terms of pseudo-spins. 
Then, using the unitary renormalization group (URG) technique recently developed by some of us~ 
%shown by Mukherjee et al. 
\cite{MukherjeeMott1,MukherjeeMott2,pal2019,MukherjeeNPB1,
MukherjeeNPB2,MukherjeeTLL}, we resolve in a step-wise manner the quantum fluctuations arising from the non-commutativity between the kinetic energy of the electrons and the inter-particle interactions. This involves
%and Pal et al. \cite{pal2019,pal2020}. In this method, at each step, they 
decoupling one electronic Fock state in the momentum space from all the other states it was connected to, such that the occupation number of the decoupled state is rendered as an integral of motion. The decoupling proceeds in a hierarchical fashion in terms of the kinetic energies of the electrons, from high (near the Brillouin zone edge, UV) to low (near the Fermi surface, IR), and an effective Hamiltonian is generated at every step. For the sake of clarity, we have encapsulated the major aspects of the URG method in Appendix \ref{section:appendix_A}.
%, via quantum fluctuations. This way at each step it generates one Integral of Motions (IOM) and one lower energy effective Hamiltonian. We keep disentangling electrons from high kinetic energy towards the low kinetic energy sector of the bare Hamiltonian until 
As shown in Sec. \ref{section:RG}, the RG flow stops at an IR fixed point, yielding a low energy effective theory of a fixed number of condensed Cooper pairs but without any breaking of the macroscopic $U(1)$ phase (i.e., with vanishing phase stiffness).
%condense shown in section\ref{section:RG}. 
\pin
We find that the emergent fixed point effective theory (eq.\eqref{hcoll01}) involves a non-local renormalised interaction between all pseudospins within the window that is emergent in $k$-space around the erstwhile Fermi surface, and described by a collective zero-mode degree of freedom comprised of these pseudospins. 
%show that the reduced BCS Hamiltonian is the low energy Hamiltonian we get. We do another second step of RG to go to even lower energy, here we find all-to-all zero-mode as a low energy fixed point solution, where all pseudo-spins are talking to each other via a constant renormalized coupling (relevant under RG) with zero kinetic energy term in the Hamiltonian. This entire process does not break any symmetry of the parent Hamiltonian. 
We call this symmetry unbroken phase of condensed Cooper pairs as the \textit{Cooper Pair Insulator} (CPI). 
%This zero-mode is an isotropic Lipkin-Meshkov-Glick (LMG) model \cite{LMG1,LMG2,LMG3}.
%Note this CPI phase is different from the insulating phase\cite{SIT_dis_1} obtained through disorder and magnetic field. 
%\pin
The RG procedure involves a novel energy scale for quantum fluctuations ($\omega$). In keeping with this, the RG phase diagram obtained in Fig.\ref{fig:rg_phase_diagram} clearly displays a quantum phase transition separating a CPI phase (at low $\omega$) and a gapless Fermi liquid metal (at higher $\omega$) for any repulsive interaction. The CPI Hamiltonian corresponds to a collective quantum rotor model coupled to an effective Aharanov-Bohm (AB) flux ($\Phi$). We demonstrate that there exist different ground states of this CPI Hamiltonian related to one another by spectral flow upon tuning $\Phi$, and displays the emergent quantisation of $\Phi$ under the RG flow. Importantly, we gain an idea of the accuracy of our method by numerically benchmarking the energy per particle of the ground state (eq.\eqref{genden}) of the CPI Hamiltonian obtained in the thermodynamic limit with that obtained from exact diagonalisation calculations (Fig.\ref{fig:rg_eg_scaling}).
%The properties of this final RG fixed point Hamiltonian is Cooper-pair number conservation, U(1) phase rotation symmetric, time-reversal invariant, translationally invariant. Thus this Cooper-bound state of CPI is different from the Cooper-pair bound state of a BCS superconductor as the former lacks inter-pair long-range phase coherence.
%\par
\pin
We have studied the topological features of the CPI Hamiltonian in Sec.\ref{section:topofeat}.
%,\ref{section:entfeat}). 
%We have shown, this URG fixed point Hamiltonian ground state energy is very close to the exact diagonalization (ED) ground state energy of the bare Hamiltonian, showing that our URG fixed point Hamiltonian captures most of the physics of the bare Hamiltonian. 
We establish first the gauge theoretic nature of the emergent effective IR theory obtained from the RG by demonstrating that the collective Hamiltonian for the CPI can be written completely in terms of a nonlocal Wilson loop operator. This is not surprising, given the zero-dimensional nature of the effective Hamiltonian, where all degrees of freedom (within the emergent IR window in momentum space) are interacting with one another. 
%In this model every pseudospin is talking to everyone else with a constant strength, thus this is effectively a zero-dimensional pseudo-spin Hamiltonian, where nodes are in momentum space. This effective hamiltonian is like a rotor problem 
The CPI Hamiltonian contains an topological $\theta$-term for a non-zero effective AB flux~\cite{fernandes,cao}. As shown by Yao and Lee~\cite{yaolee}, this zero-dimensional $\theta-$term is in correspondence to a $U(1)$ Chern-Simons topological term in two spatial dimensions. This establishes the effective theory for the CPI as the microscopic origin of the phenomenological $U(1)$ Chern-Simons gauge theories obtained by various people earlier~\cite{diamantini_jja,diamantini_bosemetal,hos,moroz}.  
%also get to similar topological Chern-Simons theory upon coupling the U(1) symmetry broken BCS superconductor with a dynamical gauge field. They find 
We then reveal the nontrivial topological degeneracy and charge fractionalization signatures of the topologically ordered CPI ground state manifold through a flux-insertion spectral flow argument~\cite{MukherjeeNPB2,pal2019,liebmattis,liebmattis_hastings}. These spectral flow arguments reveal a plateaux-like quantisation of the number of Cooper pairs ($N$, a topological quantum number) upon tuning $\Phi$ through integer values (Fig.\ref{hall}), with the passage between the plateaux signifying topological transitions (at half-integer values of $\Phi$). The collapse of the energy spectrum of the CPI phase with increasing $\Phi$ (Fig.\ref{fig:pd}) yields another view of the transition between the CPI and metallic phases. As shown in Fig.\ref{fig:hcc}, the CPI state is found to possess large helicity cross-correlations ($\Upsilon$, a signature of inter-helicity two-particle scattering processes). Finally, in Fig.\ref{fig:thermals1}, we track the passage into the CPI ground state by starting from a finite (but small) temperatures and lowering towards $T=0$.
%Here in this work, we take a different route. Following \cite{MukherjeeNPB2} \cite{pal2019,liebmattis,liebmattis_hastings} we show that our RG fixed point Hamiltonian can be written as a U(1) gauge theory\ref{section:topofeat} purely in terms of non-local Wilson loops, showing nontrivial ground state degeneracy and charge fractionalization.
\par\noindent
In Sec.\ref{section:entfeat}, we present a detailed analysis of the entanglement features of the CPI ground state. The entanglement spectrum (ES) computed for the lowest CPI ground state ($\Phi=0$) is found to be doubly degenerate for all partitions of the system (Fig.\ref{fig:01flux}), reflecting the additional particle-hole symmetric nature of this CPI ground state. This degeneracy is lifted at the first topological transition ($\Phi=1/2$, Fig.\ref{fig:01flux}). Interestingly, all other CPI ground states (corresponding to non-zero positive integer values of $\Phi$) are observed to show the degeneracy of the ES for only the equipartitioned system (Fig.\ref{fig:metalnearflux}); the degeneracy is again lifted at the transitions. We find that the bipartite entanglement entropy (EE) has a logarithmic dependence on the subsystem length $L$ (Fig.\ref{zeroplatscaling}). Further, Fig.\ref{platent} shows that the $\Phi=0$ plateau possesses the largest EE, and that this is rapidly lowered to zero as $\Phi$ is tuned through various plateaux towards the gapless metal. The topological transitions between the CPI ground states display universal signatures in the variations of their EE with system size $L$ and flux $\Phi$: EE shows a non-monotonic variation with $L$ (Fig.\ref{fig:FirstPlatTransEE}), with a peak value at $L\equiv L^{*}$ that is universal as $\Phi$ is tuned (Fig.\ref{fig:TransEEpeak}). Further, for $L < L^{*}$, the EE versus $L$ data falls onto a universal curve (Fig.\ref{fig:FirstPlatTransEE}). Finally, $L^{*}$ approaches the equipartition value at $\Phi$ is tuned close to the transition between the CPI state and the metal (Fig.\ref{fig:TransEEpeak}).
Variations of the EE for the CPI ground states and the transitions with temperature also show interesting signatures (Fig.\ref{finiteEE}). The EE versus temperature curves for all values of $\Phi$ corresponding to a given ground state finally merge at $T\to 0$. Further, the EE curves for all transitions are clearly distinct from those associated with the CPI ground states. Interestingly, for large temperatures, the dominance of thermal fluctuations in smearing out the distinction between the CPI ground states and the transitions can be seen in the fact that \textit{all} EE curves have a linear variation against temperature and with a universal slope.
\pin
Having studied the topologically ordered CPI phase in detail, in in Sec.\ref{section:cpi2bcs}, we turn to its connection with the symmetry broken BCS ground state. We note that the low energy effective theory for the CPI phase corresponds to an isotropic Lipkin-Meshkov-Glick (LMG) model \cite{LMG1,LMG2,LMG3}. Recent studies \cite{van,van_thesis} have also revealed the existence of such an effective Hamiltonian within the global $U(1)$ symmetry broken BCS phase, 
%there is a number conserving zero-mode CPI phase, that is 
arising from degrees of freedom that are singular under the Bogoliubov-Valatin transformation
%. It has been shown that this singular part contains 
and corresponding to a ``thin spectrum" or Anderson tower of states \cite{anderson_tower}. Upon taking the system size to the thermodynamic limit, the collapse of this thin spectrum is believed to engender the spontaneous breaking of the $U(1)$ symmetry. On the other hand, it is believed that in a mesoscopic superconducting grain, the presence of a large charging term helps in making the thin spectrum robust~\cite{vanwezelgrain}.
Thus, in Appendix B, 
%\ref{section:appendix_B}, 
we investigate the passage from the CPI phase to the BCS phase under RG upon adding a global $U(1)$ symmetry breaking term in the CPI Hamiltonian.  
\pin 
In keeping with this, 
%the next goal of our work is in section \ref{section:cpi2bcs}, how to go back to the BCS superconductivity from this CPI phase. We explicitly add global U(1) symmetry breaking term in the CPI Hamiltonian. We use the URG to show that this symmetry breaking term becomes RG relevant thus getting an effective symmetry broken theory of cooper-pairs.  We studied this journey towards this symmetry broken phase by tuning the strength of the symmetry breaking field. 
we show in Fig.\ref{nfluc} of Sec.\ref{section:cpi2bcs} that the $U(1)$ symmetry breaking field promotes fluctuations in the number of Cooper pairs, while lowering the fluctuations in the conjugate global phase. Further, in Fig.\ref{hcchee}, we show that symmetry breaking effectively destroys the helicity cross-correlation ($\Upsilon$) among the Cooper-pairs under RG. The BCS wavefunction is a product state in momentum space, with vanishing inter-$k$ entanglement between Cooper pairs and maximum inter-spin entanglement due to the singlet configuration of each pair~\cite{bcs_puspus,bcs_tullio}. We find that an increasing symmetry breaking field lowers the inter-$k$ entanglement between Cooper pairs present in the CPI to zero, while raising the inter-spin entanglement to its BCS value (Fig.\ref{fig:interkandinterspinEE}). 
%show that for a large enough symmetry breaking field we get BCS superconductivity with the above entanglement properties. 
%By putting this symmetry-breaking field if you study 
Further, the equal-size partitioned entanglement entropy shows a monotonic decrease of entanglement for the CPI ground states (and their intervening transitions) with an increasing symmetry breaking field (Fig.\ref{fig:EEsbroken}), while the metallic phase shows a non-monotonic variation.
\pin
%To experimentally detect this 
Another way by which to distinguish the CPI phase from the BCS superconductor lies in investigating the effects of a Josephson coupling. Thus, in a second part of Sec.\ref{section:cpi2bcs}, we consider the case of two CPI systems whose bulk is coupled via Josephson coupling (i.e., we are ignoring all effects from gapless edge states), and one of whom is placed in an $U(1)$ symmetry breaking field. In Fig.\ref{fig:phasecoherence}, we find that while a Josephson coupling between two pure CPI systems (i.e., with the symmetry breaking of the second CPI system switched off) does not lead to any phase stiffness in another another, breaking the symmetry in one of them does give a proximity induced phase coherence in the other. Indeed, an increasing symmetry breaking field together with a large Josephson coupling turn a CPI into a phase stiff ground state. The generation of a Josephson current through the phase-locking, however, requires the separate symmetry breaking of the two individual subsystems. This is demonstrated in Fig.\ref{fig:EG_vs_phi}, through the observation of a periodic variation of the ground state energy $E(\phi)$ with the phase difference $\phi$ upon the introduction of separate non-zero symmetry breaking fields in both CPI systems. In addition to the recent transport measurements of Ref.\cite{diamantini2020} on the Bose metal, these findings serve as predictions for the experimental search of the CPI state of quantum matter.
%consider the case of one CPI system coupled to a second one with a symmetry breaking field. if you break the symmetry of one CPI phase then coherence gets generated in the other one. Thus we find symmetry breaking of one phase induced symmetry breaking to another one. Then we coupled one CPI phase with the BCS phase via the same coupling. We find for a large enough Josephson coupling strength CPI also turns into a BCS superconductor. Though via induction non-zero phase correlation can be induced, no current can travel from BCS to CPI phase until you break the symmetry of the CPI phase also.
\pin
Next, in Sec.\ref{section:entrg}, we have carried out an entanglement RG using the technique developed by us in Refs.\cite{MukherjeeTLL,MukherjeeHolography}. 
%Using the Unitary operations of the URG formalism we reached obtained the RG fixed point Hamiltonian for both the symmetry unbroken and symmetry broke bare Hamiltonian. 
This strategy can be described briefly as follows. Our URG analysis has helped obtain the IR fixed point Hamiltonians for the CPI and BCS phases, their ground state wavefunctions as well as the unitary transformations of the RG flow that led to them. for corresponding CPI and BCS wavefunctions. The quantum circuits corresponding to these unitary transformations is shown for the CPI and BCS cases in Figs.\ref{qcirccpi} and \ref{qcircbcs} respectively.
%A very special feature of this URG is the availability of those unitary operators. 
Now, applying these unitary operations in reverse on a given IR wave function, we obtain a family of wave functions under RG leading towards that pertaining to the UV theory. We compute the entanglement entropy of various sizes of partitions from this set of wavefunctions, revealing the evolution of the entanglement entropy with RG. As shown in Fig.\ref{blockentcpi}, the EE for the constituent subblocks of the emergent CPI ground state are clearly distinguished from that for all other partitions: their EE varies little under RG from UV to IR, while that of all others is lowered under RG from UV to IR. Further, this analysis reveals a remarkable hierarchy of scales of entanglement possessed by the CPI ground state. This hierarchy of scales of entanglement gradually collapses upon tuning a symmetry breaking field (Fig.\ref{entRGsmallsymmbreak}), until it is no longer present in the BCS ground state (Fig.\ref{entRGbigsymmbreak}).
%of each reverse RG steps we show how UV to IR entanglement variation happens, and 
Further, we can distinguish scaling towards the BCS and CPI ground states under RG flow. 
%is different from CPI in terms of this RG scaling of entanglement. 
Finally, we conclude in Sec.\ref{section:conclusion} with a discussion of some future directions.
%
%\par
%Thus we have shown a new phase of matter CPI in this work, which is an insulating phase where full translation and time-reversal symmetry is intact. This is also susceptible to symmetry breaking as we mentioned, this leads to superconductivity. This way one can study the CPI to Insulator transition and CPI to superconductor transitions.

\section{Effective theory of the Cooper Pair Insulator (CPI)}
\label{section:RG}
\noindent
We begin by deriving an effective Hamiltonian for an insulating state of matter comprised of a fixed number of Cooper pairs (referred to as the Cooper pair insulator, or CPI, in the introduction). For this, we will carry out a renormalisation group (RG) calculation on a system of electrons in two dimensions with a generalised pairing Hamiltonian, $H=\sum_{q}H_{\textrm{pair}}^{q}$, where
%\color{blue}
\begin{eqnarray}
H_{\textrm{pair}}^{q}&=&\displaystyle\sum_{k} \epsilon_{k} n_{k}
-\displaystyle\sum_{k\neq k',\sigma} |W^q_{kk'}| c_{k-q,\sigma}^{\dagger}c_{-k,-\sigma}^{\dagger}c_{-k',-\sigma}c_{k'-q,\sigma}\nonumber\\ 
&&+~ U\sum_{k\neq k'} (n_{k}-1/2)(n_{k'}-1/2)~,
\label{pairing}
\end{eqnarray}
\color{black}
where with $q$ denotes the pair-momenta, $\epsilon_{k}$ the kinetic energy for electrons about a circular Fermi surface, $-|W^q_{kk'}|$ is the attractive pairing interaction, $U(>0)$ a repulsive density-density interaction and $n_{k}=\sum_{\sigma}c^{\dagger}_{k,\sigma}c_{k,\sigma}$. Note that the case of a constant $|W^{q}_{kk'}|~\forall (k,k',q)$ corresponds to the attractive Hubbard model~\cite{micnas1990}. We proceed by using Anderson's pseudospin contruction in the subspace $n_{k-q,\uparrow}=n_{-k,\downarrow}$~: 
\begin{eqnarray}
\vec{S}_k=\frac{1}{2} \phi_k.\vec{\tau}.\phi_k^{\dagger}~,
\end{eqnarray}
where $\vec{\tau}=(\tau^z,\tau^x,\tau^y)$ are the Pauli matrices and $\phi_k=(c_{k\uparrow},c^{\dagger}_{-k\downarrow})$.
The pseudospins obey the standard commutation relation for spin-1/2:~ $[S_k^i,S_k^j]=i\epsilon_{jkl} S_k^j$. Then, we write $H_{\textrm{pair}}^{q}$ as
%
%\begin{eqnarray}
%S_{k,q}^{+} &=& c_{-k,\downarrow}c_{k-q,\uparrow}\\
%S_{k,q}^{-} &=& c_{k-q,\uparrow}^{+}c_{-k,\downarrow}^{+}\\
%S_{k,q}^{z} &=& \frac{1}{2}\bigg( 1-n_{k-q,\uparrow}-n_{-k,\downarrow} \bigg)
%\end{eqnarray} 
%
%%\begin{eqnarray}
%%S_{k,q}^{+}=c_{k \uparrow}c_{-k+q \downarrow}
%%\end{eqnarray}
%
%Start with the general FL hamiltonian 
%\begin{eqnarray}
%H_q=\displaystyle\sum_{k} e_k \bigg(n_{k,\uparrow}+n_{k,\downarrow}\bigg)+\displaystyle\sum_{k} |V_{k,q}| c_{k-q,\uparrow}^{\dagger}c_{-k,\downarrow}^{\dagger}c_{-k,\downarrow}c_{k-q,\uparrow}\nonumber
%\end{eqnarray}
%We write down the four Fermionic interacting q-pair momenta hamiltonian $H_q$ as {Now  we can use the Anderson pseudo spin transfomrmation and rewrite the Hamiltonian as }
%Now as we are working within that subspace , we can rewrite the kinetic energy term as
%\begin{eqnarray}
%&&e_{k-q} n_{k-q,\uparrow} + e_{-k} n_{-k,\downarrow}\nonumber\\
%&=&\frac{1}{2}e_{k-q} ( n_{k-q,\uparrow} + n_{-k,\downarrow})+ \frac{1}{2}e_{-k} (n_{k-q,\uparrow}+n_{-k,\downarrow})\nonumber \\
%&=& \frac{1}{2}\bigg(e_{k-q} + e_{-k}\bigg)\bigg(n_{k-q,\uparrow}+n_{-k,\downarrow}\bigg)
%\end{eqnarray}
%Thus 
%
%\begin{eqnarray}
%\displaystyle\sum_{k}e_k \bigg(n_{k,\uparrow}+n_{k,\downarrow}\bigg)&=&\displaystyle\sum_{k}\frac{1}{2}\bigg(e_{k-q} + e_{-k}\bigg)\bigg(n_{k-q,\uparrow}+n_{-k,\downarrow}-1\bigg)+\frac{1}{2} \displaystyle\sum_k \bigg(e_{k-q} + e_{-k}\bigg)
%\end{eqnarray}
%Thus the total general Hamiltonian can be written as 
%\color{blue}
\begin{eqnarray}
H_{\textrm{pair}}^{q}&=&-\displaystyle\sum_{k}\tilde{\epsilon}_{k,q}(S_{k,q}^z-\frac{1}{2}) -\sum_{k\neq k'} \frac{|W_{kk'}^{q}|}{2}(S_{k,q}^- S_{k',q}^+ +\textrm{h.c.})\nonumber\\
&&~+U\sum_{k\neq k'}S_{k,q}^zS_{k',q}^z~, 
\label{rghamiltonian}
\end{eqnarray}
\color{black}
where $\tilde{\epsilon}_{k,q}=\epsilon_{-k}+\epsilon_{k+q}$ is the kinetic energy for a pair of electrons. In order to ensure the extensivity of the model, $|W_{kk'}^q|=|V_{kk'}^q|/N$~(where $N$ corresponds to the total number of pseudospins, and hence $2N$ the total number of electrons). The special case of $H_{\textrm{pair}}^{q=0}$ with $|W_{kk'}^{q=0}|$ and $U=0$ is called the Richardson pairing model~(see Ref.\cite{dukelsky2004colloquium} and references therein).
%(CITE the original paper by Richardson, and Dukelsky et al., Rev. Mod. Phys. {\bf 76}, 643 (2004)).
\par\noindent
Following the strategy developed in Refs.\cite{MukherjeeMott1,MukherjeeMott2,pal2019,MukherjeeNPB1,
MukherjeeNPB2,MukherjeeTLL}, we now carry out a renormalisation group analysis on $H_{\textrm{pair}}^{q}$; see appendix~\ref{section:appendix_A} for details. The RG equations obtained for $\tilde{\epsilon}$ and $|W_{kk'}^{q}|$ are
\begin{eqnarray}
\frac{\Delta \tilde{\epsilon}_{k',q}^{(j)}}{\Delta \log \frac{\Lambda_j}{\Lambda_0}} &=& ~\frac{1}{4}\frac{|W_{k_{\Lambda}k'}^{(j)}|^2}{\bigg(\omega-\frac{\tilde{\epsilon}_{k_{\Lambda},q}^{(j)}}{2} - \frac{U}{4}\bigg)}~,\\
\frac{\Delta |W_{k'k''}^{q,(j)}|}{\Delta \log \frac{\Lambda_j}{\Lambda_0}}&=&- \frac{1}{4}~\frac{|W_{k_{\Lambda}k'}^{q,(j)}||W_{k_{\Lambda}k''}^{q,(j)}|}{\bigg(\omega-\frac{\tilde{\epsilon}_{k_{\Lambda},q}^{(j)}}{2} - \frac{U}{4}\bigg)}~,
\label{rgeqs}
\end{eqnarray}
%\color{black}
%(CHANGE THE ABOVE RG DIFFERENTIAL EQUATIONS TO DIFFERENCE EQUATIONS)
%\textcolor{blue}{After disentangling one single particle state in every step of RG we are making number of pseudospin within the emergent space one less. Thus at any arbitary RG step $|W_{k'k''}^{q(j)}|=|V_{k'k''}^{q(j)}|/N^{(j)}$}, where $N^{(j)}$ is the number of pseudo spin at $j^{th}$ RG step. 
where the index $(j)$ represents the RG step number, $|W_{k'k''}^{q,(j)}|=|V_{k'k''}^{q,(j)}|/N^{(j)}$ (for $N^{(j)}$ being the number of remnant pseudospins at the $j^{th}$ RG step) and $k_{\Lambda}$ the momentum at a $k$-space window ($\Lambda$) lying on a radial and around the circular Fermi surface. The symbol $\omega$ represents an energy scale for the quantum fluctuations that lead to UV-IR mixing. Further, we note that the RG step index $j$ starts from the number of pseudospin ($N$) lying within the bare window $\Lambda_{0}$ and proceeds to smaller values. At every step of the RG, a pseudospin with momentum $k_{\Lambda}$ lying on a given direction radial to the Fermi surface is disentangled from the rest ($\forall k<k_{\Lambda}$); the $U(1)$ symmetry of the circular Fermi surface ensures that the RG is carried out simultaneously for all pseudospins with momentum $k_{\Lambda}$. Note that the repulsive coupling $U$ does not flow under RG, as two-particle quantum fluctuations do not lead to the renormalisation of this term. Instead, it appears as a Hartree-shift in the pseudospin Greens function $G_{ps}=[\omega-\frac{\tilde{\epsilon}_{k_{\Lambda},q}^{(j)}}{2} - \frac{U}{4}]^{-1}$ present in the RG equations given above~\cite{MukherjeeMott1,MukherjeeMott2,pal2019,MukherjeeNPB1,
MukherjeeNPB2,MukherjeeTLL}.
\par\noindent
It can be seen that the normalization for $\tilde{\epsilon}_{k',q}^{(j)}$ is RG relevant for $\omega>(\tilde{\epsilon}_{k_{\Lambda},q}/2+U/4)$, while that for $|W^{q,(j)}_{k'k''}|$ is RG relevant for $\omega<(\tilde{\epsilon}_{k_{\Lambda},q}^{(j)}/2+U/4)$. Given that $\epsilon_{-k}=\epsilon_{k}$ for a circular Fermi surface, it is easily seen that $\tilde{\epsilon}_{k,q}^{(j)}\geq 2\epsilon_{k}^{(j)}$. Therefore, given the denominator $\omega - \tilde{\epsilon}_{k_{\Lambda},q}^{(j)}/2-U/4$ in both RG equations, the leading RG relevant $q$-sector for the lowest quantum fluctuation energyscale ($\omega =0$) corresponds to the case of $q=0$ (i.e., Cooper pairs with zero centre of mass momentum). Thus, we will henceforth study only the case of $\tilde{\epsilon}_{k,q=0}^{(j)}\equiv \tilde{\epsilon}_{k}^{(j)}=2\epsilon_{k}^{(j)}$ and $H_{\textrm{pair}}^{q=0}\equiv H_{\textrm{pair}}^{0}$. 
%As defined $\epsilon_{k',q} = \epsilon_{k'-q}+\epsilon_{-k'} $.
\par\noindent 
%One can define $q^{th}$ mode of  as
We define mode decompositions of the dispersion $\epsilon_{k}^{(j)}$ and the pairing coupling $|V_{kk'}^{(j)}|$ as follows
\par\noindent
%Thus we get the RG stable fixed point Hamiltonian 
%\begin{eqnarray}
%H_{\textrm{pair}}&=&-2\displaystyle\sum_{k}\epsilon_{k}^*(S_{k}^z-\frac{1}{2}) -\sum_{k\neq k'} \frac{|W_{kk'}^{*}|}{2}(S_{k}^- S_{k'}^+ +\textrm{h.c.})\nonumber\\
%&&~+U\sum_{k\neq k'}(S_{k}^z-\frac{1}{2})(S_{k'}^z-\frac{1}{2})~, 
%\end{eqnarray}
%At the RG fixed point the effective Hamiltonian contains say $N^*$ number of pseudo spins and $|W_{kk'}^{*}|=|V_{kk'}^{*}|/N^*$
%a given observable $\hat{O}$ in the 2D $\vec{k}$-space as 
\begin{eqnarray}
\hspace*{-1cm}
\bar{\epsilon}_{l}^{(j)}=\frac{1}{\sqrt{N^{(j)}}}\sum_{\vec{k}}e^{ikl}\epsilon_{k}^{(j)}~,~\bar{V}_{ll'}^{(j)}=\frac{1}{N^{(j)}}\sum_{k,k'}e^{i(kl +k'l')}|V_{kk'}^{(j)}|~,
%\hat{\tilde{O}}_{q_1,q_2}=\sum_{k} e^{i\vec{q}.\vec{k}} \hat{O}_{k_1,k_2}~,
\end{eqnarray}
\color{black}
such that the RG equations for $\bar{\epsilon}_{l=0}^{(j)}$ and $\bar{V}_{l=0,l'=0}^{(j)}$ are observed to dominate under the RG flow over all other modes for a thermodynamically large system~:
\begin{equation}
\textrm{Re}\bigg(\frac{\Delta \bar{\epsilon}_{l=0}^{(j)}}{\Delta \bar{\epsilon}_{l\neq 0}^{(j)}} \bigg) > 1~,~\textrm{Re}\bigg(\frac{\Delta \bar{V}_{l=0,l'=0}^{(j)}}{\Delta \bar{V}_{l\neq 0,l'\neq 0}^{(j)}} \bigg) >1~.
\end{equation}
%Where $\vec{q}=(q_1,q_2..q_n)$ and $\vec{k}=(k_1,k_2 . . k_n)$. 
%and we are interested in $\hat{\tilde{O}}_{q_1,q_2}=\{\tilde{\epsilon}_{q},\tilde{V}_{qq'}\}$
%Here under RG one can see that the zero mode is most relevant than other non-zero modes by looking at the RG of the ration of zero mode to other non-zero modes 
%\begin{eqnarray}
%\Delta\hat{\tilde{O}}_{(0,0..0)}/\Delta\hat{\tilde{O}}_{(q_1,q_2..q_n)} > 1 
%\end{eqnarray}
%as $|V_{k,k'}|>0, \forall k,k'$. In our case . At the thermodynamic limit one can see that at $\omega=0$ the zero mode is having highest weight within the total hamiltonian. 
The zero mode $\bar{\epsilon}_{0}^{(j)}$ is related to the center of mass kinetic energy $\bar{\epsilon}^{(j)}$~: $\bar{\epsilon}_{0}^{(j)}=(\sum_k \epsilon_{k}^{(j)})/\sqrt{N^{(j)}} = \sqrt{N^{(j)}} \bar{\epsilon}^{(j)}$. 
%Here $\bar{\epsilon}=(\sum_k \epsilon_k)/N$. 
Similarly, the zero mode $\bar{V}_{00}^{(j)}$ is connected to it's center of mass value~: $\bar{V}^{(j)}=(\sum_{kk'} V_{kk'}^{(j)})/(N^{(j)})^{2}=V_{00}^{(j)}/N^{(j)}$. Thus, the RG relations of these zero modes is equivalent to the study of the center of mass degrees of freedoms
%. We get the RG equations in terms of these degree of freedoms. Here \textcolor{blue}{$\bar{W}=\bar{V}/N$} are
\begin{eqnarray}
\Delta\bar{\epsilon}^{(j)}&=& \frac{1}{4}~\frac{|\bar{W}^{(j)}|^2}{\bigg(\omega-\frac{\bar{\epsilon}^{(j)}}{2}-\frac{U}{4}\bigg)}=-\Delta|\bar{W}^{(j)}|~,
%\Delta|\bar{V}^{(j)}|&=&-\frac{|\bar{V}^{(j)}|^{2}}{\omega-\frac{\bar{\epsilon}^{(j)}}{2} + \frac{U}{4}}=-\Delta\bar{\epsilon}^{(j)}~.
%\nonumber\\
%\frac{\Delta \epsilon}{\Delta |V|} = -1 ~&,&~  \epsilon^{(j)}=C-|V|^{(j)}
\label{U1RGeq}
\end{eqnarray}
where $\bar{W}^{(j)}=\bar{V}^{(j)}/N^{(j)}$. In this way, we observe below the emergence of the well known reduced BCS model~\cite{bcs} at the stable fixed point of the RG eq.\eqref{U1RGeq}.
\par\noindent
The relation between the two RG equations (eq.\eqref{U1RGeq}) leads to a RG invariant: $\bar{\epsilon}^{(j)}+|\bar{W}|^{(j)}=C~,~C\in\mathcal{R}$. From this invariant, it can now be seen that when the kinetic energy $\bar{\epsilon}$ is RG relevant, the attractive coupling $|\bar{W}|$ is RG irrelevant and vice versa. We can now write the effective Hamiltonian obtained at the stable fixed point of the RG flow as 
%\textbf{\textcolor{blue}{Put $N ~ \rightarrow ~ N^*$}}
%in the leading order upto a constant term
\begin{eqnarray}
H_{coll}&=&-\frac{2\bar{\epsilon}^{*}}{N^{*}}\displaystyle\sum_{k}(S^z_{k}-\frac{1}{2})-\frac{\bar{V}^{*}}{2N^{*}}\displaystyle\sum_{k\neq k'}(S_{k}^{+}S_{k'}^{-} + \textrm{h.c.})\nonumber\\
&&~+U\sum_{k\neq k'}S_{k,q}^zS_{k',q}^z~,
\label{eq:hcoll}
\end{eqnarray}
where $\bar{\epsilon}^{*}$, $\bar{V}^{*}$ and $N^{*}$ are the fixed point values of $\bar{\epsilon}^{(j)}$, $\bar{V}^{(j)}$ and $N^{(j)}$ respectively reached at the endpoint of the RG flow (and $\bar{V}^{*}/N^{*}=C-\bar{\epsilon}^{*}$). Finally, by defining the composite pseudospin $\vec{S}=\sum_k \vec{S}_k$, we can rewrite the Hamiltonian $H_{coll}$ (upto additive constants) as 
\begin{eqnarray}
H_{coll}&=& -\frac{2\bar{\epsilon}^*}{N^{*}} S^z -\frac{\bar{V}^*}{2N^{*}} (S^+ S^-+ S^-S^+)+US^{z 2} \nonumber \\
&=& -\frac{2\bar{\epsilon}^*}{N^{*}} S^z -\frac{\bar{V}^*}{N^{*}} (S^2-S^{z 2})+US^{z 2}~.
\label{collham}
\end{eqnarray}
%where $\bar{\epsilon}^{*}$ and $\bar{V}^{*}/N^{*}=C-\bar{\epsilon}^{*}$ are the fixed point values reached at the endpoint of the RG flow.
%For simplicity from now onwards we call $\tilde{\epsilon}_0\rightarrow \epsilon$ and $\tilde{V}_{00}\rightarrow V$. 
While the first term arises from the electronic kinetic energy, the second the potential energy saved by the formation of pairs (i.e., the condensation energy) and the third represents the repulsive charging energy cost of the electrons that form the Cooper pairs. Note that the Hamiltonian eq.\eqref{eq:hcoll} has the global U(1) symmetry of the generalised pairing Hamiltonian eq.\eqref{pairing}. This is expected, as RG transformations are symmetry preserving.
\par\noindent
We present the RG phase diagram below in Fig.\ref{fig:rg_phase_diagram} by a numerical solution of the RG equations for the electronic dispersion along a radial to the circular Fermi surface being $\epsilon_{k}=2t\cos(k)$, a bare window near the Fermi energy $v_{F}\Lambda_{0}=0.3t$, a constant bare attractive coupling $|V^{q}_{kk'}|=4t/N$ and the total number of pseudospins $N=51$. The phase diagram is presented in the plane of the effective quantum fluctuation energy scale $\omega$ and the repulsive coupling $U$ (and both are in units of the kinetic energy bandwidth $4t$). It clearly shows that the Cooper pair insulator (CPI) is stabilised at lower values of $\omega$ for all $U$, 
%where $\bar{\epsilon}^{*}\ll\bar{V}^{*}$ (strong coupling), such 
and that a metallic phase (lying at higher values of $\omega$) is obtained through a quantum phase transition into a gapless Fermi liquid metallic phase. 
%(where $\bar{\epsilon}^{*}\gg\bar{V}^{*}$, weak coupling). 
\begin{figure}
\includegraphics[scale=0.36]{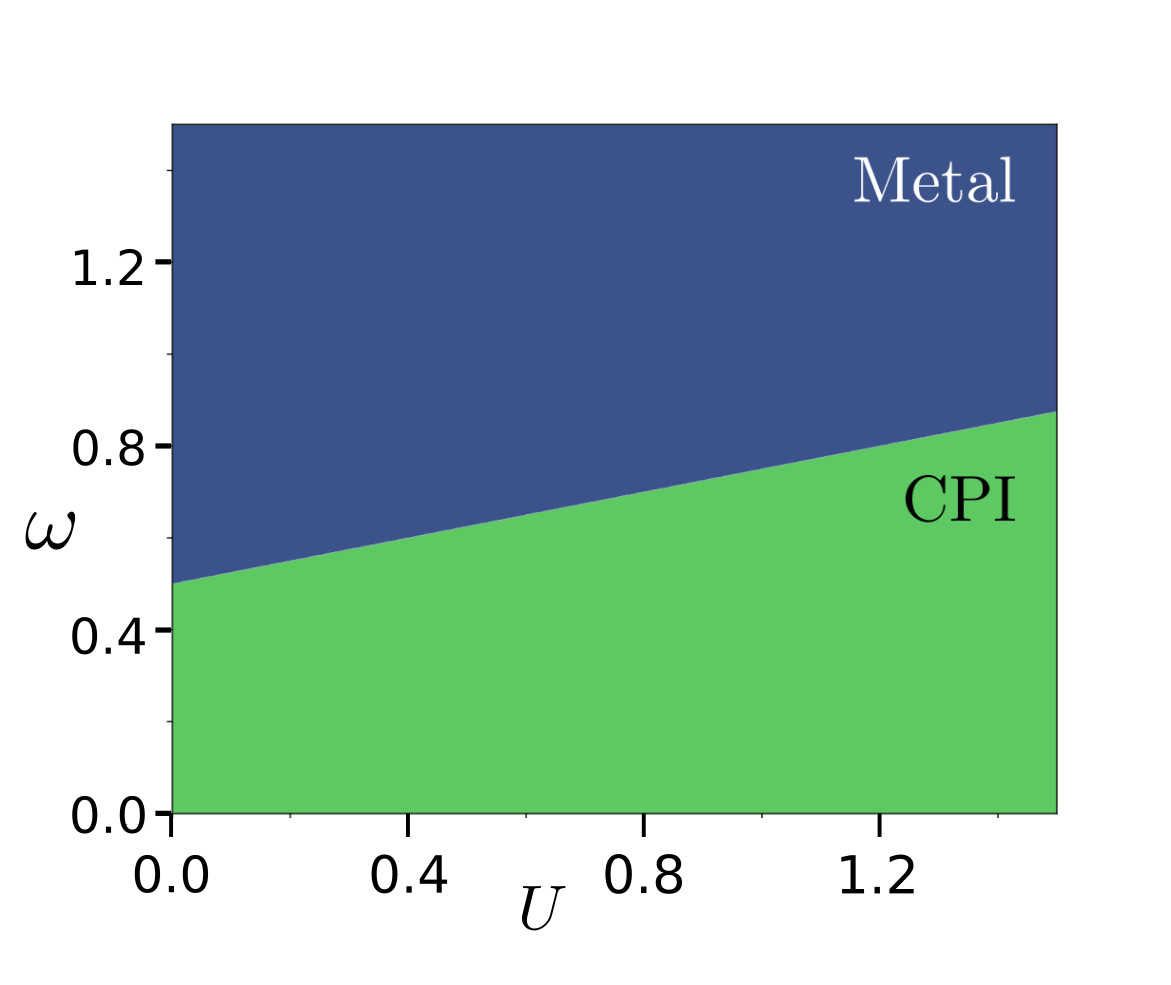}
\caption{Renormalisation group phase diagram for the effective pairing Hamiltonian given in eq.\eqref{rghamiltonian}. The y-axis represents the energyscale ($\omega$) for quantum fluctuations  that are resolved under the RG flow, while the x-axis represents the repulsive interaction ($U$) in the parent metal (whose electrons feel the additional attractive pairing). The phase diagram clearly shows the existence of an emergent Cooper pair insulator (CPI) phase at low $\omega$ (i.e., energyscales proximate to the Fermi surface of the parent metal) for all $U\geq 0$.}
\label{fig:rg_phase_diagram}
\end{figure}
\par\noindent
For the sake of simplicity, we will henceforth focus on the case of $U=0$~\cite{van,LMG1,LMG2,LMG3,vidalspectrum} 
\begin{eqnarray}
H_{coll} &=& -\frac{2\bar{\epsilon}^{*}}{N^{*}} S^{z} -\frac{\bar{V}^{*}}{N^{*}} (S^{2} - S^{z 2})~,\nonumber\\
&=& \frac{\bar{V}^{*}}{N^{*}} (S^{z} - \Phi)^{2} - \frac{\bar{V}^{*}}{N^{*}} S^{2}~,
\label{hcoll01}
\end{eqnarray}
where $\Phi\equiv \bar{\epsilon}^{*}/\bar{V}^{*}$ and we have ignored a constant ($\propto (\bar{\epsilon}^{*}/N^{*})^{2}$). Further, for $N^{*}$ pseudospins, $0\leq S (\in \mathcal{Z})\leq N^{*}/2$, $-S\leq S^{z} (\in \mathcal{Z})\leq S$ and the number of Cooper pairs is given by $N_{c}=S-S^{z}$. It is easily seen that both $S$ and $S^{z}$ commute with $H_{coll}$.  Further, $H_{coll}$ arises a global collective angular momentum degree of freedom, $S^{z}$, and possesses the form of a quantum particle whose dynamics is confined to a circle and coupled to an (effective) dimensionless Aharanov-Bohm (AB) flux $\Phi$.
As well will discuss in a later section, this points to a topological property possessed by the ground state manifold of its Hilbert space.
\par\noindent
We note here, however, that the emergent Hilbert space corresponding to $H_{coll}$ possesses the property of \textit{spectral flow}. First, observe that minimisation of the energy is achieved under RG flow for the case of $\Phi=\bar{\epsilon}^{*}/\bar{V}^{*} \to 0+$
%One can also see that 
by a ground state 
%(i.e., for a relevant coupling $V$ reaching strong coupling in the IR limit) 
possessing $S=N^{*}/2$ and $S^{z}=0$, corresponding to the largest number of Cooper pairs ($N_{c}=N^{*}/2$). Spectral flow refers to the existence of ground states, corresponding to other positive integer values of $S^{z}$, that can be reached under RG for values of the effective AB flux $\Phi\gg 1$ flowing to the same positive integer. 
%(i.e., towards weak coupling values of $V$ reached under RG). 
%Therefore, we will tune the ratio $\bar{\epsilon}/\bar{V}$ in all that follows.
As we will now see, this emergent quantisation of the effective AB flux $\Phi$ under RG also provides a relation between the RG invariant $C$ and the quantum fluctuation scale $\omega$. As shown in Appendix \ref{section:appendix_A}, 
%\textcolor{blue}{\textit{RG invariant: PUT IN THE SPECTRAL FLOW ARGUMENT HERE. USE IT TO SHOW THE EMERGENT QUANTISATION OF THE AB FLUX. }} From the denominator of the RG equations we can see, 
the final fixed point value of $\bar{\epsilon}^*$ is given by $\bar{\epsilon}^*=2\omega-\frac{U}{2}$. Using this together with the relation for the RG invariant is $C=\bar{\epsilon}^{*}+|\bar{W}|^{*}$, we find the effective flux at the IR fixed point is 
\begin{eqnarray}
\Phi &=& \frac{\bar{\epsilon}^*}{|\bar{V}^*|}=\frac{2\omega-\frac{U}{2}}{N^*|\bar{W}^*|}=\frac{2\omega-\frac{U}{2}}{N^*(C-2\omega+\frac{U}{2})} \equiv n \in \mathcal{Z}~.
\end{eqnarray}
For the case of $U=0$, this leads to 
%\begin{eqnarray}
%A&=&\frac{2\omega}{N^*(C-2\omega)}= n \in\mathbb{Z}
%\end{eqnarray}
%This requires $2\omega/(C-2\omega)=z\in \mathbb{Z}$. This quantizes $\omega$ and $C$ as following
\begin{equation}
C=2\omega (\frac{1+nN^{*}}{nN^{*}}) \to 2\omega~~\textrm{for}~n\gg 1~.
\end{equation}
%\item $$ for a fixed $\omega$, and 
%\item $\omega_z=\frac{zC}{2(1+z)}$ for a fixed $C$. 
%\end{enumerate}
%\par\noindent
%N is the number of electron within the Window W around the Fermi surface , then $N/2$ is the number of maximum possible cooper pairs.Here one can see that $S,S_z$ both commutes with the Hamiltonian. We are interested in the low energy Ground state properties.From the hamiltonian one can see that in the ground state configuration $S$ is larest and $S_z=\frac{\epsilon}{V} $. Thus one can see that for $\epsilon = 0$ the $S_z=0$ is the ground state that has the largest number of cooper pair. Now let me recall the definition of this pseudo spins
%\begin{eqnarray}
%S_k^{+}=c_{-k\downarrow}c_{k\uparrow} ~,~ S_k^z=\frac{1}{2}(1-n_{-k\downarrow}-n_{k\uparrow})
%\end{eqnarray}
%From this one can see that this pseudo spins $S_k^z$ represents the zero momentum electron pairs.
%Here we will tune the $\epsilon/V$ ratio by tuning $\epsilon$.
\par
\begin{figure}\hspace*{-0.5cm}
\includegraphics[scale=0.55]{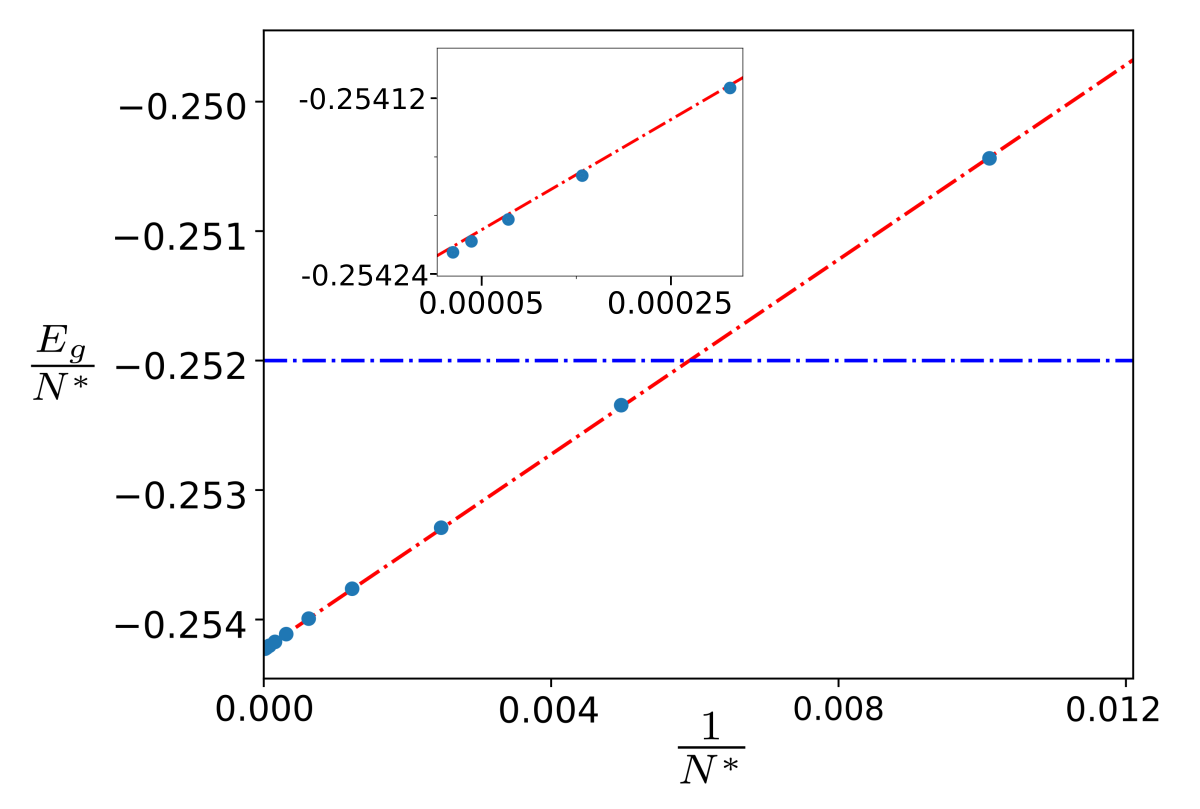}
\caption{Finite-size scaling of the ground state energy density $E_{g}/N^{*}$ under RG for system sizes $N=100, 202, 406, 814, 1600, 3200, 6400, 12800, 25600, 51200$ (blue circles from right to left). $N^{*}$ is the number of pseudospins within the emergent subspace at the stable fixed point of the RG. The red dash-dot line shows a linear fit to the finite-size scaling data. The blue dash-dot line shows the thermodynamic limit value for $E_{g}/N$ obtained from a finite-size scaling exact diagonalisation study of small systems ranging between $5-15$ pseudospins. Inset: Zoom of RG finite-size scaling data for $N=3200, 6400, 12800, 25600, 51200$.}
\label{fig:rg_eg_scaling}
\end{figure}

\noindent
%\textit{wavefunction : } In order to understand the nature of the ground state wavefunction for the VC phase ground state $S^{z}=0$ for $\epsilon=0$ , it is important to note that this is a state with a fixed number ($N_{C}=N/2$) of electron pairs.In general for arbitary nonzero $\epsilon$ the ground state will be $S_z=int (\epsilon/V + 1/2)$.Number of cooper pair is $N_c=2S-S_z$ thus the generay 
Finally, we note that the ground state wavefunction of the $U(1)$-symmetric CPI state with $S^{z}=0$ (i.e., at strong coupling) is given by
%As you can see that our Vortex Condensate phase is number fixed phase , if the number of electron pair in the momentum space window under study is  $N_c$  then the ground state wavefuction of the VCS phase looks like
\begin{eqnarray}
|\psi_{g}\rangle &=&\mathcal{N} \bigg(\displaystyle\sum_{k}c_{-k\downarrow}^{\dagger}c_{k\uparrow}^{\dagger}\bigg)^{S}|\textrm{vac}\rangle 
\label{groundstate}
\end{eqnarray}
where $|\textrm{vac}\rangle$ is the state that contains no Cooper pairs, and $\mathcal{N}$ is a normalisation factor. By acting with $H_{coll}$ on $|\psi_{g}\rangle$, we obtain the ground state energy density as
\begin{eqnarray}
\frac{E_{g}}{N^{*}} &=& -\frac{\bar{V}^{*}}{N^{* 2}}S^{2} = -\frac{\bar{V}^{*}}{N^{* 2}}\frac{N^{*}}{2}(\frac{N^{*}}{2}+1)\nonumber\\
&=&-\frac{\bar{V}^{*}}{4}(1+ \frac{2}{N^{*}})\simeq -\frac{\bar{V}^{*}}{4}~\textrm{for}~N^{*}>>1~.
\label{genden}
\end{eqnarray} 
%\textbf{Gutzwiller projection operator connection}
%\subsection{Phase diagram and Benchmarking:}
In order to gauge the accuracy of the effective Hamiltonian (eq.\eqref{hcoll01}) and ground state wavefunction (eq.\eqref{groundstate}) obtained from the RG procedure, we compare the ground energy density value obtained in the thermodynamic limit (eq.\eqref{genden}) from a finite-size scaling analysis with that obtained from a finite-size scaling for exact diagonalization (ED) studies of small systems of the bare Hamiltonian (eq.\eqref{pairing}) for $U=0, |V^{q}_{kk'}| = 2$ (in units of a hopping parameter $t$), $|W^{q}_{kk'}|=2/N$. For a $U(1)$-symmetric Fermi surface, it suffices to compare the energy density value obtained along any one diameter of the spherical Fermi volume. As is shown in Fig.\ref{fig:rg_eg_scaling}, we find excellent agreement between the results obtained in the thermodynamic limit from the two approaches: $E_{g}/N^{*}\simeq -0.254t$ from the RG, as against $E_{g}/N^{*}\simeq -0.252t$ obtained from ED. This indicates the efficiency of the RG method in preserving the spectral content during the flow towards the stable IR fixed point, and offers confidence in the analyses of subsequent sections that offer insight into the properties of the CPI phase.   
%of system size 13 using QuSpin Exact diagonalization and tried to benchmark our RG fixed point hamiltonian Ground state energy density. 
%Starting Hamiltonian 
%\begin{figure}
%\includegraphics[scale=0.46]{plt/rg_scheme}
%\caption{Choosing the window around the Fermi surface to start the Renormalization Group.  MODIFY}
%\label{fig:rg_scheme}
%\end{figure}
%\begin{eqnarray}
%H&=&-\displaystyle\sum_{k} \epsilon_k S_k^z - \frac{2}{N} \displaystyle\sum_{k\neq k'}(S_k^+S_{k'}^- +~\textrm{h.c.})\\ \nonumber 
%&+& V_{zz}\displaystyle\sum_{k\neq k'} S_k^z S_{k'}^z  ~,~~~~~~~~~~~V_{zz}>0
%\label{eq:benchmark}
%\end{eqnarray} 
%We are working with circular Fermi surface. We will choose a window around the Fermi surface $\Lambda_0$ as shown in the figure (\ref{fig:rg_scheme}), and do RG on that. We did ED on the Hamiltonian (\ref{eq:benchmark}) for the pseudo spins within the window ($\Lambda_0$).
%
%RG shows $\tilde{\epsilon}_0$ becomes RG-irrelevant and $\tilde{V}_{00}$ reaches fixed point value $\tilde{V}^*_{00}$. For $V_{zz}=0$ we see that the grouns state energy density is $EgD=\tilde{V}_{00}/4$.\\
%
%In the previous section we have shown under RG how the most Rellevent effective Hamiltonian is the all to all LMG model \cite{LMG1} .Hamiltonian is 
%\begin{eqnarray}
%H_{coll} &=& -\frac{2\epsilon}{N} S_z -\frac{V}{N} \bigg( S^2 - S_z^2 \bigg) ~,~~~ V>0
%\label{hcoll01}
%\end{eqnarray}
%
%We get the phase diagram (\ref{fig:rg_phase_diagram}). We have kept $\omega$ fixed at $\omega=0.15$. Varying the $\Lambda_0$ we draw the phase diagram. For one perticular choice of $\Lambda_0=0.122$ we do the RG and from the effective fixed-point Hamiltonian find the ground state energy density. We do the scaling of this ground state energy density shown in the figure (\ref{fig:rg_eg_scaling}). Thus we see good agreement with ED result with accuracy $<2 \% (1.8\%)$.

\section{Topological Features of CPI}\label{section:topofeat}
\noindent 
We will, in this section, study the topological properties of the many-body system described by the stable fixed point effective Hamiltonian $H_{coll}$ (eq.\eqref{hcoll01}) obtained from the RG. 

\subsection{Topological nature of the effective theory}
\noindent
We begin by showing that the effective Hamiltonian for the plateau state in the strong coupling limit (i.e., with $\bar{\epsilon}^{*}=0$) system) is purely topological. This will be done by rewriting $H_{coll} (\bar{\epsilon}^{*}=0)=-\frac{\bar{V}^{*}}{N^{*}}(S_x^2+S_y^2)$ in terms of emergent Wilson loop operators defined on a torus created by imposing periodic boundary conditions in the $\Lambda$-direction (i.e., the window in $k$-space that defines the CPI condensate, see Fig.\ref{torusgauge} below).
%in $(k_{x},k_{y})$-space within the emergent window ).  
%Thus our zero plateau Hamiltonian is $H_{coll}^{0}=-\frac{V}{N}(S_x^2+S_y^2)$. 
\par\noindent
We define the $k$-space translation 
%operators 
($\mathcal{T}_{\hat{s}}$, brown curved line in Fig.\ref{torusgauge})
%,\mathcal{T}_{\hat{s}_{\perp}}$) 
and twist operators ($\hat{O}^i_{\hat{s}}$, blue dashed line in Fig.\ref{torusgauge})
\begin{figure}
\hspace*{-0.35cm}
\includegraphics[scale=0.3]{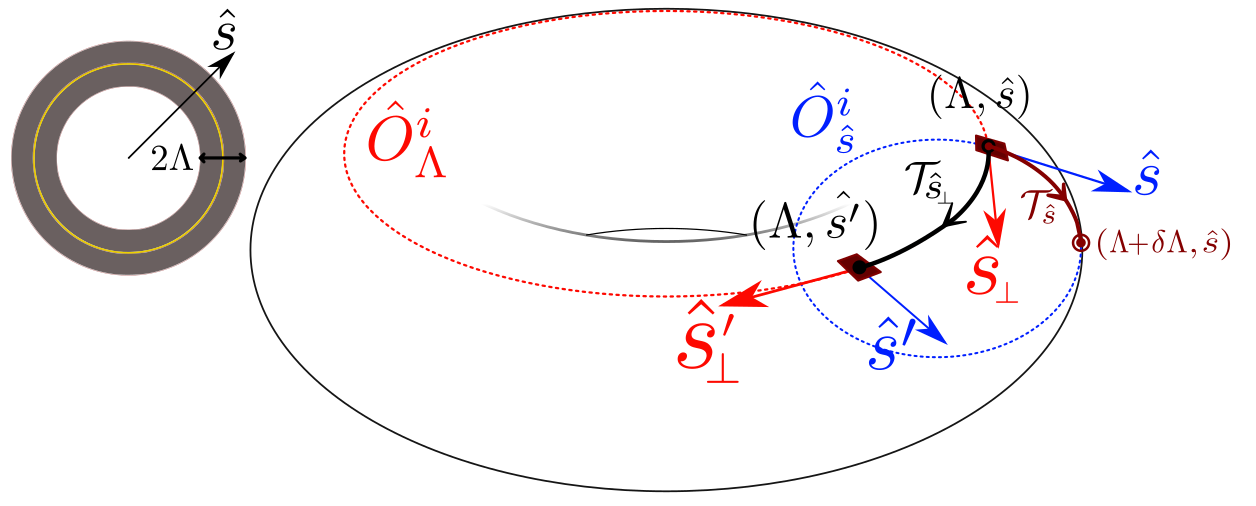}
\caption{(Left) Effective $k$-space window (dark grey) of size $2\Lambda$ around the Fermi surface (FS, yellow circle) formed under RG for the CPI phase. $\hat{s}$ represents a given direction in $k$-space normal to the FS. (Right) Construction of the respective twist ($(\hat{O}_{\hat{s}}, \hat{O}_{\Lambda}$) and translation operators ($(\hat{\mathcal{T}}_{\hat{s}}, \hat{\mathcal{T}}_{\hat{s}_{\perp}}$) defined on the torus created by imposing periodic boundary conditions on emergent window in $k$-space (i.e., on all directions $\hat{s}$).}
\label{torusgauge}
\end{figure}
\begin{eqnarray}
%&&\mathcal{T}_{\hat{s}_{\perp}} : S^i_{\Lambda,\hat{s}} \rightarrow S^i_{\Lambda,R\hat{s}} \\
&&\mathcal{T}_{\hat{s}} : S^i_{\Lambda,\hat{s}} \rightarrow S^i_{\Lambda+\delta\Lambda,\hat{s}}~,~ \hat{O}^i_{\hat{s}}=exp\bigg[\frac{2\pi}{N^{*}}i\displaystyle\sum_{n=0}^{N^{*}-1} n S^i_{n\Lambda,\hat{s}}\bigg]~,
\end{eqnarray}
where the twist operator $\hat{O}^i_{\hat{s}}$ spans all values of $\Lambda$ in the $\hat{s}$ direction, and imparts a gradual twist to the pseudospins $S^i_{n\Lambda,\hat{s}}$ such that the total twist imparted across the $\hat{s}$ direction is $2\pi$. Further, we can also define a composite twist operator that spans the entire torus shown in Fig.\ref{torusgauge}
\begin{eqnarray}
%&& \hat{O}^i_{\Lambda}=exp\bigg[\frac{2\pi}{N}i\displaystyle\sum_{n=0}^{N-1} n S^i_{\Lambda,R^n \hat{s}}\bigg]\\
&& {\tilde{O}}^i=\prod_{n=0}^{N^{*}-1} \hat{O}_{R^n\hat{s}}=exp\bigg[\frac{2\pi i }{N^{*}} \displaystyle\sum_{n,m=0}^{N^{*}-1} m S^i_{m\delta \Lambda,R^n \hat{s}} \bigg]~.
\end{eqnarray}
Then, we compute the following (nonlocal) Wilson loop operator $\mathcal{W}^{i}_{1}~(i=(x,y))$ defined in terms of $\mathcal{T}_{\hat{s}}$ and $\tilde{O}^i$
\begin{eqnarray}
\mathcal{W}^{i}_{1}=\mathcal{T}_{\hat{s}} \tilde{O}^i\mathcal{T}^{\dagger}_{\hat{s}} \tilde{O}^{i \dagger} &=& exp\bigg[\frac{2\pi i}{N^{*}} \displaystyle\sum_{m,n=0}^{N^{*}-1}S^i_{m\Lambda,R^n\hat{s}}  \bigg]\nonumber\\
&\times & exp\bigg[ 2\pi i \displaystyle\sum_{n=0}^{N^{*}-1} S^i_{\Lambda=0,R^n \hat{s}} \bigg]~,
\end{eqnarray}
where the first term on the right hand side imparts the twist to the centre of mass of the torus of pseudospins. The second denotes the trivial phase twist accumulated at a virtual boundary defined on the torus by the curve $\Lambda=0$, $exp\bigg[ 2\pi i \displaystyle\sum_{n=0}^{N-1} S^i_{\Lambda=0,R^n \hat{s}} \bigg]=1$. Thus, we obtain the composite pseudospin ($S^{i}~,i=(x,y)$) in terms of $\mathcal{W}^{i}_{1}$ as
\begin{eqnarray}
S^i=\displaystyle\sum_{m,n} S^i_{m\Lambda,R^n\hat{s}}
%=\frac{N}{2\pi i} log\bigg[\mathcal{T}_{\hat{s}} \tilde{O}^i\mathcal{T}^{\dagger}_{\hat{s}} \tilde{O}^{i \dagger}\bigg] 
= \frac{N^{*}}{2\pi} \textrm{Im}\left[\ln(\mathcal{W}^{i}_{1})\right]~,
\end{eqnarray}
such that the $U(1)$-symmetric effective Hamiltonian obtained from the RG 
%at strong coupling ($\bar{\epsilon}^{*}=0$) 
can be written purely in terms of the emergent $\mathcal{W}^{i}_1$ as
%the Wilson loop operator $\mathcal{W}^{i}_1=\mathcal{T}_{\hat{s}} \tilde{O}^i\mathcal{T}^{\dagger}_{\hat{s}} \tilde{O}^{i \dagger}$ as
\begin{eqnarray}
H_{coll}=-\frac{\bar{\epsilon^{*}}}{\pi}\textrm{Im}\left[\ln(\mathcal{W}^{z}_{1})\right]-\frac{N^{*}\bar{V}^{*}}{4\pi^2} \hspace*{-0.2cm}\displaystyle\sum_{i=x,y}\hspace*{-0.2cm}\bigg(\textrm{Im}\left[\ln (\mathcal{W}^{i}_{1})\right]\bigg)^2~.
\end{eqnarray}
This shows us that the $U(1)$ symmetry is encoded in the invariance of the Wilson loops $\mathcal{W}^{i}_{1}$ large gauge transformations~\cite{fradkin2013field,wen1992classification}, and that the nonlocal nature of their dynamics is encoded in the dependence of $H_{coll}$ on $\mathcal{W}^{i}_{1}$. In this way, we can clearly see the emergence of an effective gauge theory from the microscopic Hamiltonian eq.\eqref{pairing}.
\par\noindent
We note that $H_{coll}$ can also be written in terms of another set of emergent Wilson loop operators $\mathcal{W}^{i}_{2}$ obtained from a different pair of translation and twist operators defined on the torus (see Fig.\ref{torusgauge})
\begin{equation}
\mathcal{T}_{\hat{s}_{\perp}} : S^i_{\Lambda,\hat{s}} \rightarrow S^i_{\Lambda,R\hat{s}}~,~\hat{O}^i_{\Lambda}=exp\bigg[\frac{2\pi}{N^{*}}i\displaystyle\sum_{n=0}^{N^{*}-1} n S^i_{\Lambda,R^n \hat{s}}\bigg]~,
\end{equation}
such that we can redefine as earlier the following composite twist (${\tilde{O}}^{i}$) and pseudospin ($S^{i}$) and the Wilson loop $\mathcal{W}^{i}_{2}$ as 
\begin{eqnarray}
%&& \hat{O}^i_{\Lambda}=exp\bigg[\frac{2\pi}{N}i\displaystyle\sum_{n=0}^{N-1} n S^i_{\Lambda,R^n \hat{s}}\bigg]\\
{\tilde{O}}^{i}&=&\prod_{m=0}^{N^{*}-1} \hat{O}_{m\delta\Lambda}=exp\bigg[\frac{2\pi}{N^{*}}i\displaystyle\sum_{n,m=0}^{N^{*}-1} m S^i_{m\delta\Lambda,R^n \hat{s}} \bigg]~,\\
S^{i} &=& \frac{N^{*}}{2\pi}\textrm{Im}\left[\ln(e^{i\pi}~\mathcal{W}^{i}_{2})\right]~,~\mathcal{W}^{i}_{2} = \mathcal{T}_{\hat{s}_{\perp}} \tilde{O}^i\mathcal{T}^{\dagger}_{\hat{s}_{\perp}} \tilde{O}^{i \dagger}~.
\end{eqnarray}
We can once again write the effective Hamiltonian $H_{coll}$ as
\begin{eqnarray}
H_{coll} &=& -\frac{\bar{\epsilon}^{*}_{2}}{\pi}\textrm{Im}\left[\ln(\mathcal{W}^{z}_{2})\right]-\frac{N^{*}\bar{V}^{*}}{4\pi^2}\hspace*{-0.2cm} \displaystyle\sum_{i=x,y} \hspace*{-0.2cm}\bigg(\hspace*{-0.1cm}\textrm{Im}\left[\ln (\mathcal{W}^{i}_{2})\right]\hspace*{-0.1cm}\bigg)^2~,
\end{eqnarray}
where $\bar{\epsilon}_{2}^{*}=\bar{\epsilon}^{*}-N^{*}V^{*}/2$.
\par\noindent
The fact that the effective Hamiltonian ($H_{coll}$) obtained from the RG can be written completely in terms of global collective gauge degrees of freedom (i.e., the Wilson loops $\mathcal{W}^{i}_{1}$ and $\mathcal{W}^{i}_{2}$) is not surprising. Indeed, following Refs.\cite{diamantini_jja,hos} on the effective theory for the CPI phase being a Chern-Simons gauge field theory, we expect that the effective Hamiltonian for the CPI cannot be written in terms of local degrees of freedom.
Thus, for finite and non-zero $\bar{\epsilon}^{*}$, the association of a $U(1)$-symmetric Chern-Simons gauge field theory with the effective  quantum rotor Hamiltonian $H_{coll}$ (eq.\eqref{hcoll01}) in $0$-spatial dimensions can be argued for as follows. The action corresponding to $H_{coll}$ contains a $0$-dimensional topological $\theta$-term~\cite{fernandes} 
\begin{equation}
\theta = i\frac{\bar{\epsilon}^{*}N^{*}}{\bar{V}^{*}}\int_{0}^{\beta} d\tau \frac{\partial \phi}{\partial\tau}~, 
\end{equation}
written in terms of a global phase $\phi$ conjugate to $S^{z}$, such that 
\begin{equation}
-i\hbar\partial/\partial\phi \equiv S^{z} = -\frac{i}{\hbar}[S^{x},S^{y}] =\frac{iN^{*2}}{4\pi\hbar}[\ln W^{x}_{1},\ln W^{y}_{1}]~,
\end{equation}
and with a Berry phase given by $\gamma=2\pi\frac{\bar{\epsilon}^{*}}{\bar{V}^{*}}=2\pi\Phi$~\cite{fernandes}. 
It was shown by Yao and Lee~\cite{yaolee} that such a $\theta$-term in $0$-spatial dimensions is in precise correspondence with a $U(1)$ Chern-Simons topological term in $2$-spatial dimensions. Hansson et al.~\cite{hos} show that the Chern-Simons term encodes a topological coupling of the vorticity (or winding number part) of the global phase field $\phi$ to a field associated with the quasiparticle excitations. In this way, they show that the system, in the presence of a dynamical gauge field, possesses gauge invariance under large gauge transformations.  Further, they argue that the time-reversal invariance of the original problem necessitates that the $K$-matrix of the equivalent 2-flavour mixed Chern-Simons theory is $K=2\sigma_{x}$~ (see also Ref.\cite{moroz}). Then, the topological ground state degeneracy on a torus of genus $g$ is given by $|Det(K)|^{g}=4^{g}$~\cite{fradkin2013field}. This clarifies that the topologically ordered condensate of vortices observed in Ref.\cite{hos}, arising from the coupling the global phase of the superconducting ground state to dynamical electromagnetic gauge fields, corresponds to the Cooper pair number fixed insulating state of matter (the CPI phase) found at the stable fixed point of the RG flow. Next, we will demonstrate the 4-fold degeneracy for the special case of $\bar{\epsilon}^{*}=0=\Phi$.
%As shown just above, for a torus of genus $g=1$, a $4$-fold degeneracy is thus obtained.

\subsection{Topological degeneracy at $\Phi=0$.}
%\subsection{Twist-translation operators and ground state degeneracy}
%\subsubsection{Momentum-space translation symmetry}
%\begin{figure}[!h]
%\includegraphics[scale=0.2]{plt/TwistAndTranslation}
%\caption{Momentum translation symmetry.\textbf{Remove the $\mathcal{T}_{\hat{s}_{\perp}}$ etc. below the figure.}}
%\label{fig:momentumtranslation}
%\end{figure}
\par\noindent
In order to unveil a ground state degeneracy at $\Phi=0$, we follow the adiabatic flux insertion treatment of Oshikawa~\cite{oshikawa}. For this, we define the following momentum translation ($\hat{\mathcal{T}}_{\hat{s}_{\perp}}$)
%operator 
(see Fig.\ref{torusgauge}) 
%One can see that our Hamiltonian has momentum translation symmetry (Figure.\ref{fig:momentumtranslation}). our fixed point Hamiltonian $H^{(*)}$ commutes with this translation operator.
%This translation operator can be written as, 
%\begin{equation}
%\hat{\mathcal{T}}_{\hat{s}_{\perp}}=e^{i\hat{X}_{\hat{s}_{\perp}}}~,
%\end{equation}
and twist ($O^{\hat{s}_{\perp}}_{ph}$) operators 
\begin{eqnarray}
\hat{\mathcal{T}}_{\hat{s}_{\perp}}&=& e^{i\hat{\Theta}_{\hat{s}_{\perp}}}~,~O^{\hat{s}_{\perp}}_{ph} = \exp\bigg\{i\frac{2\pi}{N} \displaystyle\sum^{N-1}_{p=0} kS_{(R^pk_{\hat{s}})}^z\bigg\}~,\\ S_{(R^pk_{\hat{s}})}^z &=&\sum_{\Lambda_{(R^pk_{\hat{s}})}} S^z_{\Lambda_{(R^pk_{\hat{s}})}}~,
\end{eqnarray}
where $\hat{\Theta}_{\hat{s}_{\perp}}$ denotes the center of mass angular position along $\hat{s}_{\perp}$. With this, we find
\begin{eqnarray}
\mathcal{T}_{\hat{s}_{\perp}} O^{\hat{s}_{\perp}}_{ph} \mathcal{T}^{\dagger}_{\hat{s}_{\perp}}
%&=& exp \bigg[ \frac{i2\pi}{N} \bigg( 0.S^z_{1} + 1.S^z_{2} + . . . + (N-1).S^z_{0} \bigg) \bigg] \nonumber\\
&=& O^{\hat{s}_{\perp}}_{ph}~exp \bigg[ \frac{i2\pi}{N}\bigg( N S^z_{0}- \sum^{N-1}_{k_{\hat{s}}=0} S^z_{k_{\hat{s}}}
\bigg)\bigg]~,\nonumber\\
%~~~~~~~ \textrm{on $0^{th}$ plateau $\sum^{N-1}_{k_{\hat{s}}=0} S^z_{k_{\hat{s}}}=0$} \nonumber\\
&=&
%O^{\hat{s}_{\perp}}_{ph} \times exp \bigg[ i2\pi S^z_{0} \bigg] 
O^{\hat{s}_{\perp}}_{ph}~exp \bigg[ i\pi (2n+1)\bigg] 
%\nonumber\\&=&  
= O^{\hat{s}_{\perp}}_{ph}~ e^{i\pi}~, 
%\nonumber\\
\end{eqnarray}
where we have set $\sum^{N-1}_{k_{\hat{s}}=0} S^z_{k_{\hat{s}}}=0$ and $S^{z}_{0}=(2n+1)\frac{1}{2}$ (and $2n+1$ being the number of pseudospin states in the direction $\hat{s}$) in the second line in order to obtain the third. Thus,
\begin{eqnarray}
\{\mathcal{T}_{\hat{s}_{\perp}}, O^{\hat{s}_{\perp}}_{ph} \} = 0~,~[H^{(*)},\mathcal{T}_{\hat{s}_{\perp}}]=0=[H^{(*)},O^{\hat{s}_{\perp}}_{ph}]~.
\end{eqnarray}
These relations imply the existence of two degenerate states labelled by the center of mass angular position along $\hat{s}_{\perp}$ ($\Theta_{\hat{s}_{\perp}}$) 
\begin{eqnarray}
|\Theta_{\hat{s}_{\perp}}=0 \rangle~,~ |\Theta_{\hat{s}_{\perp}}=\pi \rangle~,
\end{eqnarray}
with transitions from one to the other taking place via the twist operator $O_{ph}$
\begin{eqnarray}
O^{\hat{s}_{\perp}}_{ph} ~|\Theta_{\hat{s}_{\perp}}=0 \rangle =~ |\Theta_{\hat{s}_{\perp}}=\pi \rangle
\end{eqnarray}
%\subsubsection{Helicity reversal symmetry}
\par\noindent
We now unveil another two-fold degeneracy of the ground state manifold. 
%Using the $U(1)$ symmetry of the RG fixed point Hamiltonian, we can study any one diameter of the momentum space circular strip, 
%we choose say the $k_y=0$ diameter. 
%We have defined the pseudospins as $S_{k,\eta_k}^z= \frac{1}{2}(n_{k_\uparrow}+n_{-k\downarrow}-1 )$. We choose $\eta_k=+1$ for $k_x>0$ and $\eta_k=-1$ for $k_x<0$. For simplicity of notation we have not used the $\eta$ index explicitly.
By first defining pseudospin degrees of freedom along a given direction of momentum space ($\hat{s}$) that are resolved in terms of the eigenvalue of the helicity operator ($\eta=\pm 1$)
\begin{eqnarray}
\tilde{S}^{\hat{s},z}_0 &=& \sum_{k} \delta_{\eta_k,-1} ~ S^{\hat{s},z}_{k,\eta_k}~,~~ \tilde{S}^{\hat{s},z}_1=\sum_{k} \delta_{\eta_k,+1} ~ S^{\hat{s},z}_{k,\eta_k}~,
\end{eqnarray}
we define a helicity twist operator $O^{\hat{s}}_{H}$
\begin{equation}
O^{\hat{s}}_{H} = e^{i\frac{2\pi}{2} (0.\tilde{S}_{0}^{\hat{s},z} + 1.\tilde{S}_{1}^{\hat{s},z}) }~.
\end{equation}
Then, we define the helicity inversion operator $\mathcal{T}_{H}$
\begin{eqnarray}
\mathcal{T}^{\hat{s}}_{H}\equiv e^{i\hat{H}^{\hat{s}}_e}~: S^{\hat{s},z}_{k,\eta_k} \rightarrow S^{\hat{s},z}_{k,-\eta_k}~,
%~~~~\mathcal{T}^{\hat{s}}_{H}=e^{i\hat{H}^{\hat{s}}_e}
\end{eqnarray}
where $\hat{H}^{\hat{s}}_e$ corresponds to the generator of helicity inversion  of the center of mass along $\hat{s}$. These helicity twist and translation operators follow the algebra
\begin{eqnarray}
\mathcal{T}^{\hat{s}}_{H}O^{\hat{s}}_{H}\mathcal{T}^{\hat{s},\dagger}_{H} 
%&=&  \nonumber\\
&=& O^{\hat{s}}_{H} \times exp \bigg[ \frac{i2\pi}{2} \bigg( 2 \tilde{S}^{\hat{s},z}_{0}- [\tilde{S}^{\hat{s},z}_{0}+\tilde{S}^{\hat{s},z}_{1}] 
\bigg)\bigg]  \nonumber\\
&=& O^{\hat{s}}_{H} \times exp \bigg( i2\pi   \tilde{S}^{\hat{s},z}_{0} \bigg) ~, \nonumber\\
&=&  O^{\hat{s}}_{H} \times e^{i2\pi (2n+1)\frac{1}{2}} = O^{\hat{s}}_{H} e^{i\pi}~, 
\end{eqnarray}
where we have set $[\tilde{S}^{\hat{s},z}_{0}+\tilde{S}^{\hat{s},z}_{1}]=0$ and $\tilde{S}^{\hat{s},z}_{0}=(2n+1)\frac{1}{2}$ in the second line in order to obtain the third. Thus, 
\begin{eqnarray}
\{\mathcal{T}^{\hat{s}}_{H}, O^{\hat{s}}_{H} \} = 0~,~[H^{(*)},\mathcal{T}^{\hat{s}}_{H}]=0=[H^{(*)},O^{\hat{s}}_{H}]~.
\end{eqnarray}
Again, these relation imply the existence of two degenerate states labelled by the eigenvalue of the generator of helicity inversion ($\hat{H}^{\hat{s}}_e$) of the center of mass along $\hat{s}$
%Thus we can see that the two degenerate states are labeled by 
\begin{eqnarray}
|H^{\hat{s}}_e=0 \rangle~,~ |H^{\hat{s}}_e=\pi \rangle~,
\end{eqnarray}
with transitions from one to the other taking place via the twist operator $O_{ph}$
\begin{eqnarray}
\mathcal{T}^{\hat{s}}_{H} ~|H^{\hat{s}}_e=0 \rangle =~|H^{\hat{s}}_e=\pi \rangle~.
\end{eqnarray}
\par\noindent
%\textbf{\textit{Main commutation relations:}} 
Importantly, we find that 
\begin{eqnarray}
[\mathcal{T}^{\hat{s}}_H, \mathcal{T}_{\hat{s}_{\perp}}]=&0&=[O^{\hat{s}}_H,O^{\hat{s}_{\perp}}_{ph}]~, \nonumber\\
\{\mathcal{T}^{\hat{s}}_{H}, O^{\hat{s}}_{H} \} = &0& = \{\mathcal{T}_{\hat{s}_{\perp}}, O^{\hat{s}_{\perp}}_{ph} \}~.
\end{eqnarray}
Thus, these four operators together label the four-fold degenerate ground state manifold. As noted above, this matches the result for the phenomenological BF Chern-Simons gauge field theory formulation of Hansson et al.~\cite{hos}
\par\noindent
Finally, the topological order is protected by the spectral gap 
\begin{equation}
\Delta_{top}= E_{S^{z}=1} - E_{S^{z}=0} = \frac{V^{*}}{N^{*}}
\end{equation}
separating the degenerate ground state manifold from the lowest lying excited state of the effective Hamiltonian $H_{coll}$ (eq.\eqref{hcoll01}). 
%For this, we note that the appearance of the $2\pi$ phase factor in eq.(\ref{spinback}) reflects 
%as arising from 
Further, these ground states are also separated from the single-particle excitations by a many-body gap ($\Delta_{MB}$) that arises from 
%the helicity-changing backscattering process for a pair of electrons that leads to the pairing-induced gap: $\left((k_{F},\uparrow),(-k_{F},\downarrow)\right)\leftrightarrow \left((k_{F},\downarrow),(-k_{F},\uparrow)\right)$~. 
%Thus, the term $(U_{s}^{2} + U_{s}^{\dagger 2})$ plays a role analogous to 
%This is seen in 
the helicity backscattering term, $(\bar{V}^{*}/N^{*})\sum_{k}(S_{k}^{+}S_{-k}^{-} + {\textrm h.c.})$~, contained within the effective Hamiltonian $H_{coll}$ (eq.\eqref{hcoll01})
%, and leads to an excitation gap ($\Delta$) given by
~\cite{nakamura2002lattice}
\begin{eqnarray}
\Delta_{MB} 
%&=& \langle \psi_{0}| U_{s}^{2}H_{coll}U_{s}^{\dagger 2} - H_{coll}|\psi_{0}\rangle~,\nonumber\\
&=& \langle \psi_{0}| \frac{\bar{V}^{*}}{N^{*}}\sum_{k}(S_{k}^{+}S_{-k}^{-} + \textrm{h.c.}) |\psi_{0}\rangle~.
\end{eqnarray}
%\par\noindent
%We have already argued for the association of a Chern-Simons gauge field theory with the effective Hamiltonian obtained from the RG. 
%can also be argued for as follows. 
%The action corresponding to the quantum rotor Hamiltonian eq.(\ref{lmgandpoc}) contains a $\theta$-term~\cite{fernandes} 
%\begin{equation}
%\theta = i\frac{\epsilon}{N}\int_{0}^{\beta} d\tau \frac{\partial \phi}{\partial\tau}~, 
%\end{equation}
%where $S^{z}\equiv -i\partial/\partial\phi$, leading to a Berry phase $\gamma=2\pi\frac{\epsilon}{N}$~. It was shown by Yao and Lee~\cite{yaolee} that such a $\theta$-term in $0$-spatial dimensions is in precise correspondence with a Chern-Simons topological term in $2$ spatial dimensions. 
%In this regard, Hansson et al.~\cite{hos} argue that the time-reversal invariance of the original problem necessitates that the $K$-matrix of the equivalent 2-flavour mixed Chern-Simons theory is $K=2\sigma_{x}$~ (see also Ref.(\cite{moroz})), such that the (topological) ground state degeneracy on a torus of genus $g$ is given by $|Det(K)|^{g}=4^{g}$~\cite{wenzee1990}. As shown just above, for a torus of genus $g=1$, a $4$-fold degeneracy is thus obtained.

\subsection{Spectral flow, plateau ground states and topological quantum numbers}
\begin{figure}
\includegraphics[scale=0.7]{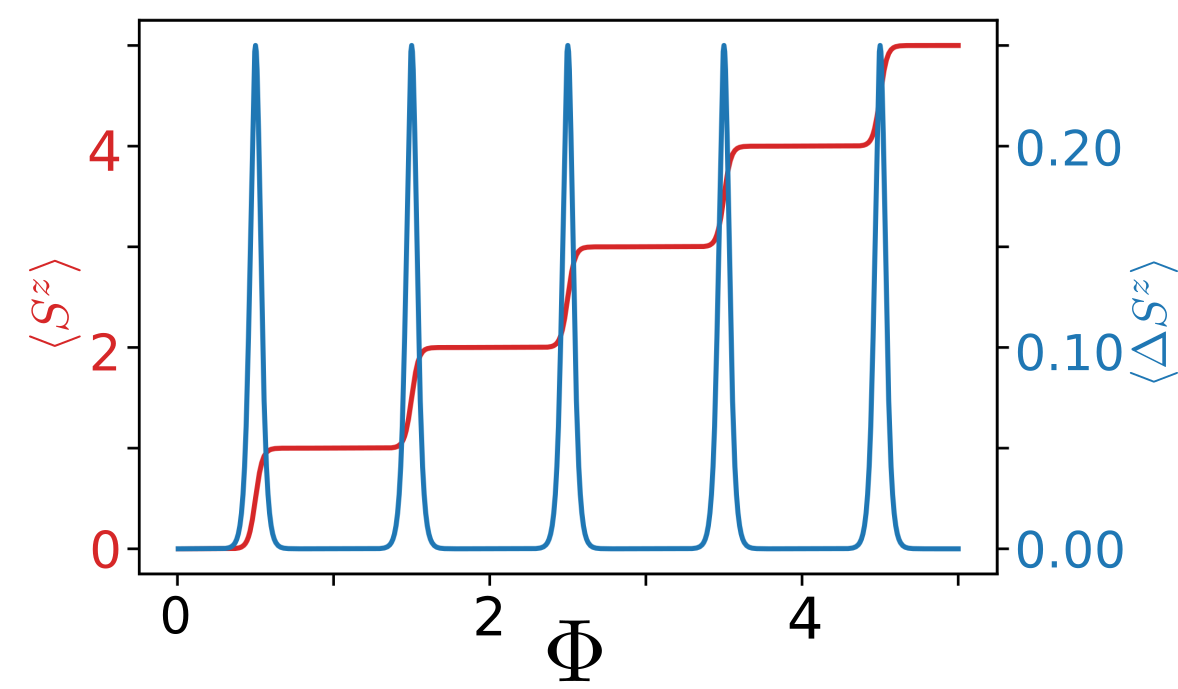}
\caption{Variation of $\langle S^{z}\rangle$ (red curve, and y-axis on left) and $\langle \Delta S^{z}\rangle$ (blue curve, and y-axis on right) with effective AB flux $\Phi$ at small temperature $T=0.01$ (in units of $k_{B}$). $\langle S^{z}\rangle$ and $\langle \Delta S^{z}\rangle$ are computed using the effective CPI Hamiltonian eq.\eqref{hcoll01}. See text for discussion.}
\label{hall}
\end{figure}
%\textbf{Entanglement Entropy connected to log(N cooperpairs ) in VC state.
% Entanglement Entropy and Number fluctuations connected to log(N electrons ) in metal, as shown by Swingle.}\\
\par\noindent
As mentioned in the previous subsection, by tuning the ratio $\Phi\equiv \bar{\epsilon}^{*}/\bar{V}^{*}$, we can access ground states with different number of Cooper pair bound states (i.e., the eigenvalue of the operator $S^{z}$).
%: increasing the parameter $\Phi\equiv \bar{\epsilon}^{*}/\bar{V}^{*}$ leads to a ground state with a decreasing number of Cooper pairs. 
We will now study the passage between these ground states, and also show the journey towards a metallic (gapless) ground state (i.e., with a vanishing number of Cooper pair bound states). 
\par\noindent
%\textit{Studying Spectral Flow :} 
%We can re write the collective Hamiltonian (\ref{hcoll01}) upto a constant as
%\begin{eqnarray}
%H_{coll}&=&-\frac{V}{N}S(S+1)+\frac{V}{N}(S^z-\frac{\Phi}{\Phi_0})^2 \nonumber \\
%E_{coll}(m,S)&=&-\frac{V}{N}S(S+1)+\frac{V}{N}\bigg(m-\frac{\Phi}{\Phi_0}\bigg)^2
%\end{eqnarray}
%Where $\Phi/\Phi_0=\epsilon/V$. 
%Total number of the electron is constant in the system and the energy is minimum for $S\rightarrow$ Maximum so $S=\frac{N}{2}$. $2N$ is the number states within gap. 
We recall that for $\Phi=0$, the ground state is given by $|\psi_{g}\rangle = |S=N^{*}/2, S^{z}=0\rangle$, i.e., a state with $N^{*}$ Cooper pairs. The action of $S^+$ on $|\psi_{g}\rangle$ is
%states are known 
\begin{eqnarray}
S^{+} \bigg |\frac{N^{*}}{2},0\bigg \rangle &=& \sqrt{(S-S^{z})(S+S^{z}+1)}\bigg |\frac{N^{*}}{2}, 1\bigg \rangle\nonumber\\ 
&=& \sqrt{\frac{N^{*}}{2}(\frac{N^{*}}{2}+1)}\bigg |\frac{N^{*}}{2}, 1\bigg \rangle~,
%S^+=\sum_k S_k^+&=&\sum_k a_{-k\downarrow}a_{k\uparrow}\\
%a_{-k\downarrow}a_{k\uparrow}|N_c\rangle&=&|N_c-2\rangle_k
\end{eqnarray}
i.e., lowers the Cooper pair number by 1. Energetically, this is equivalent to a value of the parameter $\Phi$ in $H_{coll}$ within the range $0.5< \Phi <1.5$. In the same way, $\frac{m}{2} < \Phi < \frac{m}{2} +1~(m\in\mathcal{Z}$) leads to a ground state $|N^{*}/2, m\rangle$, such that for $\Phi > (N^{*}-1)/2$, we attain a ground state with a vanishing number of Cooper pairs. This amounts to reducing the spectral gap of the CPI phase in a step-like manner, until a gapless spectrum (the ``metal") is attained. Thus, tuning the parameter $\Phi$ amounts to a process of spectral flow between various ground states. Each gapped ground state (corresponding to different values of $S^{z}$) possesses topological features (as discussed in the previous subsection).
\begin{figure}[h!]
\includegraphics[scale=0.7]{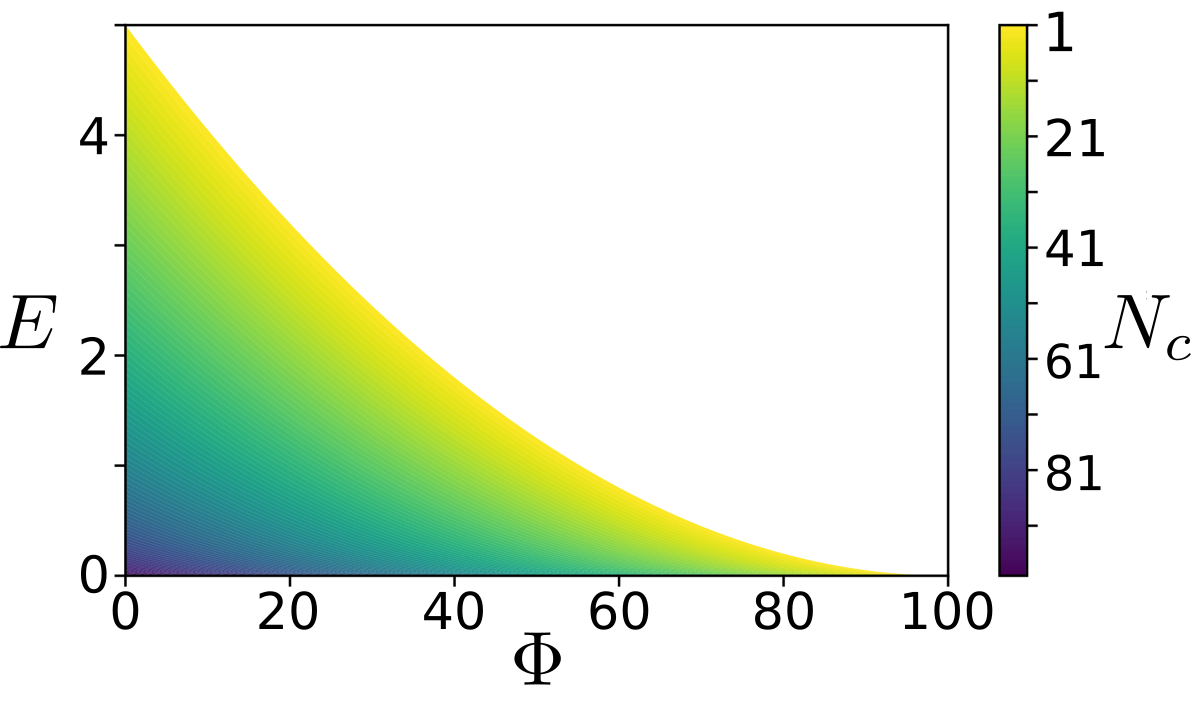}
\caption{Variation of the energy spectrum ($E$, y-axis) of the effective CPI Hamiltonian (eq.\eqref{hcoll01}) with the effective AB flux $\Phi$ (x-axis) for a system of 200 electrons. The colour scale represents the number of Cooper pairs ($N_{C}$). The bright yellow border represents the transition between the metal (white region) and CPI phases upon tuning $\Phi$.}
\label{fig:pd}
\end{figure}
\par\noindent
At zero temperature, in the presence of a pairing-induced gap $\Delta$, these gapped ground states will show plateaux in a variation of $\langle S^{z}\rangle$ with $\Phi=\bar{\epsilon}^{*}/\bar{V}^{*}$ (see 
Fig.\ref{hall} for a small $k_{B}T=0.04$). Further, it is easily seen that at various half-integer values of $\Phi=m+1/2$, the ground state becomes degenerate via level-crossings,
%: for $\Phi=m+1/2$, the ground state is 
i.e., a linear combination of $|S=N^*/2,S^{z}=m\rangle$ and $|S=N^*/2,S^{z}=m+1\rangle$
\begin{equation}
|\psi_{\textrm{PT}}^{m}\rangle = c_{1}|\frac{N^{*}}{2},m\rangle + c_{2}|\frac{N^{*}}{2},m+1\rangle~,
\label{superpos}
\end{equation}
where $|c_{1}|^{2}+|c_{2}|^{2}=1$. As shown in Fig.\ref{hall}, these correspond to transitions between plateaux in $\langle S^{z}\rangle$ and lead to large fluctuations ($\langle \Delta S^{z}\rangle$) in $S^{z}$. It can be shown that the largest $\langle \Delta S^{z}\rangle$ is obtained for $c_{1}=1/\sqrt{2}=c_{2}$. As noted above, the final level-crossing is attained at $\Phi=(N^{*}-1)/2$. In Fig.\ref{fig:pd}, we show a variation of the energy for excited states obtained from $H_{coll}$ (computed with respect to the ground state energy) with the parameter $\Phi$, and where the colour scale denotes the number ($N_{c}$) of Cooper pairs in a given state. The plot clearly shows the collapse of the excitation spectrum of the gapped plateaux as $\Phi$ is tuned towards passage from the final plateau into the gapless metal (white space in Fig.\ref{fig:pd}).
%We end this section with a discussion of the CPI at finite temperatures. Our $H_{coll}$ is exactly solvable model, it's already in diagonal form, Here we have taken the eigen spectrum of this model and simulated to finite temperature using standard partition function method. As can be seen from the plot (\ref{fig:sd}) that at finite temperature sharpness of the plateau decreases as temperature fluctuation starts dominating.
%\subsection{Helicity Cross-Correlations}
\begin{figure}
\includegraphics[scale=0.76]{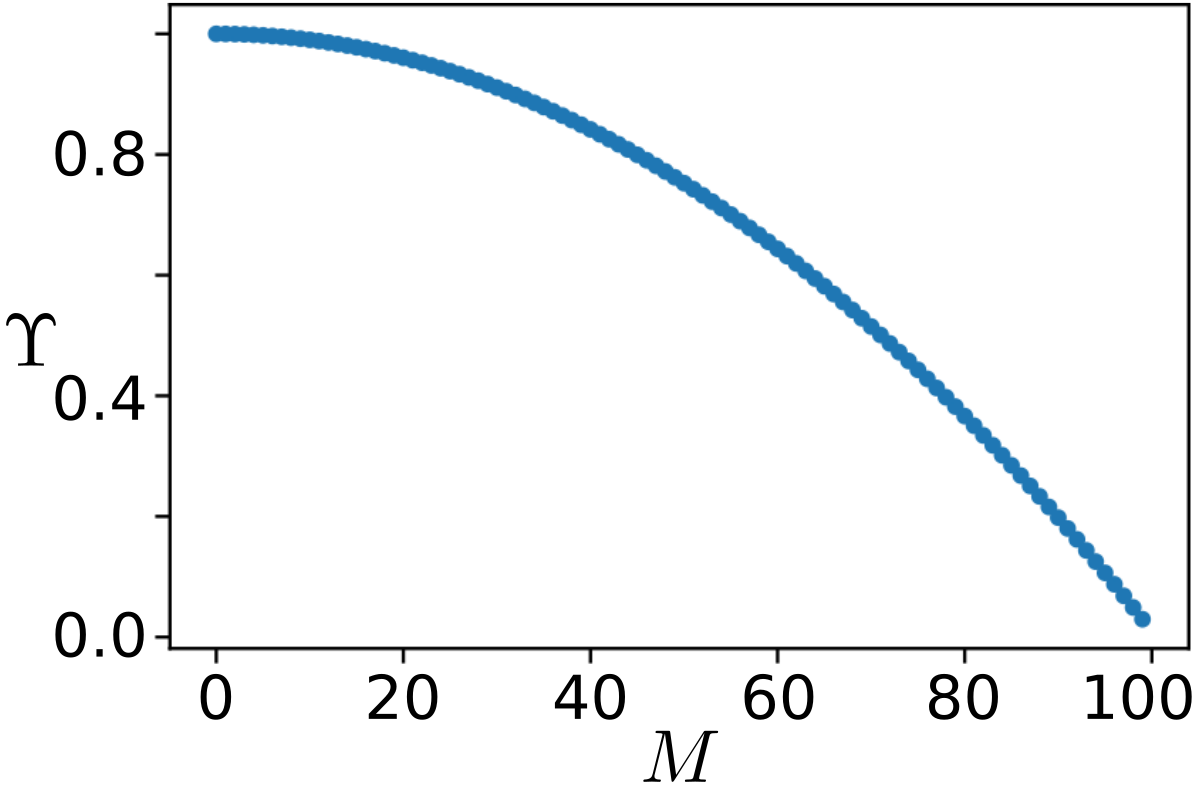}
\caption{Variation of the normalized helicity cross correlation ($\Upsilon$) with the ground state eigenvalue of $S_{z}$ ($M$) for a system of 200 electrons. $M$ is increased by increasing the flux $\Phi$. The large values of $\Upsilon$ for $M\to 0$ characterises the stable CPI phase, and the curve represents the passage to the metal ($\Upsilon\to 0$ as $M\to 100$).}
\label{fig:hcc}
\end{figure}
\begin{figure}[h!]
\includegraphics[scale=0.5]{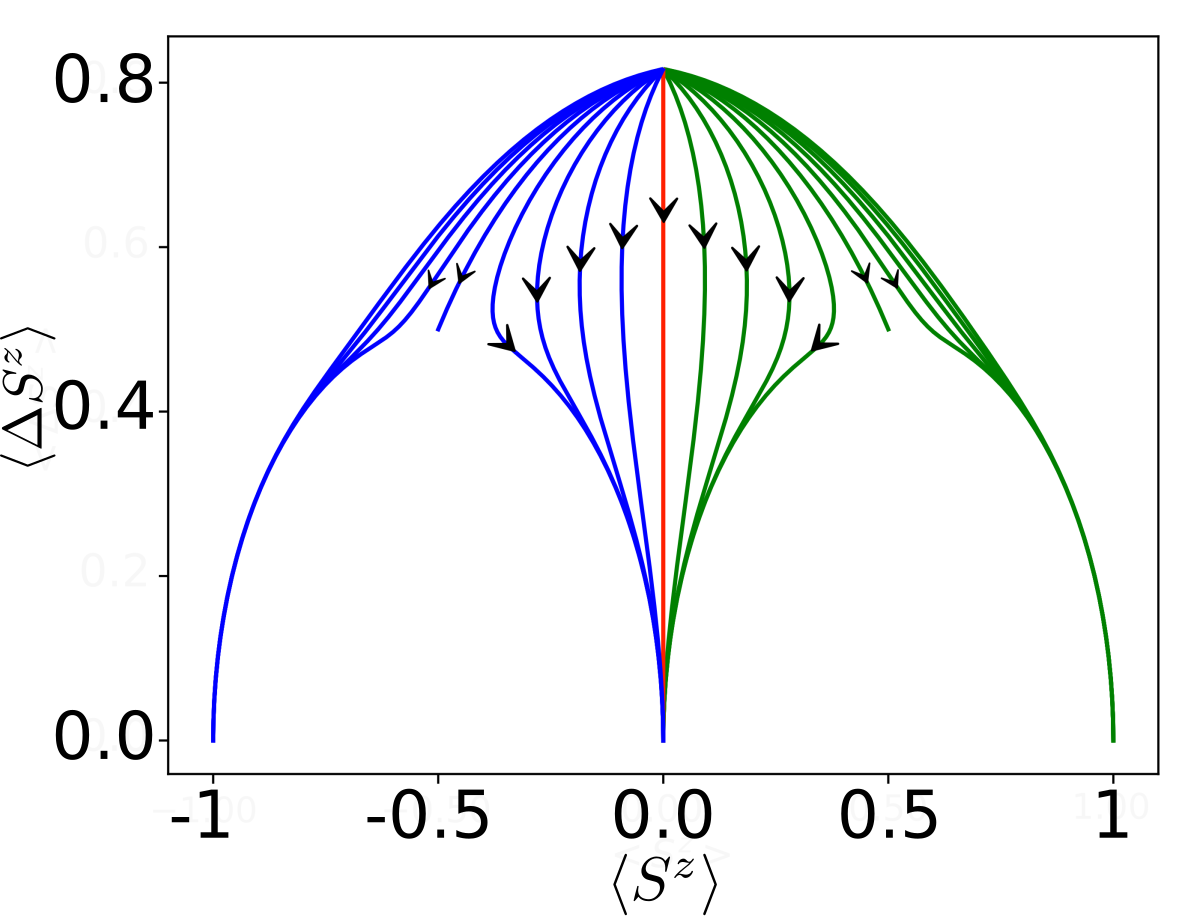}
\caption{Variation of $\langle S_{z}\rangle$ and $\rangle \Delta S_{z}\rangle$ with lowering temperature (from $T=1000$ to $T=0$ (in units of $k_{B}$), and indicated through arrows) and flux $\Phi=0$ towards the three CPI ground states $S_{z}=0,\pm 1$ that are reached for the $S=1$ system. $\langle S_{z}\rangle$ and $\rangle \Delta S_{z}\rangle$ are computed using the CPI Hamiltonian (eq.\eqref{hcoll01}). See text for discussion.
}
\label{fig:thermals1}
\end{figure}
\par\noindent
This is reinforced by a study of the helicity cross correlation ($\Upsilon$), i.e., inter-helicity two-particle scattering, defined as
%Now we will calculate a two particle entanglement in Vortex Condensate phase.Helicity cross correlation is a measure of an entanglement developed due to the Inter Helicity scattering, In the vortex condensate Hamiltonian $V/N(S^2-S_z^2)$ term is the source of this kind of entanglement .Here we will calculate how this Helicity Cross Correlation (H.C.C.) is related to the Vortex condensate gap . Definition of H.C.C. is 
\begin{eqnarray}
\Upsilon = \langle S^+S^- \rangle + \langle S^-S^+ \rangle - 2\langle S^+ \rangle \langle S^- \rangle~,
%\langle S^- S^+ \rangle - \langle S^- \rangle \langle S^+ \rangle~,
\label{HCC1}
\end{eqnarray}
where the expectation value is taken with respect to the ground state. 
For $|\psi_{g}\rangle=|S,S^{z}=M\rangle$, it can be shown that $\Upsilon = 2 (S^2+S-M^2)$. A plot of $\Upsilon$ versus $M$ in Fig.\ref{fig:hcc} shows that the strength of inter-helicity scattering gradually reduces as $M$ increases, i.e., the parameter $\Phi$ is tuned towards the gapless metal.
%In the Vortex Condensate phase if A is the flux then ground state pseudospin magnetization is $M= int(A+1/2)$ .Thus we can see that the in the state $|M \rangle$ the H.C.C. is $H.C.C(M) = 2 (S^2+S-M^2) $. In the figure (\ref{fig:hcc}) we have plotted the normalized H.C.C defined as $HCC_n=(S(S+1)-M^2)/S(S+1)$.
%Thus approximately density of states say constant g(0) thus is the Vortex Condensate spin gap is $\Delta_{VC}^s$ then Number of electrons in the gap is $N_e=\Delta_{VC}^sg(0)$, $N_e=4S-2M$ .Thus one gets from the equation (\ref{HCC}) 
%\begin{eqnarray}
%\Delta^s_{VC}(M)=\frac{1}{g(0)} N_e = \frac{1}{g(0)} (4S-2M)
%\end{eqnarray}
%\begin{eqnarray}
%\Delta^s_{VC}&=&\frac{4S}{g(0)} \bigg[ 1\pm \sqrt{\frac{(1+S)}{4S}-\frac{H.C.C.(M)}{8S^2}} \bigg]
%\end{eqnarray}
%Thus one can see from the figure (\ref{fig:hccc}) how the Helicity Cross Correlatin and the  vorex condensate charge gap is connected.Thus one can see that this charge gap vanishes at the last plateau .Thus at this last plateau $S_z=S$ neither it has charge gap nor spin gap.
\par\noindent
Upon increasing the temperature, the plateaux are steadily degraded and the fluctuations $\langle\Delta S^{z}\rangle$ at the transitions increase in strength (Fig.\ref{hall}). In Fig.\ref{fig:thermals1}, we show a plot of $\langle \Delta S^{z}\rangle$ against $\langle  S^{z}\rangle$ obtained from $H_{coll}$ for different values of the parameter $\Phi$ and temperature $T$ for the case of $S=1, S^{z}=0,\pm 1$. The blue curves are for $-1\leq \Phi < 0$ (with $-1\leq \langle S^{z}\rangle <0$) and the green curves are for $0< \Phi \leq 1$ (with $0< \langle S^{z}\rangle \leq 1$). The direction of the arrows denote the lowering of temperature. Fig.\ref{fig:thermals1} shows that lowering $T$ generically leads to a plateau ground state ($\langle S^{z} \rangle = 0,\pm 1~,~ \langle \Delta S^{z}\rangle = 0$). On the other hand, there also exist special cases when lowering $T$ leads to (unstable) ground states located precisely at the plateau transitions ($\langle S^{z} \rangle = 1/2 =\pm \langle \Delta S^{z}\rangle$). While the figure shows the numerical computation for $H_{coll}$ with $S=1$, we have observed that a similar plot for a much larger value of $S$ also shows the same ``dome"-like structure of the curves.

\section{Entanglement Features of CPI}\label{section:entfeat}
\subsection{Entanglement Spectrum}

\begin{figure}
\includegraphics[scale=0.5]{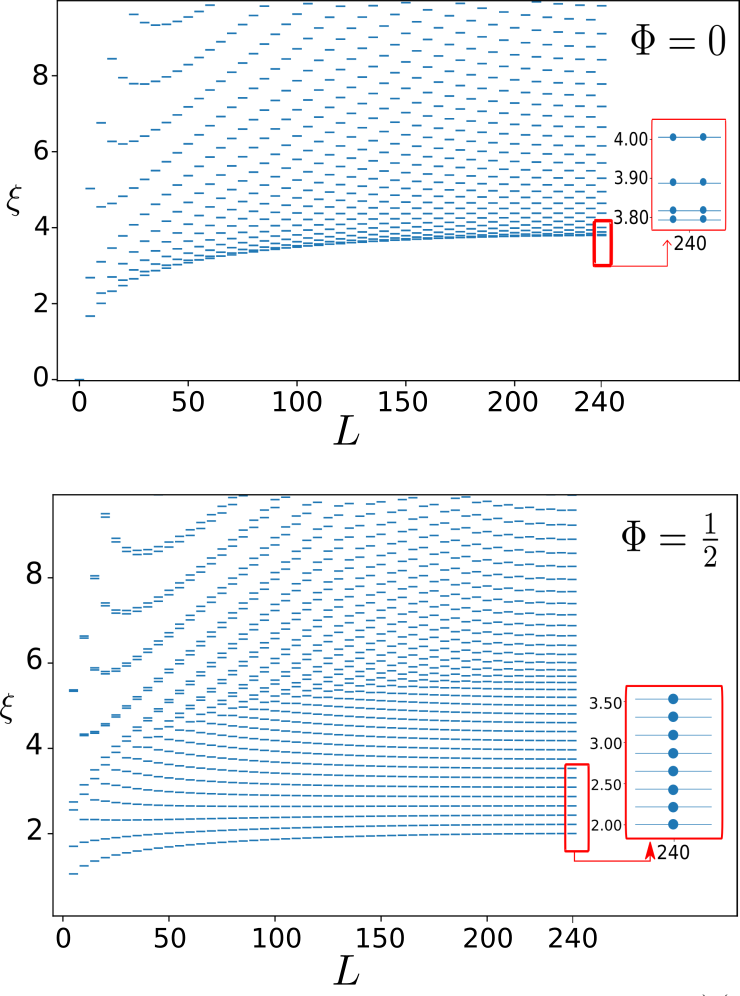}
\caption{Entanglement spectrum (ES, $\xi_i=-log_2\lambda_i$) of a subsystem of size $L$ in the ground state wavefunction at $\Phi=0$ (eq.\eqref{groundstate}, upper plot) and at $\Phi=1/2$ (eq.\eqref{superpos}, lower plot) for a system of $N^{*}=488$ Cooper pairs and as a function of the subsystem size $L$. The index $i$ labels the ES eigenvalues. The upper inset shows the double degeneracy for all levels, while the lower inset shows that the degeneracy is lifted at $\Phi=1/2$.
%These plots are for $N=488$, $\xi_i=-log_2\lambda_i$ vs $L$ (A) $\epsilon/V=0$ (B) $\epsilon/V=0.5$
}
\label{fig:01flux}
\end{figure}

\begin{figure}
\includegraphics[scale=0.65]{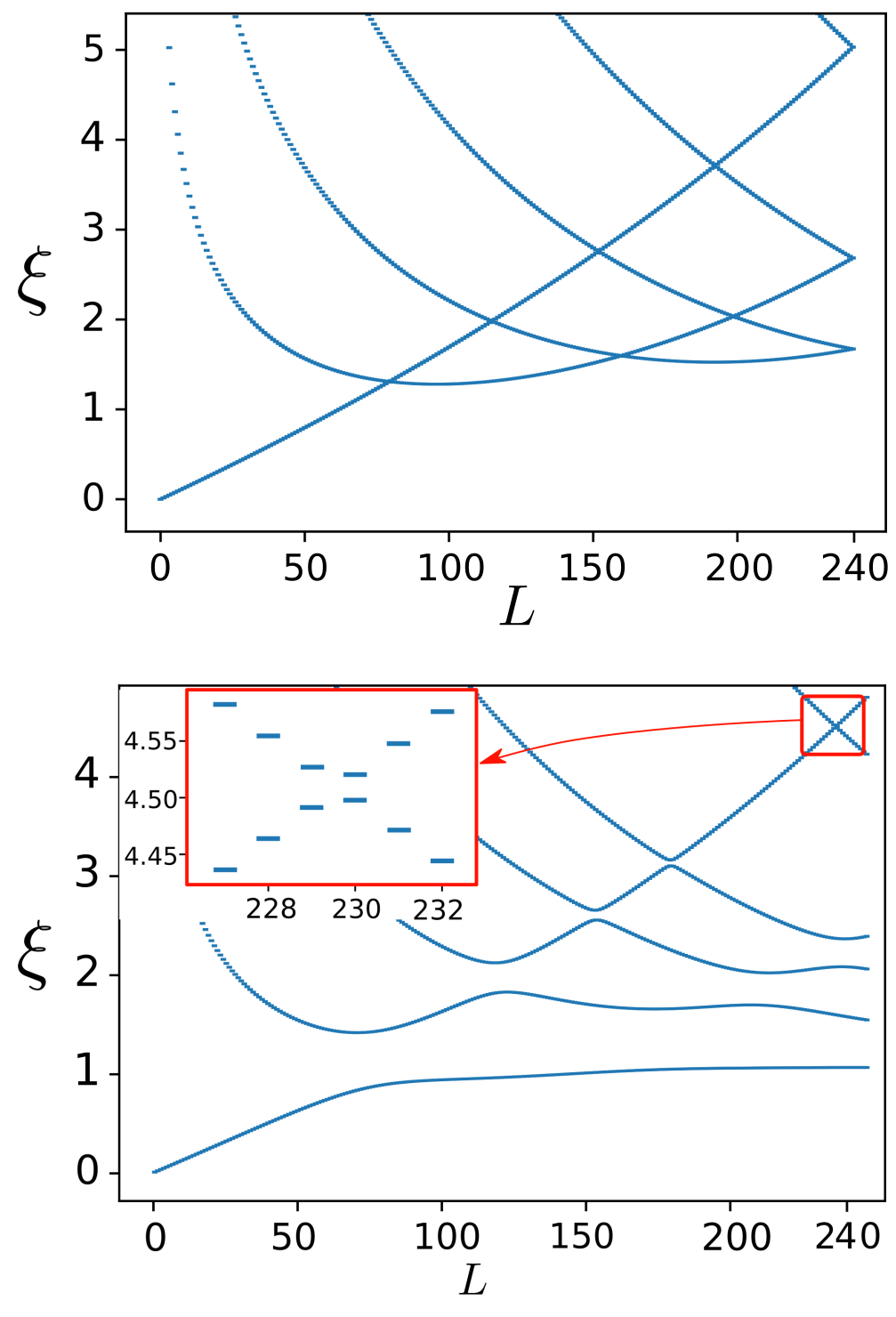}
\caption{Entanglement spectrum (ES, $\xi_i=-log_2\lambda_i$) of a subsystem of size $L$ in the ground state wavefunction at $\Phi=240$ (eq.\eqref{groundstate}, upper plot) and at $\Phi=240.5$ (eq.\eqref{superpos}, lower plot) for a system of $N^{*}=488$ Cooper pairs and as a function of the subsystem size $L$. The index $i$ labels the ES eigenvalues. The inset in the lower figure shows the lifting of degeneracies observed in the upper plot.
%These plots are for $N=488$, $\xi_i=-log_2\lambda_i$ vs $L$ (A) $\epsilon/V=240$ (B) $\epsilon/V=240.5$ \textbf{Remove the (a) and (b) labels. Increase the x axis tick size for upper figure.}
}
\label{fig:metalnearflux}
\end{figure}
\par\noindent
We begin our discussion of the entanglement features of the CPI phase with an investigation of the entanglement spectrum for the plateau ground states, whose wavefunction is given by $|\psi_{\textrm{P}}\rangle=| S,S_z \rangle$ for $\Phi= m~,m\in\mathcal{Z}$. We will also present the entanglement spectrum at the plateau transitions where the ground state is degenerate, $\psi_{\textrm{PT}}=\frac{1}{\sqrt{2}}(|S,S_z\rangle+|S,S_z+1\rangle)$ for $\Phi= m/2~,m\in\mathcal{Z}$. 
%We have $N=2S$ number of anderson pseudo spins in the system. 
We first 
%obtained from the plateau ground state of $N$ pseudospins at a given $\Phi$, 
Schmidt decompose
the state $|\psi_{P}\rangle$ (with $n=S^{z}+N^{*}/2$ $\uparrow$-pseudospins) into subsystems of length $L$ and $N^{*}-L$ (with $l$ and $n-l$ $\uparrow$-pseudospins respectively)
\begin{equation}
|\psi_{P}\rangle = \sum_{l=l_{min}}^{l_{max}} \lambda_{l}^{n} |L,l\rangle \otimes | N^{*}-L,n-l \rangle~,
\label{basicpartition}
\end{equation} 
where $\lambda_{l}^{n}$ are the Schmidt coefficients. The number $l$ ranges within $l_{min}$ and $l_{max}$ given by 
\begin{eqnarray}
l_{max}&=&n~~\textrm{for}~~ n\le L~~\textrm{and}~~
%\quad  
\\
%n > L \quad l_{max}l_{max}
&=&L ~~\textrm{for}~~n > L~,\\
l_{min} &=&0 ~~\textrm{for}~~ n\le (N^{*}-L) ~~\textrm{and}\\
&=&n-(N^{*}-L)~~\textrm{for}~~n\le (N^{*}-L)~.
% n\le (N-L) \quad l_{min}&=&0  \nonumber\\
%n>(N-L)  \quad l_{min}&=&n-(N-L)~.
\end{eqnarray}
From the pure state density matrix $\rho_{\textrm{P}}=|\psi_{\textrm{P}}\rangle \langle\psi_{\textrm{P}}|$, we can then obtain a reduced density matrix $\rho_{\textrm{P,L}}(n)$ for a subsystem of $L$ pseudospins
%(by tracing over $N-L$ pseudospins) 
%Further, we Schmidt decompose $\rho_{\textrm{P,L}}$ 
%(and where the state $|\psi\rangle_{P}$ had $n=S^{z}+N/2$ $\uparrow$-pseudospins)~(CITE)
%On the plateau the density matrix can be written as 
\begin{eqnarray}
\rho_{\textrm{P,L}}(n)&=&\sum_{l=l_{min}}^{l_{max}} (\lambda_{l}^{n})^{2} |L,l\rangle\langle L,l|~.
%\otimes | N-L,n-l \rangle~,
\label{reddenmat}
\end{eqnarray}
The Schmidt coefficients ($\lambda_{l}^{n}$) are determined by the combinatorial factor that specifies the number of ways one can choose $l \uparrow$ spins from $n \uparrow$ spins:
\begin{equation}
(\lambda_{l}^{n})^{2}=\frac{C^{L}_{l}C^{N^{*}-L}_{n-l}}{C^{N^{*}}_{n}}=\frac{L!(N^{*}-L)!n!(N^{*}-n)!}{l!(L-l)!(n-l)!(N^{*}-L-n+l)!N^{*}!}~.
\label{schmidtcoeff}
\end{equation}
\par\noindent
At the plateau transitions, we start with a pure state density matrix obtained from the linear superposition state $|\psi_{\textrm{PT}}\rangle$, 
$\rho_{\textrm{PT}} = |\psi_{\textrm{PT}}\rangle \langle\psi_{\textrm{PT}}|$,
%$\rho_h=\frac{1}{2}(|N,n\rangle\langle N,n |+|N,n+1\rangle\langle N,n+1 |$\\$+|N,n\rangle\langle N,n+1|+|N,n+1\rangle\langle N,n |)$, 
and proceed identically as above to obtain the reduced density matrix $\rho_{\textrm{PT,L}}(n)$. %Then we trace out $N-L$ spins out of this $N$ spins and calculate the reduced density matrix $\rho_L=Tr_{L'} (~\rho)$ for different L. 
The entanglement spectrum (ES) is obtained from the Schmidt eigenvalues, $\xi_{i}^{n}=-log_2(\lambda_{i}^{n})$, for $\rho_{\textrm{P,L}}(n)$ and $\rho_{\textrm{PT,L}}(n)$ 
%gives the entanglement spectra (ES) on the plateau and plateau transition respectively 
with given values of $N^{*}$ and $L$.
\par\noindent
For a fixed $N^{*}=488$ pseudospins, 
%that is the eigenvalues values ($\lambda_i$) of the reduced density matrix.
we plot the ES for various values of the reduced partition size $L$ for the case of the plateau at strong coupling $\Phi=0$ and the first plateau transition at $\Phi=1/2$ in Fig.\ref{fig:01flux}(a) and (b) respectively. The double degeneracy of all levels in the ES for all $L$ at $\Phi=0$ is revealed by the small splitting revealed at $\Phi=1/2$. This double degeneracy for the \textit{entire spectrum} reflects the additional particle-hole symmetric nature of the CPI ground state at $\Phi=0$, and corroborated by the degeneracy lifting precisely at the transition point ($\Phi=1/2$). 
%This double degeneracy of the plateau ground state is related to the two topologically degenerate sectors in the ground state with different eigenvalues of the non-local operator $X$ given earlier, $X=\pm 1$, as the splitting in the ES at the plateau transition ($\Phi=1/2$, Fig.\ref{fig:01flux}(b)) clearly arises from the degeneracy of two ground states with different values of $Z$. 
We have checked that a similar degeneracy of the ES is revealed for all other plateau ground states at $\Phi=m,~m\in\mathcal{Z}, m>0$ for \textit{only} a bi-partitioning of the system $L=N^{*}/2$. For instance, in Fig.\ref{fig:metalnearflux}, we present the ES at a weak coupling plateau $\Phi=240$ and plateau transition $\Phi=240.5$ for a system of $N^{*}=488$ pseudospins. Here too, the plots clearly show the double degeneracy of the plateau and the degeneracy lifting at the transition. The restricted degeneracy of the ES for all bi-partitioned CPI plateaux ground states with $\Phi >0$ may be associated with their topological order. However, this requires further investigation. 
%At zero flux value every entanglement spectrum is doubly degenerate. 
%At half-flux value the state become degenerate thus zero cost tunneling excitation makes those degenerate entanglement spectrums gapped as shown in the figure \ref{fig:01flux}(A).\\
%\par\noindent
%Near metal,\ref{fig:metalnearflux}). 
%Here one can see that at Zero flux every entanglement spectrum si doubly degenerate. 

\subsection{Entanglement Entropy of the plateau ground states and transitions} 
%and Entanglement Specrtum}
%As we tune the parameter $\phi$, we pass through differnt $S_z$ ground states. Here 
We now compute the bipartite entanglement entropy in momentum space for various plateau ground states obtained by tuning the parameter $\phi$. As before, we take a system where $N^{*}$ and $L$ are the total number of pseudospins and number of pseudospins within the reduced subsystem, while $n$ and $l$ are the number of $\uparrow$-pseudospins within $N^{*}$ and $L$ respectively. 
%Say the size of the momentum space is N (number pseudo spins), L is the size of the sub system. If $N_{\uparrow}=n$ within the whole system N and l is the number of up spins within the subspace L. 
%Then we can write the configurations as 
\begin{figure}
\includegraphics[scale=0.5]{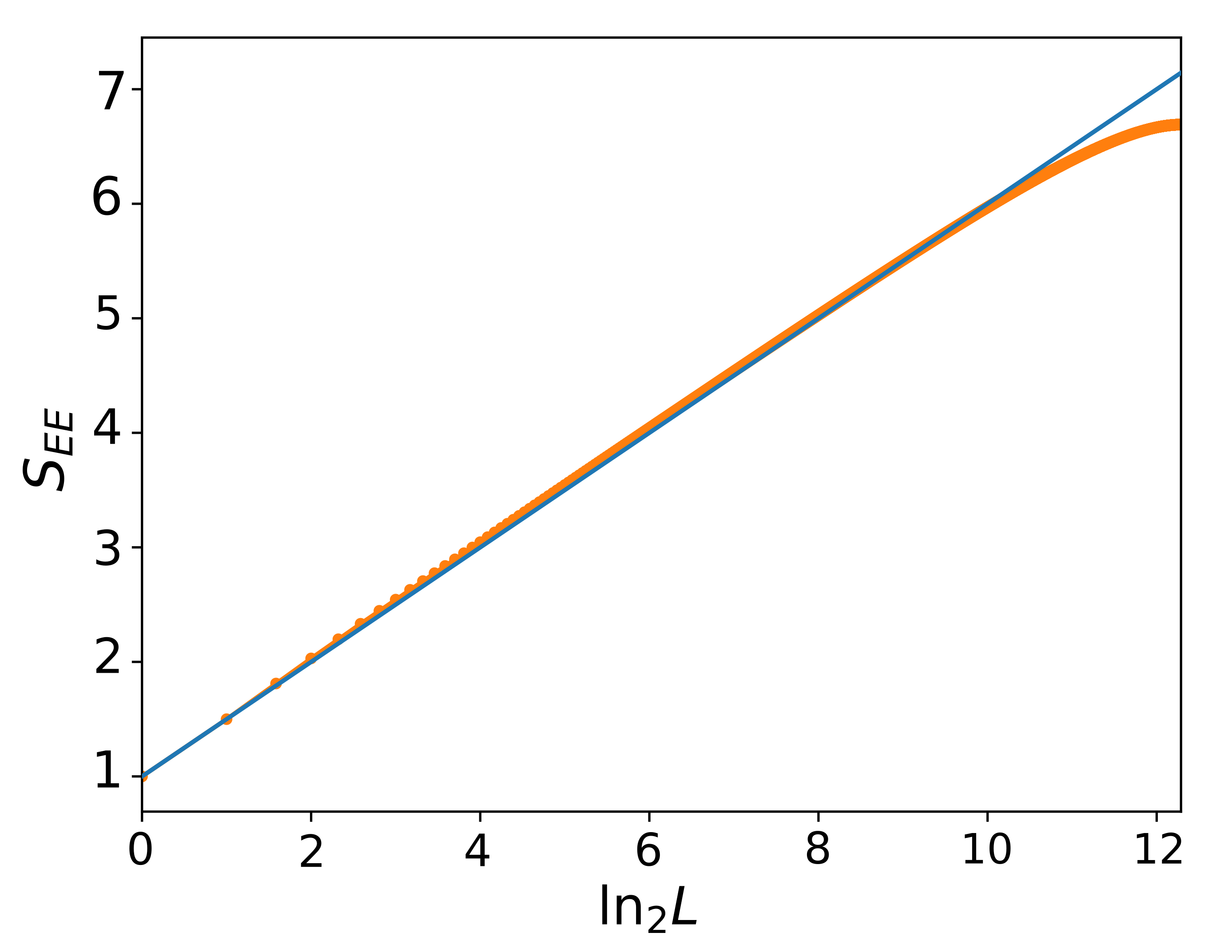}
\caption{The red curve shows the variation of the entanglement entropy (EE) with $\ln_{2}(L)$ ($L$ is the subsystem size) for the CPI ground state at $\Phi=0$ for a system of $N^{*}=488$ Cooper pairs. The blue curve is a linear fit to the form $1/2\ln_{2}(L) + 1$. See text for discussion.}
\label{zeroplatscaling}
\end{figure}
%\begin{figure}
%\includegraphics[scale=0.8]{plt/EEcontour}
%\caption{Contourplot of EE in L-Flux plane.}
%\label{fig:eecplot}
%\end{figure}
%\begin{figure}
%\centering
%\includegraphics[scale=0.68]{plt/ESsplit}
%\caption{Entanglement content for each state for different Flux}
%\label{fig:ESsplit}
%\end{figure}
The bipartite entanglement entropy ($S_{EE}$) can simply from the Schmidt coefficients ($\lambda_{l}^{n}$) via the following formula
\begin{equation}
S_{EE} (n,L) = -\sum_{l=l_{min}}^{l_{max}}  d_{l} |\lambda_{l}^{n} (L)|^{2}log_{2}|\lambda_{l}^{n} (L)|^{2}~, 
\label{basicentangentr}
\end{equation}
where $d_{l}$ is the degeneracy factor for the $l$th state of the entanglement spectrum. We have observed earlier in Fig.\ref{fig:01flux}(a) that for the case of the strong coupling ground state at $\Phi=0$, $d_{l}=2~\forall l$ due to the two topologically distinct sectors $X=\pm 1$. The appearance of the constant $d_{l}=2$ thus signifies the influence of the topological nature of the ground state manifold on $S_{EE}$. In Fig.\ref{zeroplatscaling}, we see that $S_{EE}$ varies linearly with $\textrm{log}_{2}(L)$ for $N^{*}=10000$ for a large range of $L$, departing from the linear variation only very near the equipartitioning value of $L=N^{*}/2$. We have found the value of the slope to be a simple number ($1/2$) as $\Phi$ is varied through the first 2000 plateaux. However, it is not clear whether this indicates a universality of plateaux ground states observed at strong coupling with those at intermediate coupling. Unlike the observation of logarithmic scaling of $S_{EE}$ with subsystem size in 1+1D quantum critical systems~(see Ref.\cite{eisert2010colloquium} and references therein), the log-scaling observed by us in Fig.\ref{zeroplatscaling} is indicative of %criticality. We are instead discussing the entanglement entropy of 
the physics of a gapped ground state of the effectively zero-dimensional Hamiltonian $H_{coll}$ (eq.\eqref{hcoll01}) obtained from the RG~\cite{vidalee}. 
%at present, the significance of this observation remains unclear.  
\par\noindent
The value of the intercept ($1$, in units of $\textrm{log}_{2}(2)$) is the entanglement entropy of a subsystem size of $L=1$ and corresponds to a maximally mixed pseudospin. The intercept is, however, observed to decrease steadily beyond the first 200 plateaux as $\Phi$ is varied, indicating that a single pseudospin's entanglement with the rest of the system within a plateau ground state is lowered as $\Phi$ is tuned towards weak coupling.
\begin{figure}
\includegraphics[scale=0.5]{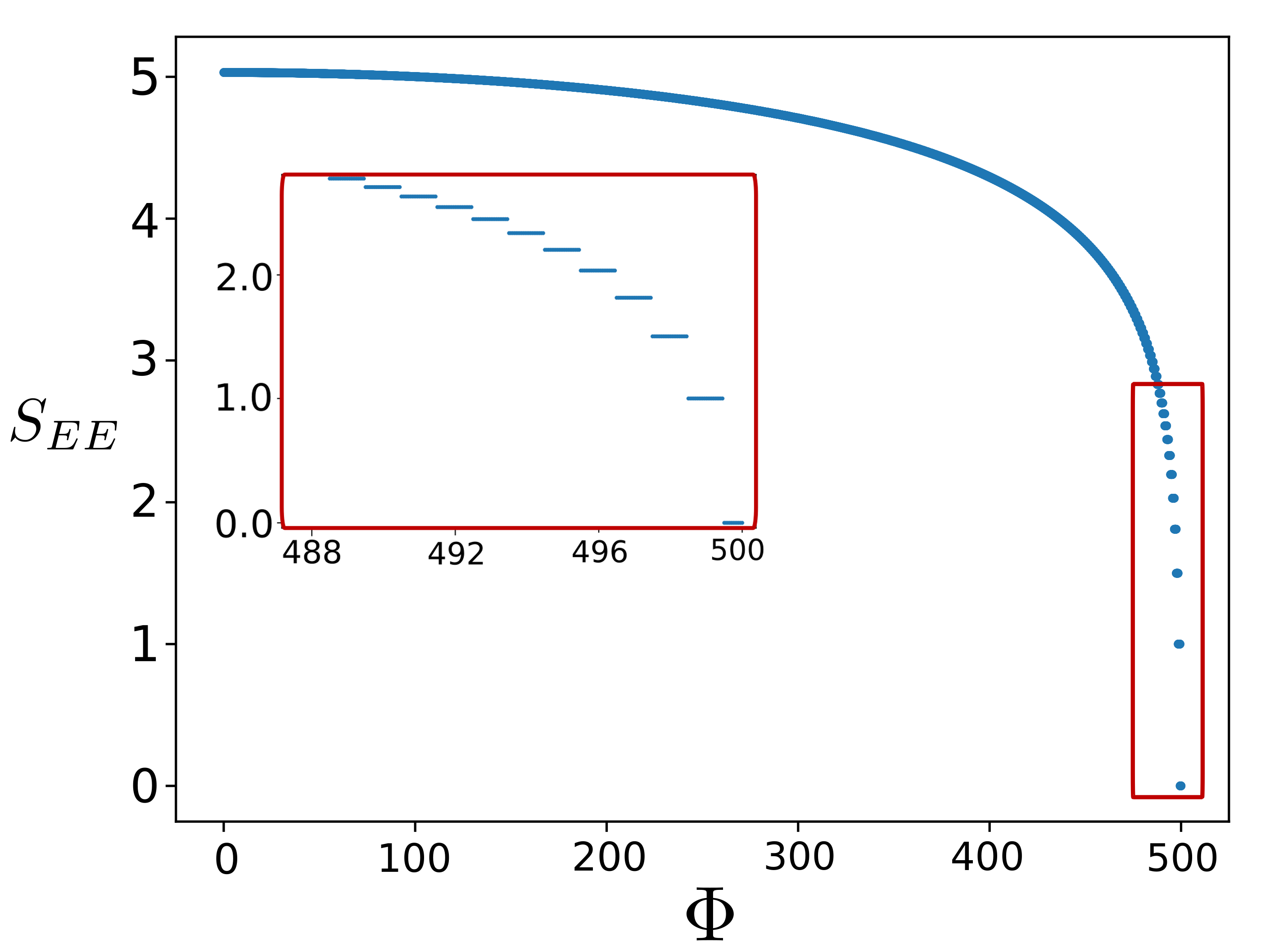}
\caption{Plot of bipartite entanglement entropy for various CPI ground states corresponding to different various flux $\Phi$ of a system of $N^{*}=500$ Cooper pairs. Inset shows the rapid fall of the bipartite EE upon approaching the CPI to Metal ($\Phi=500$) transition.}
\label{platent}
\end{figure}
%(PRESENT PLOT FOR (TOPOLOGICAL) S-EE OF ZEROTH PLATEAU HERE.)
%\begin{eqnarray}
%|n>&=&\displaystyle\sum_{l=l_{min}}^{l_{max}}\sqrt{ p_l } |l>\otimes |n-l>\\
%n\le L \quad l_{max}&=&n \nonumber\\
%n > L \quad l_{max}&=&L  \nonumber\\
% n\le (N-L) \quad l_{min}&=&0  \nonumber\\
%n>(N-L)  \quad l_{min}&=&n-(N-L)
%\end{eqnarray}
%Thus this $p_l$ is totally determined by the commutorial factor , how many ways one can choose l up spins from n up spins as following.
%\begin{equation}
%p_l=\frac{C^{L}_{l}C^{N-L}_{n-l}}{C^{N}_{n}}=\frac{L!(N-L)!n!(N-n)!}{l!(L-l)!(n-l)!(N-L-\{n-l\})!N!}
%\end{equation}
%We have partitioned the whole momentum space in two sub-system A of size $L$ and B of size $(N-L)$ and we can now find the entanglement among them.Thus the density matrix can be written as
%\begin{eqnarray}
%\rho &=&|n><n|=\displaystyle\sum_{l,\bar{l}=l_{min}}^{l_{max}}\sqrt{ p_l }\sqrt{ p_{\bar{l}} } |l_A,n-l_B><\bar{l}_A,n-\bar{l}_B|\nonumber \\
%\rho_B &=& Tr_A \rho=\displaystyle\sum_{m=l_{min}}^{l_{max}}{ p_m }|n-m><n-m|
%\end{eqnarray}
%Thus the Entanglement Entropy for  this momentum space partitioned system is .
%\begin{equation}
%S(L)=-\displaystyle\sum_{l=l_{min}(L)}^{l_{max}(L)}p_l log_2(p_l)
%\end{equation}
%Thus the Entanglement Spectrum is the $p_l$'s itself. Here one can see the as you approach the last plateau Entanglement Entropy vanishes. 
Further, in Fig.\ref{platent}, we present the variation of the bipartite $S_{EE}$ with the parameter $\Phi$ for a system with $N^{*}=1000=2L$. The plot clearly shows that 
%$S_{EE}$ has a step-like variation, %and falls rapidly from a maximal value at strong coupling ($\Phi=0$) to zero as $\Phi$ is tuned towards the gapless metal. This shows clearly that and 
the $\Phi=0$ plateau possesses the largest entanglement content, and that this is rapidly lowered to zero as $\Phi$ is tuned through various plateaux towards the gapless metal~\cite{vidalee}.
%\subsection{Entanglement entropy at the plateau transitions}
\par
\begin{figure}
\includegraphics[scale=0.5]{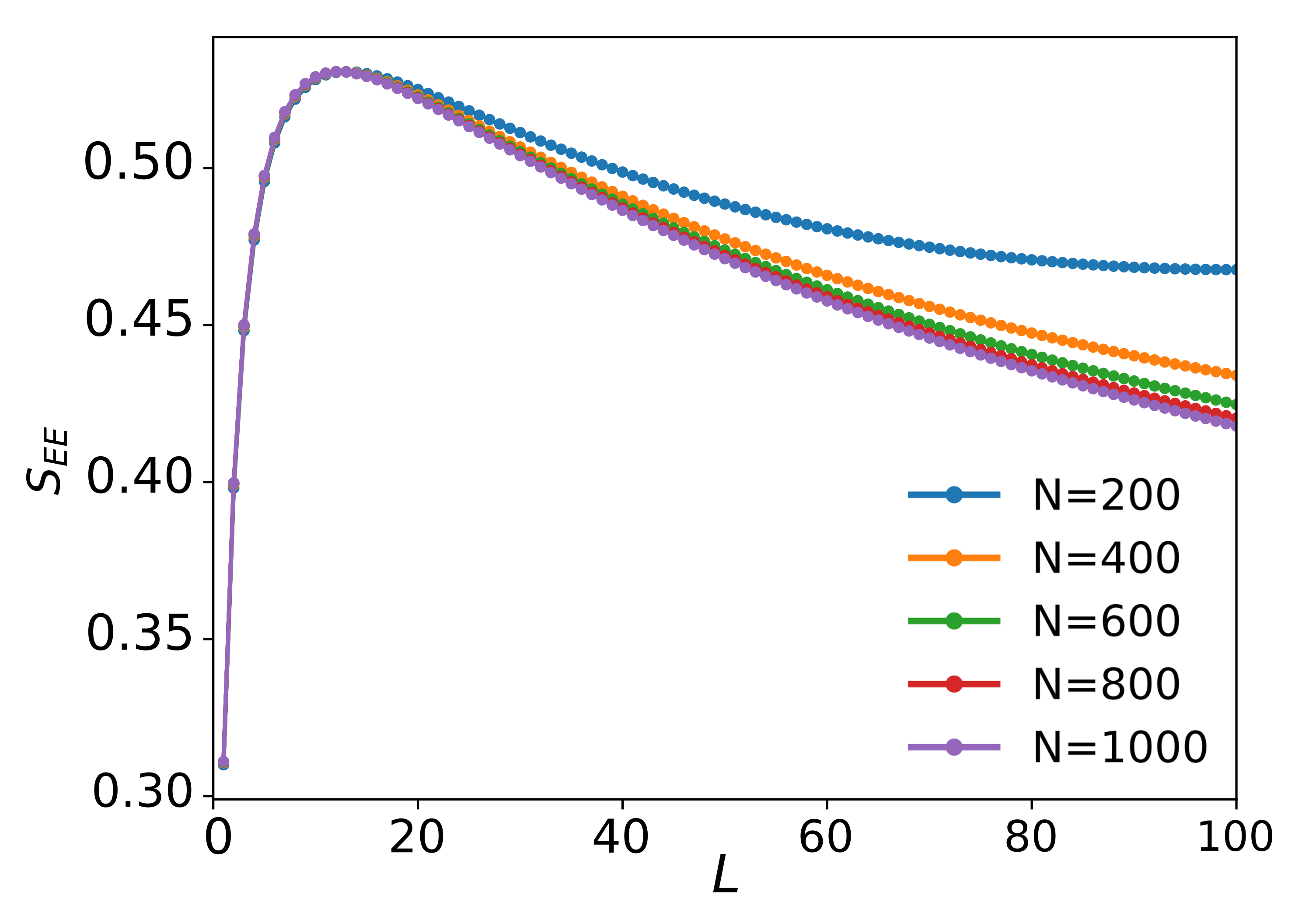}
\caption{Variation of the entanglement entropy ($S_{EE}$) with subsystem size $L$ at the first plateau transition ($\Phi=0.5$) for various system sizes in the range $100\leq N^{*}\leq 1000$ Cooper pairs.}
\label{fig:FirstPlatTransEE}
\end{figure}

\begin{figure}
\includegraphics[scale=0.5]{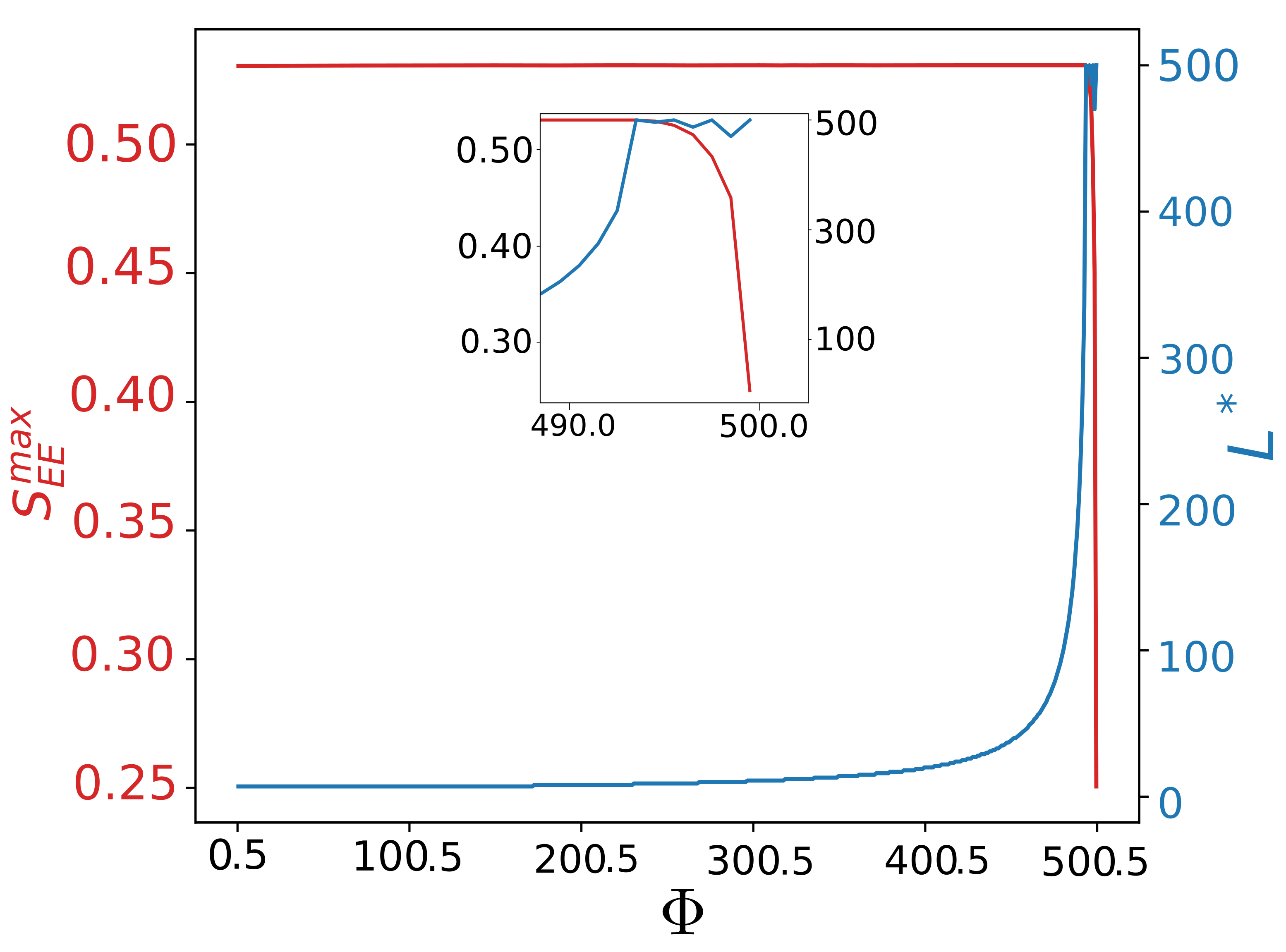}
\caption{Plot of the maximum entanglement entropy ($S_{EE}^{max}$, red curve, left y-axis) and the corresponding subsystem size ($L^{*}$, blue curve, right y-axis) at the transitions between various CPI ground states as $\Phi$ is varied for a system of $N_{C}=500$ Cooper pairs. Inset: Rapid variation of $S_{EE}^{max}$ and $L^{*}$ with $\Phi$ upon approaching the CPI to Metal ($\Phi=500$) transition.}
\label{fig:TransEEpeak}
\end{figure}

\noindent
%At the plateaux transitions, the Cooper pair number has large fluctuations (see $\langle \Delta S^{z}\rangle$ in Fig.\ref{hall}).  %rapidly because of the fluctuation in $S_z$. 
In contrast to Fig.\ref{zeroplatscaling} above, the entanglement entropy at the first transition ($\Phi=0.5$, see Fig.\ref{fig:FirstPlatTransEE}) computed using eq.\eqref{superpos} for $c_{1}=1/\sqrt{2}=c_{2}$ for various system sizes $N^{*}$ displays a non-monotonic variation of $S_{EE}$ with the subsystem size $L$. Remarkably, $S_{EE}$ displays a common peak at $L^{*}=7$ for system sizes ranging between $100\leq N^{*}\leq 1000$, with all the curves collapsing onto a universal curve for $L\leq L^{*}$. This suggests that the entanglement content at the first plateau transition is dominated by small subsystem size. While the last data point in Fig.\ref{fig:FirstPlatTransEE} corresponds to the equipartition $S_{EE}(L=N^{*}/2)$ for $N^{*}=100$, we have checked that $S_{EE}(L=N^{*}/2)$ falls logarithmically with $N^{*}$. As shown in Fig.\ref{fig:TransEEpeak}, we have also observed that the maximum value of $S_{EE}$ observed in Fig.\ref{fig:FirstPlatTransEE} remains unchanged (as $\Phi$ is tuned across various plateau transitions for a system of size $N^{*}=1000$) until almost the very last few transitions, where it falls rapidly to zero. On the other hand, $L^*$ increases gradually with $\Phi$, climbing rapidly at the last few transitions. This clearly demonstrates that the peak in $S_{EE}$ is a universal feature of the plateau transitions. We have also observed that the $S_{EE}$ for a single pseudospin (i.e., $L=1$) falls to zero gradually from its value at the first transition as $\Phi$ is varied.
%
%Entanglement Entropy $S=-\lambda_+ log \lambda_+ -\lambda_- log \lambda_-$. Here one can see that due to this complicated structure of this Entanglemetn entropy , minima of the entanglemetn of entropy is not at $D_1=D_2$. This entanglement is minima for $D_2=D^*$ . This $D^*=D^*(M) $ is a funtion of M the plateau value shown in the figure (\ref{fig:EEmin}). 
%The plot shows that $c{1}=1/\sqrt{2}=c_{2}$ is the coefficient that minimises $S_{EE}$ as $\Phi$ is varied between the lowest 75 plateau. for the first one reaches the metal state where the minimum Entanglement vanishes.
%\subsection{EE at finite temperatures}
%Here we will study the $H_{coll}$ hamiltonian.
\par\noindent 
Finally, we present the computation of the entanglement entropy of the plateau ground states and transitions at finite temperature. 
The thermal density matrix for the plateau ground state $|N^{*}=2S,n=S^{z}+\frac{N^{*}}{2}\rangle$ %with $n$ number of $\uparrow$-pseudospins 
can be written as 
\begin{eqnarray}
\rho (\beta)&=&\sum_{n} \frac{e^{-\beta E_{n}}}{Z} |N^{*},n\rangle \langle N^{*},n |~,%\nonumber\\
%| n \rangle &=& \sum_{l=l_{min}}^{l_{max}} D^n_l(N,L) ~~ | l \rangle_A \otimes | n-l \rangle_B
\end{eqnarray}
where $\beta$ is the inverse temperature and $Z$ the partition function. Equipartitioning the system precisely as described earlier in eqs.\eqref{reddenmat}-\eqref{schmidtcoeff} (with $L=N^{*}/2$), we obtain the thermal reduced density matrix as
%In this system we will be calculating equipartitoned entanglement. Our partitioning scheme is same as previous case \ref{section:zero_teperature}. Where $D^n_l(N,L)$ is defined in the equation (\ref{communatorial}), m is the $S_z$ eigen value and l and m-l are $S_A^z$,$S_B^z$ eigenvalues respectively ,the density matrix is shown below . Then we calculate the reduced density matrix $\rho_A$.
\begin{eqnarray}
\rho_{L} (\beta) =\sum_{l,n} \frac{e^{-\beta E_{n} }}{Z}  (\lambda_{l}^{n})^{2} |L, l \rangle \langle L, l|~.
\end{eqnarray}
The reduced density matrix is easily seen to be diagonal
\begin{equation}
[\rho_{L}(\beta)]_{l,l'} = \delta_{l,l'}\sum_{n} (\lambda_{l}^{n})^2 \frac{e^{-\beta E_{n}}}{Z}~,
\end{equation}
such that the entanglement entropy at a non-zero temperature is obtained as
\begin{eqnarray}
S_{EE}(L,\beta) &=& - \sum_{l} [\rho_{L}(\beta)]_{l,l}~log_2([\rho_{L}(\beta)]_{l,l})~.
\end{eqnarray}
Precisely the same formalism can also be carried out with the state equal admixture state at the plateau transition ($|\psi\rangle_{PT}$).
\begin{figure}[htb]
\includegraphics[scale=0.5]{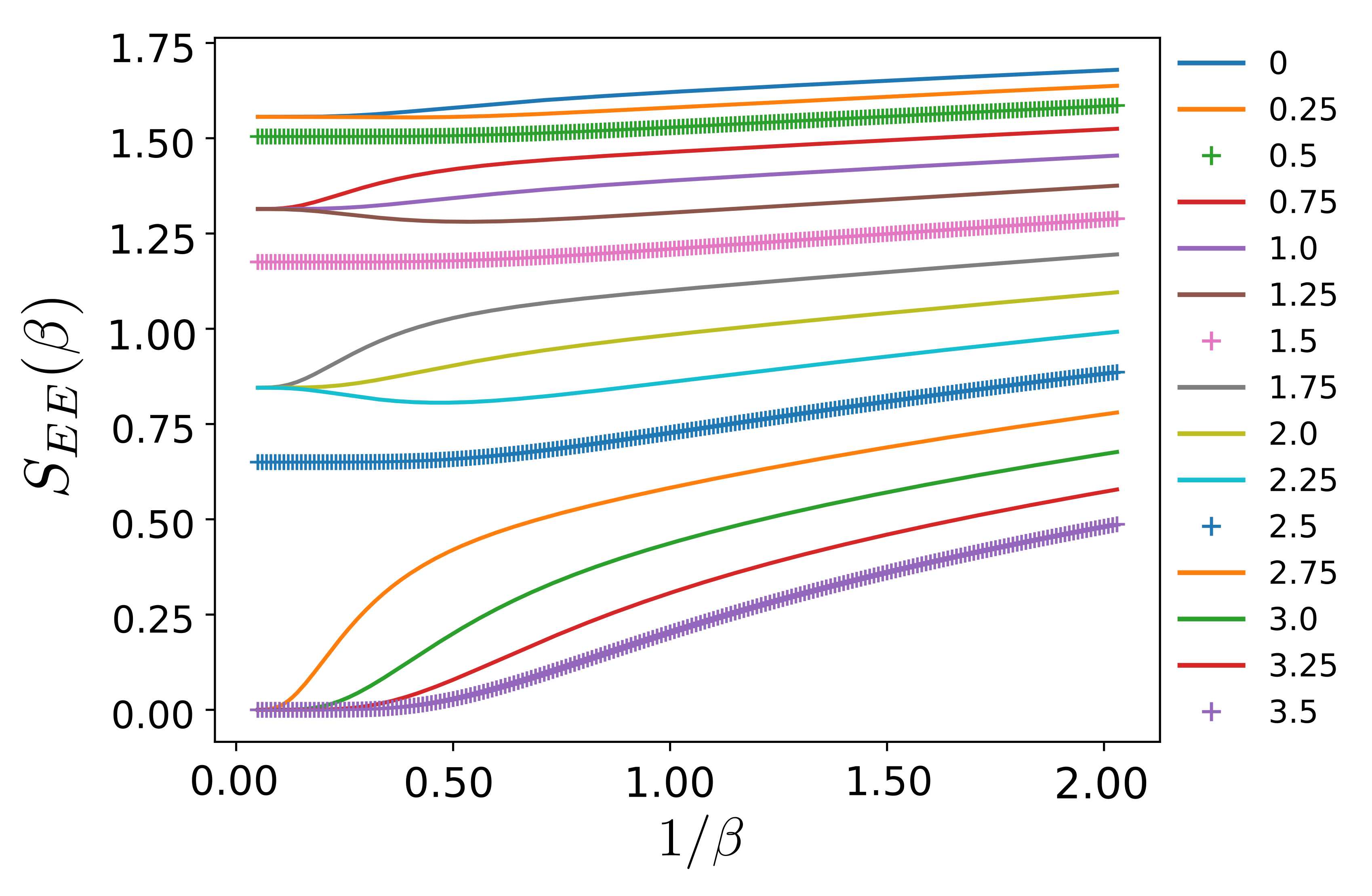}
\caption{Variation of the equipartition entanglement entropy ($S_{EE}(\beta)$) with temperature ($\beta^{-1}$) for a subsystem size $L=4$ Cooper pairs. Various coloured curves correspond to the CPI ground states and transition ground states at different values of the flux $\Phi$. See text for discussion.}
\label{finiteEE}
\end{figure}
\par\noindent
In Fig.\ref{finiteEE}, we present a numerical evaluation of $S_{EE}(L,\beta)$ for a subsytem of $L=4$ pseudospins with varying $k_{B}T=\beta^{-1}$ for the first four plateaus (coloured curves centered about $\Phi=0, 1, 2$ and $3$) and transitions ($+$ symbol curves centered about $\Phi=0.5, 1.5, 2.5$ and $3.5$). While we have chosen a small system here ($N^{*}=8$) for the sake of visual clarity, we have checked that all features of the plot are qualitatively unchanged for larger $N^{*}$. Remarkably, the plot shows that the $S_{EE}$ corresponding to the transitions clearly separates all curves arising from neighbouring plateaus for temperatures $k_{B}T << V^{*}$. The $S_{EE}$ curves for all $\Phi$ corresponding to a plateau collapse to a universal value at $T=0$ characteristic of that plateau. Thermal fluctuations are observed to affect the curves of both the plateaus and the transitions in the same manner. For instance, the position of the divergence of the curves for a given plateau as $T$ is increased suggests the robustness of that plateau to thermal transitions. Clearly, the $\Phi=0$ plateau (strong coupling) is the most robust, the $\Phi=1$ slightly less and so on, ending at the last plateau (at $S_{EE}(T=0)=0$) beyond which lies the gapless metal. Similarly, the curves for transitions associated with higher plateaus depart from their initial flat behaviour at lower temperatures in comparison to that for lower plateaus. The domination of thermal fluctuations as $T$ is raised is also clearly observed: various $S_{EE}$ curves corresponding to a particular $T=0$ plateau show a linear increase with $T$ asymptotically, and with a slope common to that of the $S_{EE}$ curve for the transition that leads to the next plateau (e.g., the slopes of the $S_{EE}$ curves for $\Phi=0, 0.25$ and $0.5$ are the same for large $T$ etc.).

\section{Passage to the BCS ground state}
\label{section:cpi2bcs}
\noindent
In order to chart the passage from the number-fixed CPI ground state to the conjugate phase-fixed BCS ground state, we will carry out the RG analysis upon adding a global $U(1)$-symmetry breaking term ($-2B\sum_{k}S^{x}_{k},~B>0$) to the collective Hamiltonian $H_{coll}$ (eq.\eqref{eq:hcoll}) with the repulsive density-density interaction $U=0$. The $B$ field represents a Josephson coupling to an external phase-fixed BCS superconductor. We will show that at large $B$, the RG flow leads to a BCS-like ground state. This will also be reinforced by studying the variation of several quantities with the symmetry-breaking field $B$, e.g., inter-$k$ entanglement, helicity-partitioned entanglement, helicity cross-correlation, pair number fluctuation etc. 
%\subsection{Renormalization Study of the symmetry broken system}
%Here under RG we will show how inpresence of symmetry breaking field Hamiltonian $H_{sb}$ reaches fixed point solution. This time we willbe adding a U(1) symmetry breaking term 
\par\noindent
Thus, we begin with the Hamiltonian
\begin{eqnarray}
H_{SB}=-\frac{2\bar{\epsilon}}{N}\displaystyle\sum_{k}S_{k}^{z}-\frac{\bar{V}}{2N}\displaystyle\sum_{kk'}(S_{k}^{+}S_{k'}^{-}+\textrm{h.c.}) - 2|B|\displaystyle\sum_{k}S_{k}^{x}~.~~~~
\end{eqnarray}
The RG equations for $\bar{\epsilon}$ and $\bar{V}$ are those given earlier in eq.\eqref{U1RGeq} (with $\bar{W}=\bar{V}/N$), while the RG for the symmetry-breaking field $B$ is found to be
%zero mode Hamiltonian is 
\begin{eqnarray}
%\epsilon^{(j+1)}-\epsilon^{(j)}=\frac{|V^{(j)}|^2}{\omega-\epsilon^{(j)}}\nonumber\\
%|V^{(j+1)}|-|V^{(j)}|=-\frac{|V^{(j)}||V^{(j)}|}{\omega-\epsilon^{(j)}}\nonumber\\
\frac{\Delta |B^{(j)}|}{\Delta \log\frac{\Lambda_j}{\Lambda_0}}=-\frac{1}{2}\frac{|\bar{W}^{(j)}||B^{(j)}|}{\bigg(\omega-\frac{\epsilon^{(j)}}{2}-\frac{U}{4}\bigg)}~.
\label{b.eq}
\end{eqnarray}
In the regime $\omega<\frac{\epsilon^{(j)}}{2}+\frac{U}{4}$, both $\bar{W}$ and $B$ are found to be RG relevant. From the RG eqs.\eqref{U1RGeq} and \eqref{b.eq}, we find
\begin{eqnarray}
\frac{\Delta(\epsilon)}{\Delta|\bar{W}|}=-1,~ \frac{\Delta(\epsilon)}{\Delta B}=-\frac{|\bar{W}|}{2B}~\Rightarrow~\frac{\Delta|\bar{W}|}{\Delta B}=\frac{|\bar{W}|}{2B}~, \label{vb}
\end{eqnarray}
indicating that while both $|\bar{W}|$ and $B$ grow to strong coupling under the RG flow, the ratio $|\bar{W}|/B$ remains invariant. This shows that while the original BCS mean-field Hamiltonian~\cite{bcs} is achieved only in the limit of the RG invariant $|\bar{W}|/B \to 0$, a $U(1)$-symmetry broken BCS-like effective Hamiltonian is emergent from the RG flow at strong coupling  
%One can see the BCS meanfield hamiltonian is a very special hamiltonian where the ration .Thus one can see that simplicity of the zero mode study , easy to solve analytically. So our symmetry brokne hamiltonian is 
\begin{eqnarray}
H_{SB} = H_{coll}-B^{*}S_{x} ~,
\label{eq:hsb}
\end{eqnarray}
where $H_{coll}$ is given in eq.\eqref{collham}. Further, in appendix \ref{section:appendix_B}, we show that the familiar form of an exponentially small spectral gap is obtained from the RG flow to strong coupling in $B$. In considering the effective Hamiltonian obtained from the RG, we will henceforth drop the $*$ symbol from all couplings. Clearly, as $[S^{z},H_{SB}]\neq 0$, the total Cooper pair number operator (proportional to $S^{z}$) is no longer a conserved quantity. Further, the topological order parameter $Z$ encountered earlier is no longer a good order parameter, as $[Z,H_{SB}]\neq 0$: the effective Hamiltonians $H_{coll}\equiv H_{SB}(B=0)$ and $H_{SB}(B\neq 0)$ are topologically inequivalent.
We will demonstrate below that, as $B$ is tuned to larger values, the Cooper pair number fluctuations increase rapidly while the fluctuations in the conjugate $U(1)$ global phase is lowered. This indicates that a BCS-like ground state is attained under the RG flow of $B$ to strong coupling.
\par\noindent
In commonality with the BCS ground state, the ground state wavefunction for $H_{SB}$ is given by a linear superposition of states with different $S^{z}$ (eq.\eqref{groundstate})
%The wavefunction of this phase would be 
\begin{eqnarray}
|\psi (B)\rangle &=& \sum_{S^{z}=-S}^{S} \alpha_{B}(S^{z}) |S,S^{z}\rangle \nonumber\\
&=&\sum_{S{z}=-S}^{S} \alpha_{B}(S^{z}) \bigg(\sum_{k} c^{\dagger}_{k\uparrow}c^{\dagger}_{-k\downarrow}\bigg)^{S-S^{z}} |vac\rangle ~,
\end{eqnarray}
where the (normalised) coefficients $\alpha_{B}(S^{z})$ are functions of the symmetry-breaking field $B$. 
%While the precise form of this ground state waveunction is different from the BCS ground state, $|\psi_{BCS}>=\prod_k^{\otimes} ( u_k+v_k a_{k\uparrow}^{\dagger}a_{-k \downarrow}^{\dagger} )|vac>$, 
We will also show below that several properties of $|\psi (B)\rangle$ closely resemble those of $|\psi_{BCS}\rangle$ as $B$ is tuned to large values. For instance, we will show that at large $B$, $|\psi (B)\rangle$ leads to vanishing inter-$k$ entanglement. We recall that a vanishing inter-$k$ entanglement entropy is a special property of the BCS ground state, arising from the fact that different $k$-momenta electron-pair states are decoupled from one another. 
%Here we can see that $S_z,Z$ does not commutes with the Hamiltonian H anymore but only $S$ does. Now here we get $[S^2,H_{coll}]=0,~[S_z,H_{coll}]\neq 0,~[Z,H_{coll}]\neq 0,~[Z,\rho]\neq 0,~[Z,S_z]\neq 0$. These properties shows the presence of BCS like local ordere parameter.
%Thus this two situations $B=0$ and $B\neq 0$ are topologically inequivalent.Now question is how to connect these two topologically inequivalent phases,only can be connected using non-unitary operation. By applying a projection operator $P=e^{-BS_x}$ on the ground state of $H_{coll}$ one can get back symmetry brokne ground state.
%\begin{eqnarray}
%&&e^{2BS_x} |S,0>=\sum_{S_z} \alpha_{S_z}(B) |S,S_z>
%\end{eqnarray}
\subsection{Properties of the ground state}
\noindent
In order to obtain various properties of the ground state of the effective Hamiltonian $H_{SB}$, we carry out exact diagonalization computations for system sizes of $N^{*}=50$ Cooper pairs. 
%the hamiltonain $H_{sb}(B)$, calculate the entire eigenvalues and eigenspectra. 
We compute various quantities related to the ground state, e.g., fluctuation in the Cooper pair number and the conjugate global phase, helicity cross correlations, various measures of entanglement etc. 
%For different B we know the ground state wavefunction $|\psi_g(B)>$\\
\par
\begin{figure}
\includegraphics[scale=.65]{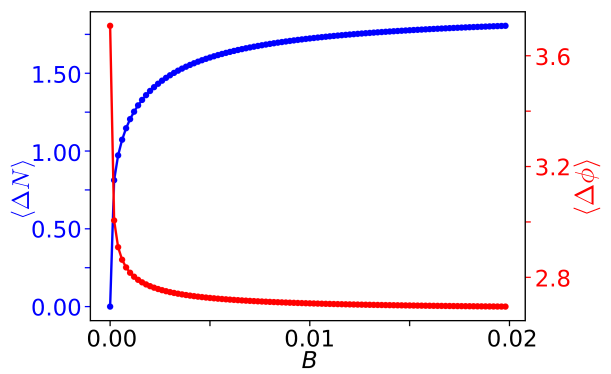}
\caption{Variation of fluctuation in number of Cooper pairs ($\langle \Delta N\rangle$, blue curve, left y-axis) and global phase ($\langle\Delta\phi\rangle$, red curve, right y-axis) with the global $U(1)$ symmetry breaking field $B$ for a system of $N^{*}=50$ Cooper pairs.
%\textbf{Increase the x-axis tick size and drop the third decimal place. Increase the sizes of the two sets of y-axis labels and ticks.}
}
\label{nfluc}
\end{figure}

\noindent
%\begin{figure}
%\includegraphics[scale=0.4]{plt/gap}
%\caption{Plot of the ratio of the energy gap density and $B$ ($\Delta E (B)/NB$) versus the $U(1)$-symmetry breaking field $B$.}
%\label{gapdensity}
%\end{figure}
%In Fig.\ref{gapdensity}, we present the variation of the ratio of the energy gap density $B$ ($\Delta E (B)/NB$) obtained from $H_{SB}$ as the field $B$ is varied. The plot shows the rapid fall of $\Delta E (B)/NB$ with $B$ to a value exponentially small compared to .
%We numerically diagonalize the Hamilonian $H_{sb}$ (\ref{eq:hsb}) and from that we calculate the gap $\Delta E$. We calculate the normalized energy gap density as $EgD=\frac{\Delta E}{N B}$ varying B.
In Fig.\ref{nfluc}, we present the variation of the fluctuations in the Cooper pair number ($\langle \Delta N\rangle \equiv \langle \Delta S^{z}\rangle$, blue curve) and conjugate global phase ($\langle \Delta\phi\rangle$, red curve). The plot clearly shows the rapid decline in $\langle \Delta\phi\rangle$ as $B$ is increased, together with an equally rapid growth in $\langle \Delta N\rangle$. We have also observed that a plot of the number fluctuations $\langle \Delta S^{z}\rangle$ versus $\langle S^{z}\rangle$ for different values of $\Phi$ and $B$ is strikingly similar to Fig.\ref{fig:thermals1}. Further, we shall see below that the entanglement content of the fluctuations induced by a non-zero $B$ are very different from that induced by thermal fluctuations.
%\textit{Number Fluctuation :} Here we will calculate the electron-pair number flucuation .We know that the number of cooper pair is related to $S_z$ value as $N_c=S-S_z$.Thus number fluctuation in the ground state can be is euqual to $S_z$ fluctuation.Thus Number fluctuation is 
%\begin{eqnarray}
%\Delta N_B=\sqrt{<\psi_g(B)|S_z^2|\psi_g(B)>-(<\psi_g(B)|S_z|\psi_g(B)>)^2}\nonumber
%\end{eqnarray}  
%\textit{Phase Fluctuation :} Phase is here dual to the Number operator $[N,\phi]\neq 0$. We can get eh phase flcutuation information by looking at $<\psi_g(B)|S^++S^-|\psi_g(B)>$. This quantity is not exactly the phase but we can get the phase fluctuation information by looking at it's fluctuation. Thus phase fluctuation is $\sqrt{<\psi_g(B)|\hat{P}_f|\psi_g(B)>-(<\psi_g(B)|\hat{P}_f|\psi_g(B)>)^2}$, where $\hat{P}_f=S^++S^-$.
\par
\begin{figure}
\includegraphics[scale=.48]{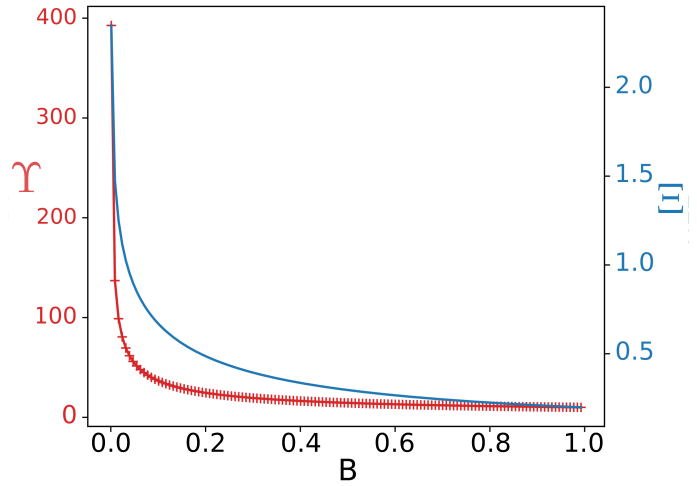}
\caption{Variation of helicity cross correlations ($\Upsilon$, red curve, left y-axis) and helicity partitioned entanglement entropy ($\Xi$, blue curve, right y-axis) in number of Cooper pairs ($\langle \Delta N\rangle$, blue curve, left y-axis) with the global $U(1)$ symmetry breaking field $B$ for a system of $N^{*}=50$ Cooper pairs.
%\textbf{Increase the x axis tick sizes. Also, change HCC to $\Upsilon$ and HEE to $\Xi$.}
}
\label{hcchee}
\end{figure}

\noindent
The red curve in Fig.\ref{hcchee} shows the variation of the helicity cross correlations ($\Upsilon$, eq.\eqref{HCC1}) with $B$. The plot displays the rapid decline of $\Upsilon$ that is characteristic of the CPI ground state towards zero as $B$ is tuned to large values. This is expected, as $\Upsilon$ vanishes for the BCS ground state. The blue curve in Fig.\ref{hcchee} shows the variation of the helicity-partitioned entangled entropy ($\Xi$) with $B$. $\Xi$ is derived from the reduced density matrix obtained by tracing out one of the two helicities ($\eta_{\pm}=sgn(k)sgn(\sigma)=\pm 1$).
%\textit{Helicity Cross Correlation :} In our problem we have studied the properties of the electron-pairs that helped us to form zero-momentum anderson pseudo-spin. We used Anderson pseudo spin \\
%$S_k^+=a_{-k\downarrow}a_{k\uparrow}$ , $S_k^-=a_{k\uparrow}^{\dagger}a_{-k\downarrow}^{\dagger}$ , $S_k^z=\frac{1}{2}[1-a_{k\uparrow}^{\dagger}a_{k\uparrow}-a_{-k\downarrow}^{\dagger}a_{-k\downarrow}]$. Helicity $\eta=sgn(k)sgn(\sigma)$.\\
%We are studying here 2D system and BCS type model that means in very low filling that means circular fermi sea. We make our convension for helicity as : momentum in the upper half-circle is $\eta=+1$ and lower one $\eta=-1$. In the VCS as $S_z=S_z^{\eta=+1}+S_z^{\eta=-1}$ is a good quantum number thus in ground state these two species of different helicities are highly entangled. Here we sill calculate the 
%\begin{eqnarray}
%H.C.C=<S_{\eta=+1}^+S_{\eta=-1}^->-<S_{\eta=+1}^+><S_{\eta=-1}^->+h.c.\nonumber
%\end{eqnarray}
The helicity-partitioned ground state wavefunction can be written as follows 
\begin{eqnarray}
&&|\psi_g(B)\rangle=\sum_{m=-S}^{S} C_m(B)~|S,S^{z}=m \rangle~,\nonumber\\
|S,m \rangle &=&\displaystyle\sum_{m_{\eta_{+}},m_{\eta_{-}}}D^{m}_{m_{\eta_{+}},m_{\eta_{-}}} |S/2, m_{\eta_{+}};S/2, m_{\eta_{-}}\rangle~,
\label{eq:symmetrybrokenstate}
\end{eqnarray} 
where $D^m_n$'s are Clebsch-Gordon coefficients given by
\begin{eqnarray}
&& D^{m}_{m_{\eta_+},m_{\eta_-}}=\delta_{m,m_{\eta_+}+m_{\eta_-}}~\times\nonumber\\
&& \sqrt{\frac{(S!)^2(S+m)!(S-m)!}{(2S)!(\frac{S}{2}-m_{\eta_+})!(\frac{S}{2}+m_{\eta_+})!(\frac{S}{2}-m_{\eta_-})!(\frac{S}{2}+m_{\eta_-})!}}~.~~~~~
\end{eqnarray}
%where $m_{\eta=-1}=m-m_{\eta=+1}$.
In the ground state, the total spin ($S$) is maximised, ensuring that the total spin within each helicity sector is also maximised (and taken to be $S/2$). 
%As we are only in the ground state and $S_x$ term does not switches on tunneling among different total spin states so can safely choose each helicity sector total spin is $S/2$. These $D^m_n$'s are the Clebsch–Gordan coefficients. 
%The helicity-partitioned entanglement entropy then offers a quantification of the entanglement between the opposite helicities: 
%creation and annihilations are correlated.\\
%\textit{Helicity partitioned Entanglement entropy :} We have already calculated the correlation between two helicity by the helicity Cross Correlation Calculation now here we willbe showing how much different helicity are entangled  with each other. Here we can take the wavefunction (\ref{eq:symmetrybrokenstate}) and construct the density matrix $\rho=|\psi_g(B)><\psi_g(B)|$ and partial trace on one helicity say $\eta=+1$. 
By tracing over a certain helicity ($\eta_{+}$), we obtain a reduced density matrix $\rho_{\eta_{-}}=Tr_{\eta_{+}}|\psi_g(B)\rangle\langle\psi_g(B)|$. %Due to the  Clebsch–Gordan coefficients this $\rho_{\eta=-1}$ is off-diagonal so using numpy we foudn it's 
The set of eigenvalues ($\{\lambda_i\}$) btained by diagonalising $\rho_{\eta_{-}}$ then gives the helicity-partitioned entanglement entropy $\Xi$ (i.e., a measure of the entanglement between the opposite helicities)
\begin{eqnarray}
\Xi (B)=-\displaystyle\sum_{i} \lambda_i(B) log_2(\lambda_i(B))~.
\end{eqnarray}
The variation of this entanglement entropy with $B$ is shown via the blue curve in Fig.\ref{hcchee}, displaying a rapid decline towards zero in the entanglement between the helicities $\eta_{+}$ and $\eta_{-}$ as $B$ is tuned to large values. This is consistent with the fact that the BCS ground state does not possess helicity entanglement; this arises simply from the fact that the BCS ground state wavefunction is a direct product state of pairs of electronic momenta $(k,-k)$.   
\color{black}
\begin{figure}
\includegraphics[scale=0.8]{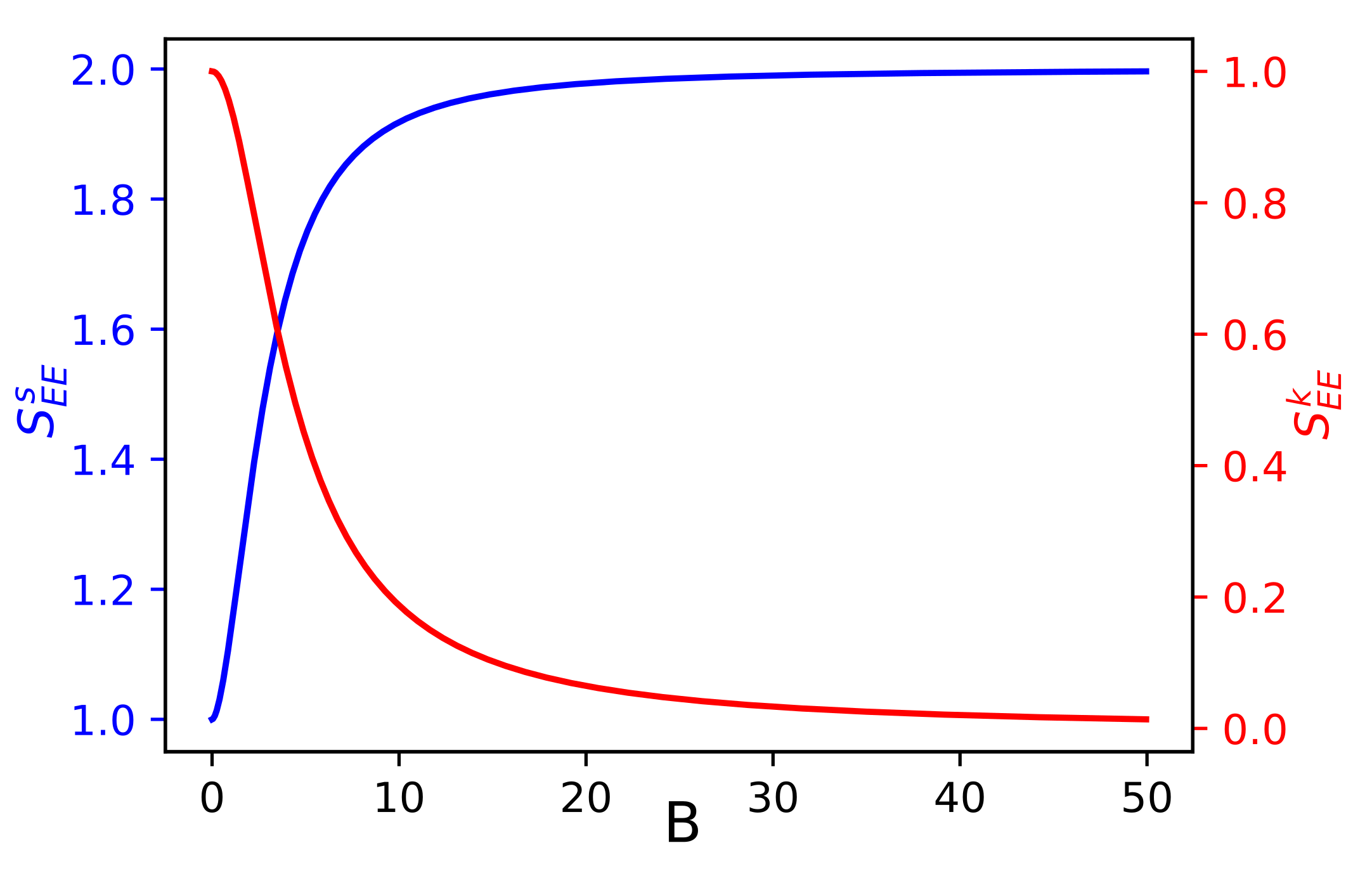}
\caption{Variation of spin partitioned entanglement entropy ($S^{S}_{EE}$, blue curve, left y-axis) and momentum partitioned entanglement entropy ($S^{k}_{EE}$, red curve, right y-axis) with the global $U(1)$ symmetry breaking field $B$ for a prototypical system of $N^{*}=2$ Cooper pairs.}
\label{fig:interkandinterspinEE}
\end{figure}
%\begin{figure}
%\includegraphics[scale=0.6]{plt/EEsEEm}
%\includegraphics[scale=0.4]{plt/EnergyFermi}
%\caption{Showing the inter-k and inter spin Entanglement entropy.}
%\label{fig: EEsEEm1}
%\end{figure}
%\textit{Variation of the spin- and momentum-partitioned entanglement entropy with the symmetry breaking field :}\\
%\par
\par\noindent
On the other hand, as Cooper pairs in the s-wave BCS state are spin singlets, there is a non-zero entanglement between the two spins of a Cooper pair~\cite{bcs_puspus}. Thus, in order to distinguish the CPI and BCS ground states further, we compute the entanglement entropies $S^{k}_{EE}$ and $S^{S}_{EE}$ by partitioning the ground state $|\psi_{g} (B)\rangle$ for an analytically tractable system of two Cooper pairs ($(k_1\uparrow,-k_1,\downarrow)$ and $(k_2\uparrow,-k_2\downarrow)$) in the momentum variable $(k_{1},k_{2})$ and the spin variable $(\uparrow,\downarrow)$ respectively for the case of $\bar{\epsilon}=0$ (strong coupling limit) in the Hamiltonian $H_{coll}$ (eq.\eqref{collham}). 
%As we have already seen that here $\uparrow$ and $\downarrow$ electrons are locked creating an effective spin half pseudo spin. Thus these two spins are not acting inedependently as  they have already formed bound state with each other. If you study the zero temperature symmetry breaking phase $H_{sb}$. 
%For simplicity we will study here two pseuso-spin system that means four electron system. Here pseudo-spin 1 : $k_1,\uparrow$ and $-k_1,\downarrow$, pseudo-spin 2 : $k_2,\uparrow$  and  $ -k_2 ,\downarrow $. 
For this system of two coupled pseudospins, $S=1$ 
%Here for this hamiltionian 
and the ground state wavefunction is 
\begin{eqnarray}
|\psi_{g} (B)\rangle &=& \sum_{\alpha=-1}^{1}C_{\alpha}(B)|S=1,S^{z}=\alpha \rangle~,
% + C_{1} |S_z=0 \rangle + C_2 |S_z=+1 \rangle
\end{eqnarray}
where the coefficients $C_{\alpha}$ are a function of the field $B$ and the coupling $V/N$ given by 
%such that at $B=0$, $C_{-1} = 0 = C_{1}~,~C_{0}=1$, and all $C_{\alpha}$ have finite non-zero values in the limit $B>>1$ (UPDATE WITH ANALYTIC EXPRESSIONS).
%Eigenvalues and eigenvectors are: 
\begin{equation}
\{C_{\alpha}\} = \mathcal{N}\bigg({1, -\frac{\alpha + \sqrt{\alpha^2 + 8 \beta^2}}{2 \beta}, 1}\bigg) \stackrel{\beta >>1}{\longrightarrow} \bigg(\frac{1}{2},-\frac{1}{\sqrt{2}},\frac{1}{2}\bigg)~,
\end{equation}
where $\mathcal{N}$ is the normalisation factor, $\alpha=\frac{V}{N}$ and $\beta=\frac{B}{\sqrt{2}}$.
%\begin{eqnarray}
%&& -\alpha~~~~~~~~~~~~~~~~~~~~~,~~~~ \mathcal{N}(-1,0,1) \\\nonumber
%&& \frac{-3 \alpha - \sqrt{\alpha^2 + 8 \beta^2}}{2}~,~~~~\mathcal{N}\bigg({1, -\frac{\alpha + \sqrt{\alpha^2 + 8 \beta^2}}{2 \beta}, 1}\bigg) \\\nonumber
%&& \frac{-3 \alpha + \sqrt{\alpha^2 + 8 \beta^2}}{2}~,~~~~\mathcal{N}\bigg({1, -\frac{\alpha - \sqrt{\alpha^2 + 8 \beta^2}}{2 \beta}, 1}\bigg)
%\end{eqnarray}
%
%Where $\mathcal{N}(\Psi)=\frac{\Psi}{\sqrt{|\Psi|}}$ and $\alpha=\frac{V}{N},~~\beta=\frac{B}{\sqrt{2}}$.
%
%Ground state Energy is $\frac{-3 \alpha - \sqrt{\alpha^2 + 8 \beta^2}}{2}$. The eigenvector is 
%\begin{eqnarray}
%\mathcal{N}\bigg({1, -\frac{\alpha + \sqrt{\alpha^2 + 8 \beta^2}}{2 \beta}, 1}\bigg)
%\end{eqnarray}
%At large $B/\alpha \gg 1 $  limit the wavefunction is 
%\begin{eqnarray}
%\mathcal{N}\bigg(1,-\sqrt{2},1\bigg)=\frac{1}{2}\bigg(1,-\sqrt{2},1\bigg)
%\label{eq:wavefunctioncoeff}
%\end{eqnarray}
%
%Thus ~$C_{-1}=\frac{1}{2}~,~~C_{0}=-\frac{1}{\sqrt{2}}~,~~C_{1}=\frac{1}{2}$. 
%The state can be written as 
%\begin{eqnarray}
%|\Psi_{g}(B)\rangle &=& \displaystyle\sum_{\alpha\in\{-1,0,1\}}C_{\alpha}~ |S=1,S^z=\alpha\rangle \\\nonumber
%&=& \frac{C_0}{\sqrt{2}} \bigg[|1_{\uparrow}1_{\downarrow}\rangle_k \otimes |0_{\uparrow}0_{\downarrow}\rangle_{k'} + |0_{\uparrow}0_{\downarrow}\rangle_k \otimes |1_{\uparrow}1_{\downarrow}\rangle_{k'} \bigg]\nonumber\\
%&+& C_{-1} |1_{\uparrow}1_{\downarrow}\rangle_k \otimes |1_{\uparrow}1_{\downarrow} \rangle_{k'}  + C_1 |0_{\uparrow}0_{\downarrow}\rangle_k \otimes |0_{\uparrow}0_{\downarrow}\rangle_{k'}~,~~~~
%\end{eqnarray}
By writing the states $|S, S^{z}\rangle$ in the basis of $|n_{k\uparrow} n_{-k \downarrow}\rangle\otimes |n_{k'\uparrow} n_{-k' \downarrow}\rangle$, 
%e.g; $|0_{\uparrow}0_{\downarrow}\rangle_{k}$ represents $n_{k\uparrow}=0$,~ $n_{-k \downarrow}=0$ state represented by pseudo spin configuration $|S=1,S^z=1\rangle$. Thus one can write down the density matrix in the basis \\
i.e., the states $\{|0_{\uparrow}0_{\downarrow}\rangle_k \otimes |0_{\uparrow}0_{\downarrow}\rangle_{k'},~|0_{\uparrow}0_{\downarrow}\rangle_k \otimes |1_{\uparrow}1_{\downarrow}\rangle_{k'},~|1_{\uparrow}1_{\downarrow}\rangle_k \otimes |0_{\uparrow}0_{\downarrow}\rangle_{k'},~|1_{\uparrow}1_{\downarrow}\rangle_k \otimes |1_{\uparrow}1_{\downarrow}\rangle_{k'}\}$, the density matrix $\rho(B) = |\Psi_{g}(B)\rangle\langle\Psi_{g}(B)|$ is found to be 
\begin{eqnarray}
\rho(B) 
%&=& |\Psi_{g}(B)\rangle\langle\Psi_{g}(B)| \\\nonumber
&=&
\begin{bmatrix}
    |C_1|^2 & \frac{C_0C_1}{\sqrt{2}} & \frac{C_0C_1}{\sqrt{2}} & C_1C_{-1} \\
    \frac{C_0C_1}{\sqrt{2}} & \frac{|C_0|^2}{2} & \frac{|C_0|^2}{2} & \frac{C_0C_{-1}}{\sqrt{2}} \\
    \frac{C_0C_1}{\sqrt{2}} & \frac{|C_0|^2}{2} & \frac{|C_0|^2}{2} & \frac{C_0C_{-1}}{\sqrt{2}} \\
    C_1C_{-1} & \frac{C_0C_{-1}}{\sqrt{2}} & \frac{C_0C_{-1}}{\sqrt{2}} & |C_{-1}|^2 
\end{bmatrix}~.
\label{eq:spin1densitymatrix}
\end{eqnarray}
\par\noindent
The momentum-partitioned reduced density matrix is then obtained by tracing out the $k'$ pseudo-spin from the density matrix (\ref{eq:spin1densitymatrix}). The reduced density matrix $\rho^{k}$ is written in the basis $\{|0_{\uparrow} 0_{\downarrow}\rangle_{k},~|1_{\uparrow}1_{\downarrow}\rangle_{k} \}$
\begin{eqnarray}
\rho^{k}&=&Tr_{k'} \rho(B)\\\nonumber
&=&
\begin{bmatrix}
|C_1|^2+\frac{C_0^2}{2}& \frac{C_0C_1}{\sqrt{2}}+\frac{C_{0}C_{-1}}{2}\\
\frac{C_0C_1}{\sqrt{2}}+\frac{C_0C_{-1}}{\sqrt{2}}&|C_{-1}|^2+\frac{C_0^2}{2}
\end{bmatrix}
\end{eqnarray}
The inter-k entanglement is $S^k_{EE}$ calculated from the density matrix $\rho^{k}$. As shown via the red curve in Fig.\ref{fig:interkandinterspinEE}, $S^{k}_{EE}$ reduces monotonically from its largest value at $B=0$ as $B$ is increased, displaying the destruction of the inter-$k$ entanglement of the CPI ground state in the passage towards the BCS ground state.
\par\noindent
Similarly, for the spin-partitioned entanglement entropy, we trace out a given spin sector, say $\downarrow$~. Then, the reduced density matrix in the basis 
%\\
%$(n_{k\uparrow}n_{k'\uparrow})\equiv
$\{|0_k0_{k'}\rangle_{\uparrow},~|0_k1_{k'}\rangle_{\uparrow},~|1_k0_{k'}\rangle_{\uparrow},~|1_k1_{k'}\rangle_{\uparrow} \}$ is 
\begin{eqnarray}
\rho^{\uparrow} &=& Tr_{\downarrow} \rho(B)\\\nonumber
&=& 
\begin{bmatrix}
    |C_1|^2 & 0 & 0 & 0 \\
    0 & \frac{|C_0|^2}{2} & 0 & 0 \\
    0 & 0 & \frac{|C_0|^2}{2} & 0 \\
    0 & 0 & 0 & |C_{-1}|^2 
\end{bmatrix}
\label{eq:spinpart}
\end{eqnarray}
The spin-partitioned entanglement entropy $S^k_{EE}$ is obtained from $\rho^{\uparrow}$. As shown via the blue curve in Fig.\ref{fig:interkandinterspinEE}, $S^{\uparrow}_{EE}$ increases steadily from its smallest value at $B=0$ as $B$ is increased, and saturates as $B>>1$. This shows the growth of the inter-spin entanglement of the BCS ground state in the passage from the CPI ground state. The limiting values of $S^{\uparrow}_{EE}$ at $B=0$ and $B>>1$ observed in Fig.\ref{fig:interkandinterspinEE} can be understood as follows. At $B=0$, we get $C_0=1,~ C_{-1}=C_1=0$, giving $S^{\uparrow}_{EE}$ for the CPI ground state as
%the spin-partitioned entanglement entropy from the reduced density matrix \ref{eq:spinpart} is 
$log_2 2=1$, as seen in Fig.\ref{fig:interkandinterspinEE}. Similarly, the diagonal elements of the diagonal density matrix (\ref{eq:spinpart}) at large $B$ become $(\frac{1}{4},\frac{1}{4},\frac{1}{4},\frac{1}{4})$. This shows that $\rho^{\uparrow} (B>>1)$ becomes maximally mixed in nature, leading to $S^{\uparrow}_{EE}(B>>1)$ 
%the Entanglement entropy at large B 
saturating to the value seen in Fig.\ref{fig:interkandinterspinEE}
\begin{eqnarray}
S=-4\times\frac{1}{4} log_2 \frac{1}{4}=2log_2 2=2~.
\end{eqnarray}
This is a clear signature of the maximal entanglement of the Cooper pair singlets in the BCS ground state.
%
%After partial tracing out all the up spins we have only the down spins, but still there is a non-zero inter-k entanglement. This inter-k entanglement tens to vanish for large B as can be seen from the plot (\ref{fig: EEsEEm1}) showing the signature of BCS superconductivity.
\par\noindent
Another diagnostic of the difference between the ground states at $B=0$ and $B>>1$ lies in the occupation for the $k$-momentum electron 
\begin{eqnarray}
\langle n_k \rangle = Tr (\hat{n}_{k} \rho^{\uparrow}) = \frac{|C_0|^2}{2} + |C_{-1}|^2~.
\label{occupation}
\end{eqnarray}
While at $B=0$, $\langle n_k \rangle \equiv \langle n_k \rangle (\alpha,\beta)$, $\langle n_k \rangle \to 1/2$ as $B>>1$. Given that $\langle n_k \rangle$ follows the Fermi-Dirac distribution for the BCS ground state~\cite{bcs_puspus}, the result of $\langle n_k \rangle=1/2$ obtained at $B>>1$ indicates that the 
%at the Fermi surface. We are studying the case $\bar{\epsilon}=0$ (strong coupling(V) limit). From the wavefunction(\ref{eq:wavefunctioncoeff}) at large B one can see 
%$\langle n_k \rangle$ approaches to $\frac{1}{2}$. Thus shows that our 
collective Hamiltonian $H_{SB}$ (eq.\eqref{eq:hsb}) in the presence of a large $U(1)$ symmetry breaking coupling $B$ describes the BCS superconductor near the Fermi surface.

\begin{figure}
\hspace*{-1cm}
\includegraphics[scale=0.8]{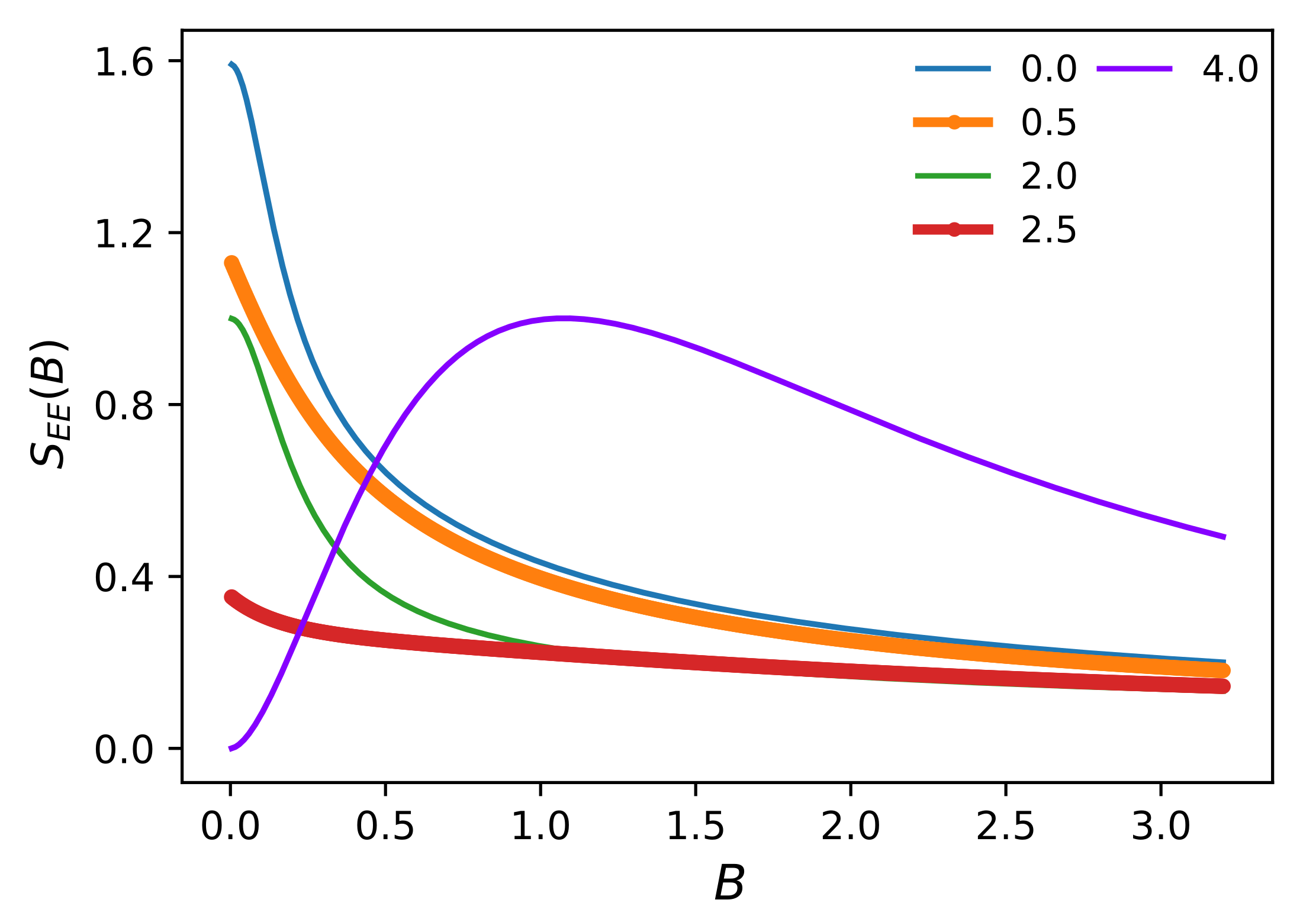}
\caption{Variation of the equipartition entanglement entropy for different CPI ground states and transition ground states (corresponding to different integer and half-integer values of $0\leq \Phi\leq 4$) with the global $U(1)$ symmetry breaking field $B$ for a system of $N^{*}=8$ Cooper pairs. Note that the non-monotonic behaviour of the purple curve arise from the fact that this corresponds to the ground state precisely at the transition from the CPI to the parent metal.}
\label{fig:EEsbroken}
\end{figure} 
\par\noindent
We now present the entanglement entropy ($S_{EE}$) computed from partitioning the ground state $|\psi_{g}\rangle$ (with spin $S$) of Hamiltonian $H_{SB}$ into two equal subsystems A and B such that $S=S_A+S_B,~S_A=S/2=S_B$ using the strategy adopted in eqs.\eqref{basicpartition}-\eqref{basicentangentr}. In Fig.\ref{fig:EEsbroken}, we present a variation of $S_{EE}$ with $B$ computed for a system of 8 pseudospins, and for ground states %that correspond to some of the plateaus and transitions 
of $H_{SB}(B=0)\equiv H_{coll}$ at various values of $\Phi=\epsilon/V$. The plot shows the monotonic decrease for $S_{EE}$ with $B$ for all ground states with a non-zero number of Cooper pairs ($0\leq \Phi<4$), while the $S_{EE}$ computed for the gapless ground state at $\Phi=4$ shows a non-monotonic variation with $B$. The latter case corresponds to the entanglement related to superconducting phase fluctuations in a  mean-field BCS Hamiltonian, i.e., a Hamiltonian $H_{SB}$ in which the $BS^{x}$ term induces pairing in the gapless spectrum of $H_{coll}$. While the BCS ground state corresponding to a vanishingly small $S_{EE}$ is obtained for all these curves in the limit of large $B$, the approach of the mean-field ground state is clearly different from those with pre-existing Cooper pair bound states: the peak in the curve for $\phi=4$ likely arises due to the creation of Cooper pairs in a gapless system.  

\subsection{The effect of a Josephson coupling}
\pin
We end with a brief presentation of the effects of a Josephson coupling between the bulk of two CPI systems A and B (i.e., we are ignoring all effects from gapless edge states), each of which is modelled by $H_{SB}$ (eq.\eqref{eq:hsb})
\begin{eqnarray}
H_{\mu}&=&-\frac{2\epsilon_{\mu}}{V_{\mu}} \displaystyle\sum_{k\in {\mu}}  S_k^z -\frac{V_{\mu}}{2N_{\mu}}\displaystyle\sum_{k\neq k'\in {\mu}}\bigg(S_{k}^+S_{k'}^- + \textrm{h.c.}\bigg) \nonumber\\
&-& B_{\mu} \displaystyle\sum_{k\in {\mu}}S^x_k ~,~~\textrm{$\mu$=A,B}~~\nonumber\\
H_{AB}&=& T\displaystyle\sum_{k\in A, k'\in B}\bigg(e^{i\phi} S_{k}^+S_{k'}^-+ e^{-i\phi}S_{k}^-S_{k'}^+\bigg)~,~~~~~~~~~~~~~~~
\label{josephson}
\end{eqnarray}
where $H_{AB}$ is the Josephson coupling between the systems $A$ and $B$, with the phase $\phi$ dependent on the externally applied voltage difference between the two systems~\cite{van}. We have simulated the equations in eq.\eqref{josephson} for two systems comprised of 4 pseudospins each. 
%contains 4 pseudo spins with different k-momentum, labled by $A\equiv\{0,1,2,3\}$ and B also contains 4 pseudo spins labled by $B\equiv\{4,5,6,7\}$ .
\begin{figure}
\includegraphics[scale=0.64]{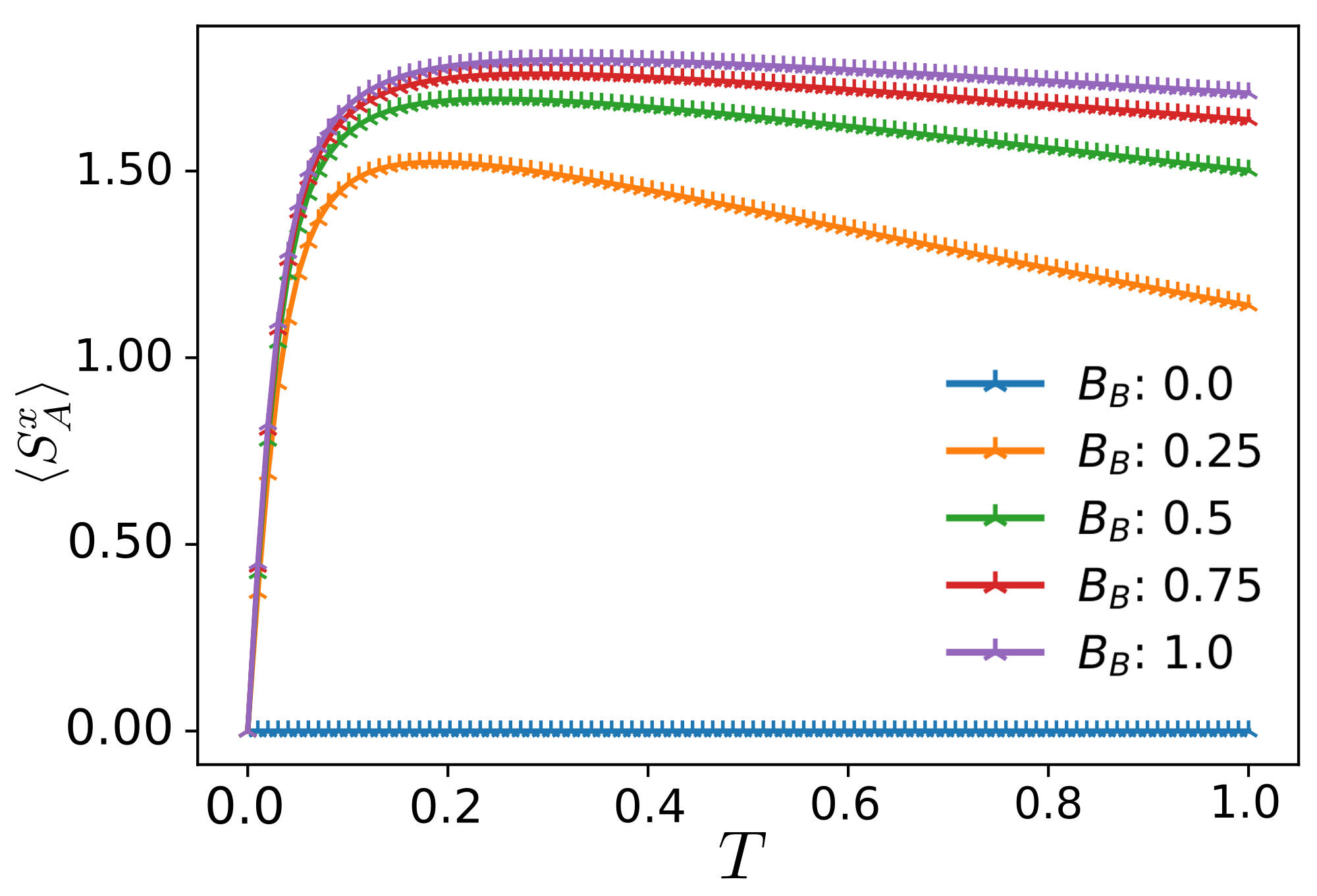}
\caption{Plot of the induced phase stiffness $\langle S^{x}_{A}\rangle$ in CPI system $A$ due to a Josephson coupling (with strength $T$) to CPI system $B$ (in presence of a global $U(1)$ symmetry breaking field ($B_{B}$)). Various curves correspond to different values of $B_{B}$.}
\label{fig:phasecoherence}
\end{figure}
\par
\begin{figure}
\includegraphics[scale=0.52]{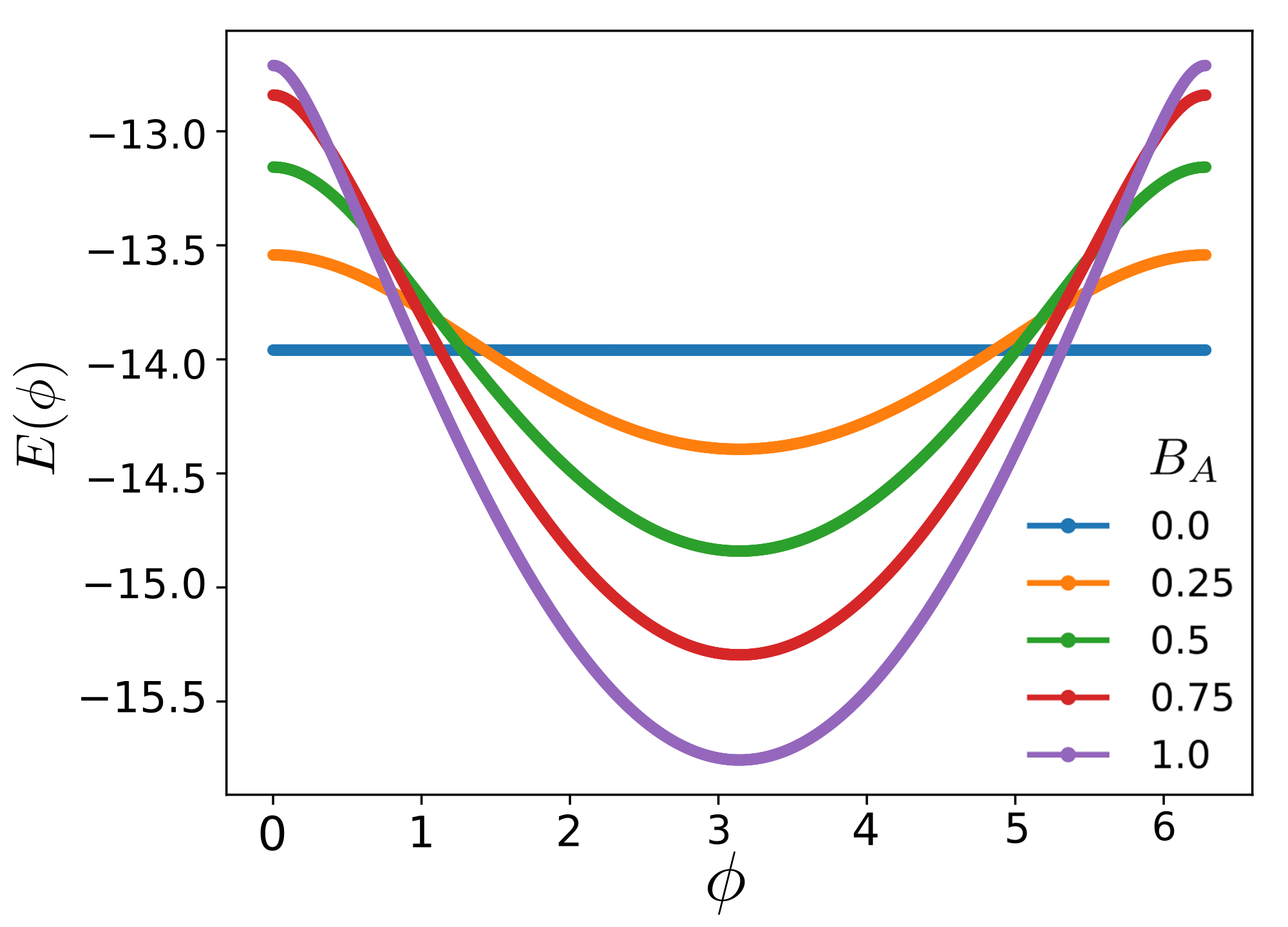}
\caption{Plot of the total ground state energy $E(\phi$) for a CPI system (placed in a gradually increasing $U(1)$ symmetry breaking field $B_{A}$) coupled to a BCS superconductor through Josephson tunneling as a function of their phase difference $\phi$. The various curves correspond to different values of $B_{A}$.}
\label{fig:EG_vs_phi}
\end{figure}

\noindent
First, we set the field $B_{A}=0$, such that system A is in a U(1) symmetric CPI phase and couple it with the system B ($H_{B}$) for several values of the field $B_{B}$ and the Josephson coupling $T$. The values of the parameters $\epsilon_{A}=\epsilon_{B}=$, $V_{A}/N_{A}=V_{A}/N_{B}=1$ and $\phi=\pi$. In Fig.\ref{fig:phasecoherence}, we study the phase coherence being generated in the system A by computing $\langle S^{x}_{A}\rangle$ in the ground state of the total system ($H=H_{A}+H_{B}+H_{AB}$). The blue line in Fig.\ref{fig:phasecoherence} clearly shows that a Josephson coupling between two systems that are individually in CPI phases ($B_{A}=0=B_{B}$) cannot lead to phase coherence being induced in either system A or B. On the other hand, for non-zero values of $B_{B}$, the other curves in Fig.\ref{fig:phasecoherence} shows that as system B already possesses some degree of phase coherence, an increasing non-zero phase coherence is induced in system A via the Josephson coupling with increasing $B_{B}$. Note, however, that while this demonstrates the breaking of the U(1) symmetry of the system A via the Josephson coupling to the symmetry-broken system B, the phases of the two systems are locked to one another with zero relative phase difference~\cite{van}. This is demonstrated in a plot of the total ground state energy $E(\phi)$ as a function of the phase $\phi$:  the blue line in Fig.\ref{fig:EG_vs_phi} clearly shows that a Josephson current ($J\propto \partial E(\phi)/\partial\phi$) cannot be generated in the coupled system for $B_{A}=0$. 
%One can see that for $B_B=0$ the CPI-CPI coupling does not induce phase coherence in system A for any value of T. Larger the value of $B_B$ larger the phase phase coherence been generated in the other system A. \par \noindent
On the other hand, in the presence of a non-zero symmetry breaking field $B_{A}$, $E(\phi)$ shows a cosinusoidal variation with $\phi$ in Fig.\ref{fig:phasecoherence}. This shows that when the symmetry is separately broken in the two systems, a Josephson coupling certainly induces a Josephson current. The results of this subsection serve as predictions for the experimental search of systems in the CPI ground state.
%Thus to detect the CPI phase one can couple it to a superconductor via the Josephson coupling ($H_{AB}$) and measure the phase coherence generated in the CPI as one tunes T. This will show signatures like fig.{\ref{fig:phasecoherence}}.

\par\noindent
%Similarly we calculated the ground state energy $E(\phi)$ as function of $\phi$. Now we keep the $B_B=1.0$ fixed and tune the value of $B_A$. In the fig.\ref{fig:EG_vs_phi} one can see that the $E(\phi)$ has a periodic cosine like dependence with $\phi$. $E(\phi)\propto \alpha~cos(\phi) $. Josephson current, $I_J\propto \frac{dE(\phi)}{d\phi}\propto \alpha$. From the fig.\ref{fig:EG_vs_phi} one can see that for $B_A=0$(CPI phase A) $\alpha=0$ thus no Josephson Current. Due to the huge phase fluctuation in the CPI phase A ($B_A=0$) Josephson Current cannot penetrate the system. But if you increase $B_A$ to non-zero value that generates non-zero phase coherence in the system A, Josephson current becomes non-zero. 

\section{Entanglement Renormalisation}
\label{section:entrg}
%We showed using coupling Forward RG, how we get a zero mode Hamiltonian at the RG fixed point starting from reduced BCS Hamiltonian. We studied that fixed point Hamiltonian. Studied Entanglement properties within the emergent subspace.\\
\par\noindent
Having explored the entanglement features of the topologically ordered CPI and symmetry broken BCS ground states at some length in previous sections, we now present an analysis the $T=0$ RG evolution of the many-particle entanglement content of these ground states. For this, we follow the strategy for entanglement renormalisation that was developed in Refs.\cite{mukherjee2020,MukherjeePatraLal2020}. For the sake of completeness, we outline briefly the strategy below.
\par
\begin{center}
\begin{figure}[!h]
%\hspace{-1.0cm}
\includegraphics[scale=0.13]{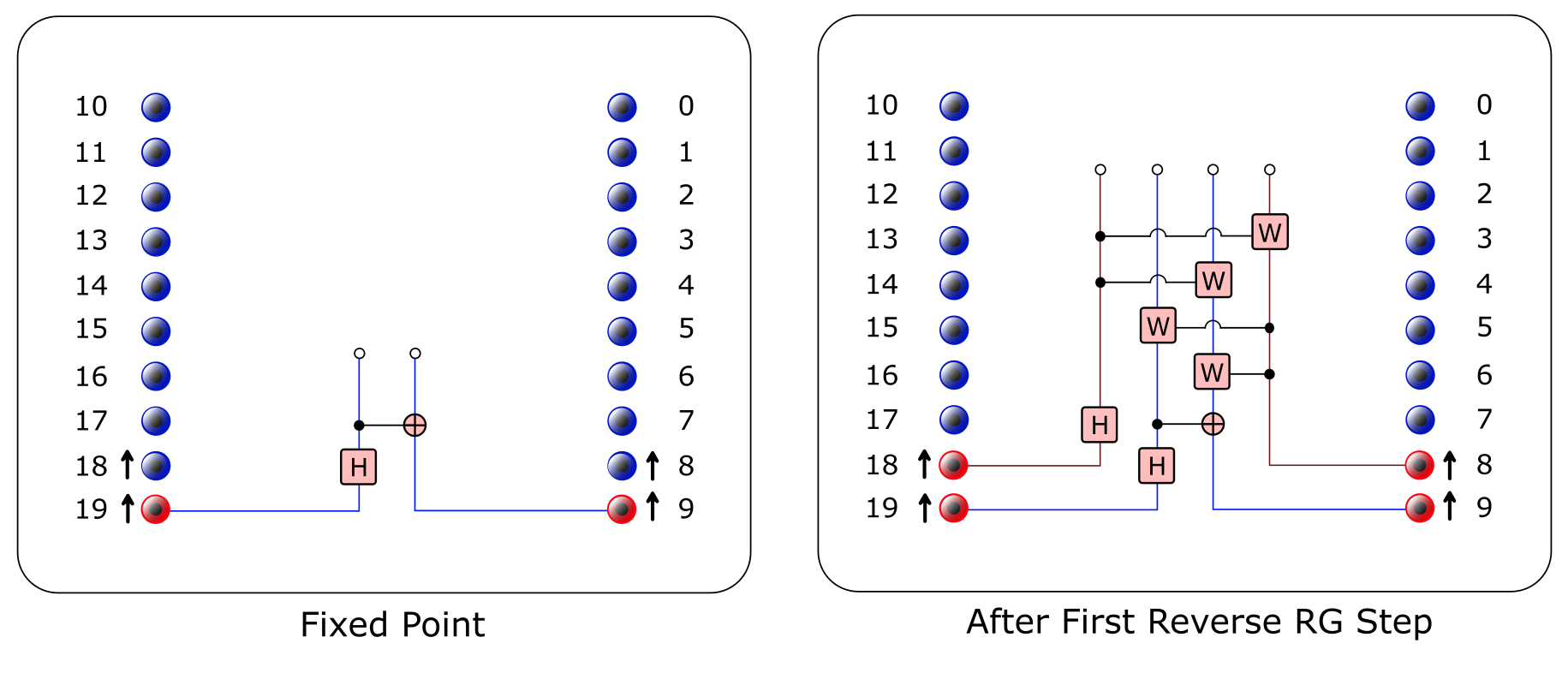}
\caption{Quantum circuit representation of (left panel) the ground state of the fixed point CPI Hamiltonian and (right panel) the ground state after the first step of the reverse unitary RG step. Both are for a system of $N^{*}=2$ Cooper pairs.}
\label{qcirccpi}
\end{figure}
\end{center}

\begin{center}
\begin{figure}[!h]
%\hspace*{-1.0cm}
\includegraphics[scale=0.13]{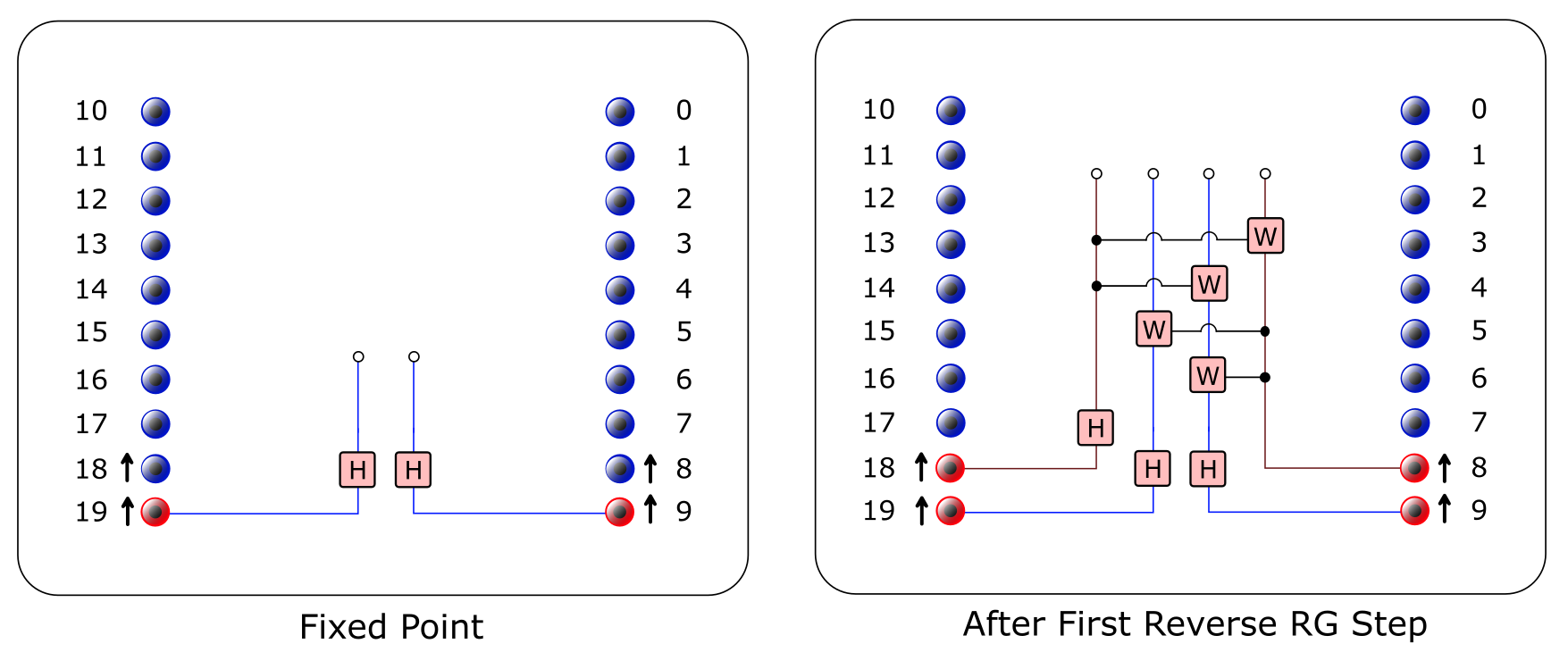}
\caption{Quantum circuit representation of (left panel) the ground state of the fixed point BCS Hamiltonian and (right panel) the ground state after the first step of the reverse unitary RG step. Both are for a system of $N^{*}=2$ Cooper pairs.}
\label{qcircbcs}
\end{figure}
\end{center}

\noindent
As we have seen earlier, the URG proceeds by disentangling electronic states sequentially from the UV towards the IR by the application of many-particle unitary transformations ($U$, see Appendix~\ref{section:appendix_A} for further details). At the IR stable fixed point, we have identified the ground state wavefunction. Now, by reversing the RG flow through the sequential applications of the appropriate $U^{\dagger}$s, we generate a family of ground state wavefunctions ranging towards the UV. This allows for the computation of several entanglement features from each member of the family of wavefunctions, thereby generating the RG flow of these entanglement features. As discussed in detail in Refs.\cite{mukherjee2020,MukherjeePatraLal2020}, the unitary operators $U$ of the URG method can be implemented as a quantum circuit, i.e., in terms of a combination of universal 2-qubit gates (e.g., Hadamard, C-NOT and phase-shift gates). Below, in Figs.\ref{qcirccpi} and \ref{qcircbcs}, we show the quantum circuit realisations that implement the reverse URG flow along one radial direction in $k$-space for the CPI and BCS wavefunctions respectively.
%\begin{enumerate}
%\item 
%\begin{eqnarray}
%H^{(*)}_{CPI}=-\frac{V}{2N} \displaystyle\sum_{i\neq j} \bigg(S_i^+ S_j^- + h.c.  \bigg)
%\end{eqnarray}
%Now I will study the Wavefunction RG. Starting with the Fixed point wavefunction using the Reverse RG, first generate the wavefunctions of different reverse RG steps.
%
%\item 
%\begin{eqnarray}
%H^{(*)}_{BCS}=-\frac{V}{2N} \displaystyle\sum_{i\neq j} \bigg(S_i^+ S_j^- + h.c.  \bigg)- \frac{B}{2}\displaystyle\sum_{i} (S_i^+  +h.c.)
%\end{eqnarray}
%Similarly studied the forward RG with symmetry beaking field $BS^x$. Here also starting with the fixed point wavefunction we do the reverse RG to generate the wavefunctions.
%
%\end{enumerate}
%\newpage
\par
\begin{center}
\begin{figure}[!h]
\hspace{-0.5cm}
\includegraphics[scale=0.24]{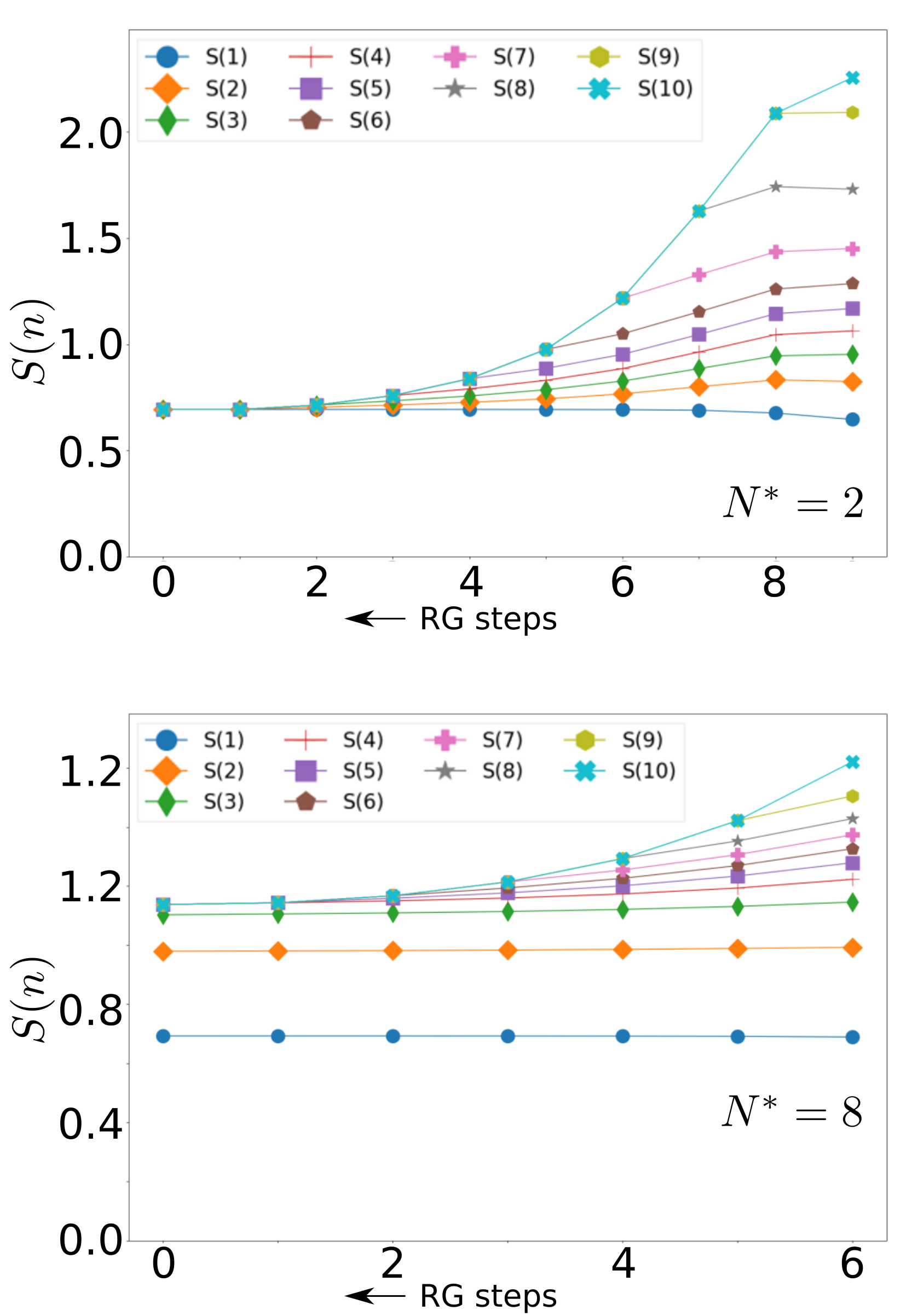}
\caption{Plot for the RG variation of the entanglement entropy $S(n)$ for various block sizes $1\leq n\leq 10$ (in different colours) for a CPI system with $N^{*}=2$ (upper panel) and $N^{*}=8$ (lower panel) Cooper pairs. See text for discussion.}
\label{blockentcpi}
\end{figure}
\end{center}

\noindent
As shown in Fig.\ref{qcirccpi}, the nodes 9 and 19 refer to the fermion states residing just outside and just inside the Fermi surface respectively. The distance from the Fermi surface increases with passage between states 9 to 0 (all outside the Fermi surface), and with passage between states 19 to 10 (all inside the Fermi surface). As indicated by the quantum circuit diagrams, the reverse RG flow starts from the emergent CPI phase (described by the effective Hamiltonian in eq.\eqref{hcoll01}) obtained at the stable fixed point and with a window of electronic states given by $N^{*}$ ($N^{*}=2$ in the figures). The reverse RG flow proceeds by the re-entangling of two electronic states lying outside the window at each step of the RG. We now present the results of the RG evolution for the entanglement entropy of a block in $k$-space along a given radial direction for CPI and the symmetry broken phases. In Fig.\ref{blockentcpi}, we present the RG variation of the entanglement entropy computed for a block (lying outside the Fermi surface) of varying size ranging from one to ten fermionic states. The two plots are for different sizes of the window ($N^{*}$) for the emergent CPI phase: the upper plot is for $N^{*}=2$ (i.e., comprised of states 9 and 19 only), while the lower is for $N^{*}=8$ (i.e., comprised of state 6-9 and 16-19). Further, the reverse RG process increases stepwise from step $0$ (in the IR) towards the UV. 
%First, 
%we show the results for the Block entropy for CPI and the symmetry broken phase.
%\begin{enumerate}
%\item CPI case:
\par\noindent
The upper panel of Fig.\ref{blockentcpi} shows that block entropy for all block sizes terminates at a universal value of $S=0.693=\ln 2$, corresponding to the entanglement for the $N^{*}=2$ pseudospins that form the emergent CPI window in the IR. Further, the plots demonstrate that the block entanglement entropy of block size 1 (i.e., for the state 9, one of the two states that form the CPI ground state in the IR) increases slowly with the RG flow from UV to IR. On the other hand, the block entropy of all other block sizes (greater than one) decreases with the stepwise decoupling of electronic states. Further, the entanglement entropy of the blocks varies non-linearly with the RG steps. Additionally, in the block entropy plots for $N^{*}=8$ (lower panel of Fig.\ref{blockentcpi}), we see that the entanglement entropy for all block sizes less than 4 (i.e., the size of the four states 9-6 that are part of the CPI ground state in the IR) are affected very little by the RG flow. This is a remarkable display of the fact that the entanglement of the electronic states proximate to the Fermi surface (and that eventually form a part of the emergent window) is quite robust under RG evolution, and distinguishes them from those that are decoupled along the flow. Further, the CPI ground state possesses a hierarchy of scales of entanglement defined by the various block sizes.
%\begin{center}
%\begin{figure}[!h]
%\includegraphics[scale=0.26]{plt/FermiSpaceBlocks}
%\end{figure}
%\end{center}
%\textbf{Comments:} 
%\begin{itemize}
%\item Those isolated lines shows the Block entropy of blocks smaller than the fixed point emergent space size. Does not changes much.
%\item Block entropy of size greater than the fixed point emergent space size shows variation with RG steps.
%\item $S_{BE}(L\geq L_{emrgnt})\propto (RG-step/\alpha)^2$
%\item $S_{BE}(L< L_{emrgnt})\propto Constant$
%\end{itemize}
%\textbf{Important feature to note:} Even at large RG steps back one can detect what was the emergent space size by measuring these Block entropy.
%\newpage
%\item Symmetry Broken case with small B field:
\begin{center}
\begin{figure}[!h]
\hspace{-0.5cm}
\includegraphics[scale=0.24]{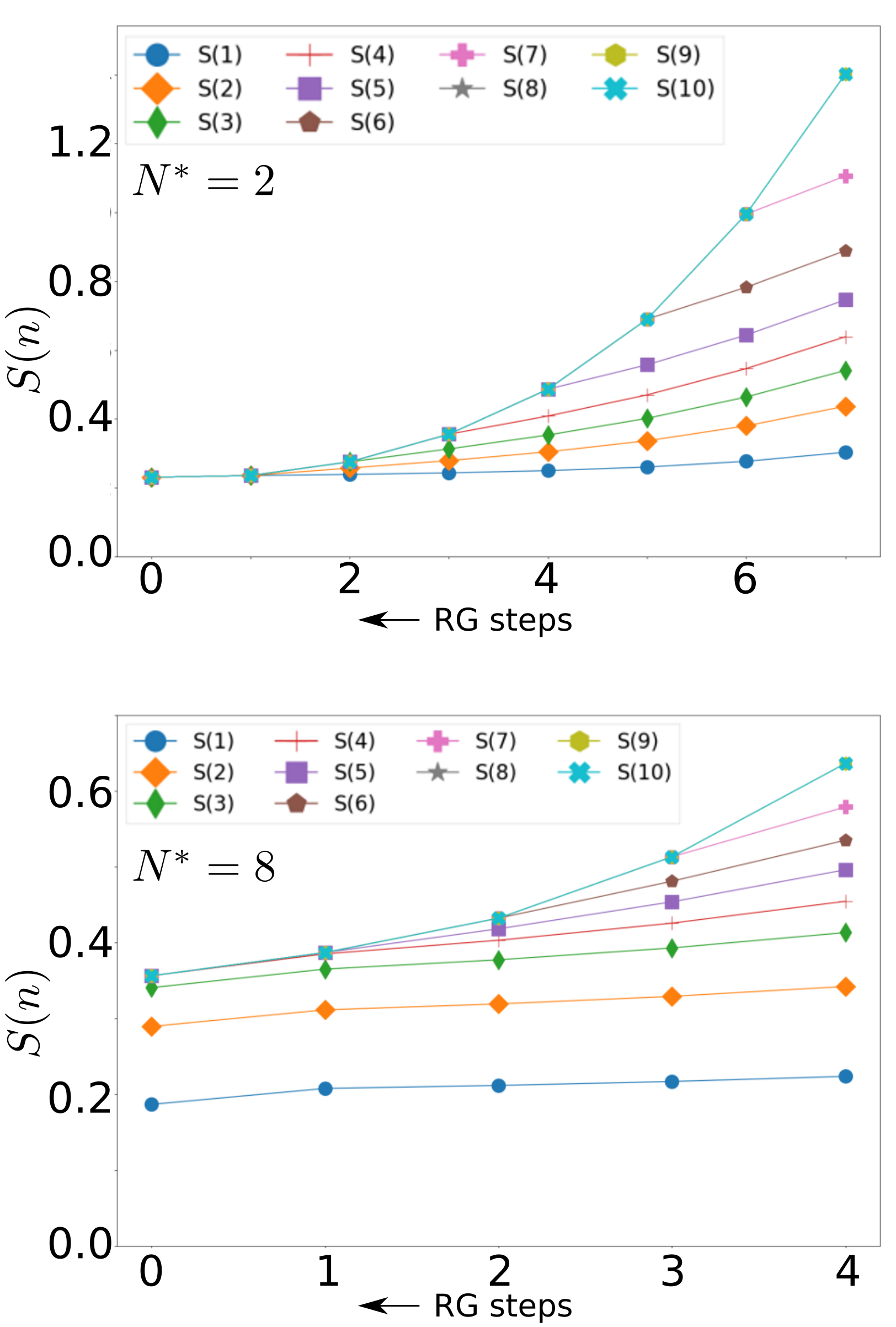}
\caption{Plot for the RG variation of the entanglement entropy $S(n)$ for various block sizes $1\leq n\leq 10$ (in different colours) for a CPI system with $N^{*}=2$ (upper panel) and $N^{*}=8$ (lower panel) Cooper pairs in the presence of a weak global $U(1)$ symmetry breaking field $B=5\times 10^{-5}$ (in units of the attractive pairing coupling $V$). Note that the curves for $n=8, 9$ and $10$ are lying on top of one another. See text for discussion.
%\textbf{Keep only the $N^{*}=2$ and $N^{*}=8$ figures. Change the labels from $N_{emr}$ to $N^{*}$. Increase all axes labels, ticks and the legends.}
}
\label{entRGsmallsymmbreak}
\end{figure}
\end{center}
\par
\begin{center}
\begin{figure}[!h]
\includegraphics[scale=0.2]{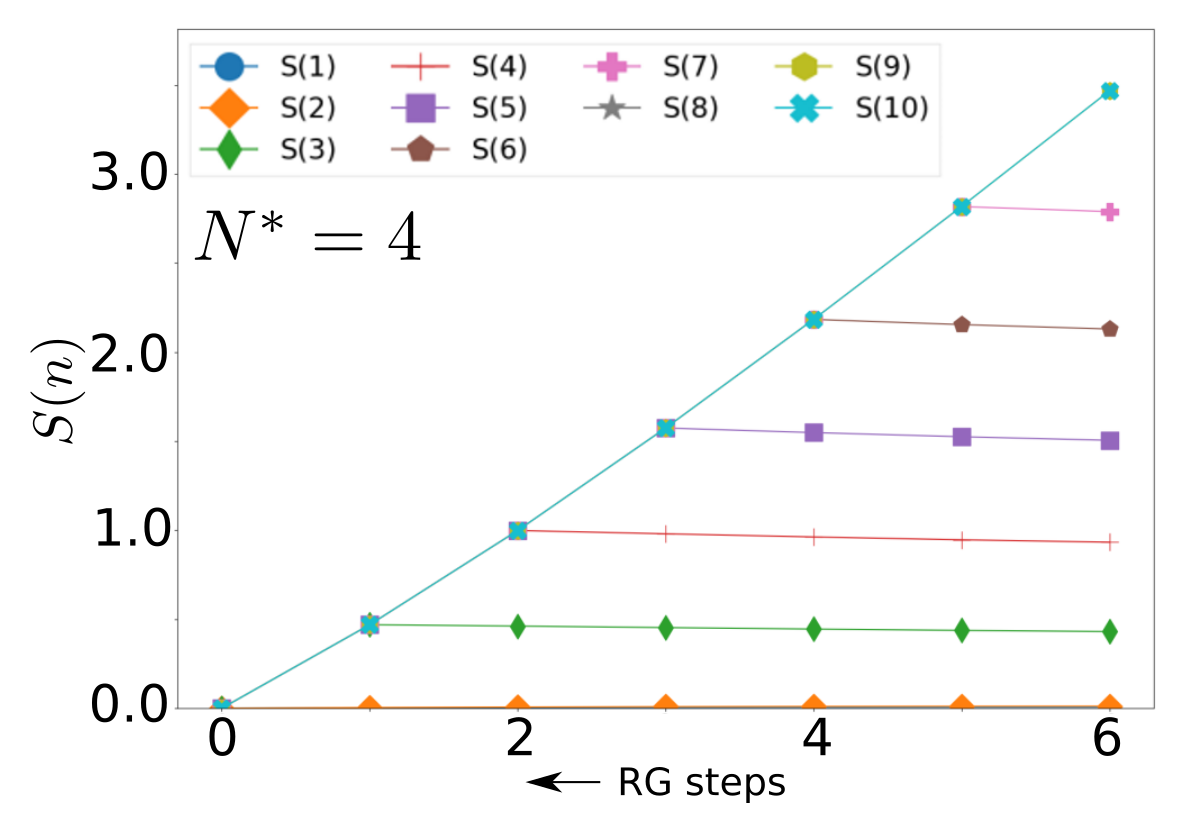}
\caption{Plot for the RG variation of the entanglement entropy $S(n)$ for various block sizes $1\leq n\leq 10$ (in different colours) for a CPI system with $N^{*}=2$ (upper panel) and $N^{*}=8$ (lower panel) Cooper pairs in the presence of a strong global $U(1)$ symmetry breaking field $B=25\times 10^{-3}$ (in units of the attractive pairing coupling $V$). Note that the curves for $n=8, 9$ and $10$ are lying on top of one another. See text for discussion.
%\textbf{Keep only the $N^{*}=4$ figure. Change the labels from $N_{emr}$ to $N^{*}$. Increase all axes labels, ticks and the legends.}
}
\label{entRGbigsymmbreak}
\end{figure}
\end{center}

\noindent
Next, we present the entanglement RG results for a system in the presence of a bare symmetry breaking field ($B$). In Fig.\ref{entRGsmallsymmbreak}, we see that the presence of a very weak bare $U(1)$ symmetry breaking field $B\sim 5\times 10^{-5}$ (in units of $V$), the entanglement RG flows from UV to IR are very similar to those shown in Fig.\ref{blockentcpi} above for the CPI (i.e. for the case of $B=0$), with only one difference: the final value of the block entropies in the IR here is reduced with respect to those obtained for the CPI. This indicates a gradual collapse of the hierarchy of scales of entanglement of the CPI upon tuning a symmetry breaking field. Finally, in Fig.\ref{entRGbigsymmbreak}, we present that the entanglement RG flows for a system with $N^{*}=4$ 
%(upper panel) and  $4$ (lower panel) 
for the case of a slightly larger (but still weak) bare $U(1)$ symmetry breaking field $B\sim 25\times 10^{-3}$ (in units of $V$). Here, we find that the entanglement curves for various block sizes is very different to those obtained for the CPI (see Fig.\ref{blockentcpi}). For instance, the block entropy for both block sizes one and two (i.e., corresponding to the two possible subblocks of the emergent BCS ground state in the IR) have zero entanglement entropy throughout the RG. 
%Similarly, for $N^{*}=4$, the block entropies for block sizes one and two are all identically zero throughout the RG flow: 
There is, thus, no longer any way to distinguish between the constituent blocks of the BCS ground state under the RG. Further, the entanglement entropy varies linearly with the RG steps for block sizes $\geq 8$.   
%\textbf{Comments:} 
%For small B we don't see much difference in the plots.
%\begin{itemize}
%\item The overall entanglement lowers.
%\item The isolated lines has slight curvature near RG-step$=0,1$, closer to zero than higher RG-step values.
%\end{itemize}
%\newpage
%\item Symmetry Broken case with LARGE B field:
%\end{enumerate}
%\textbf{Comments:} 
%\begin{itemize}
%\item This time Block entropy starts from zero at the RG fixed point.
%\item In this case all Block entanglement of smaller size than the emergent space size is zero and stays zero for all RG steps.
%\item $S_{BE}(L\geq L_{emrgnt})\propto (RG-step/\alpha)$
%\item $S_{BE}(L< L_{emrgnt})=0$
%\item One can note here that the Block entropy envelop is linear 
%\end{itemize}
%\textbf{Important feature to note:} Here also one can detect the emergent space size. But this time one cannot distinguish between different size of the blocks as all blocks sizes shows vanishing entanglement.

\section{\label{sec:conc}Conclusions and Discussions}
\label{section:conclusion}
\par\noindent
A body of theoretical work has proposed~\cite{diamantini_jja,hos,BoseMetal1}, on phenomenological grounds, the existence of a topologically ordered counterpart of the superconductor. This novel state of quantum matter, which we call the Cooper pair insulator (CPI) is expected to be a condensate of a fixed number of Cooper pairs, but without any phase stiffness. Instead, the CPI would correspond to a gapped system in the bulk and with gapless states at the boundaries. In keeping with this proposal, recent experimental studies of the superconductor to insulator transition (SIT) in thin films suggests the existence of such a CPI lying precisely at the transition~ \cite{SIT_dis_1,SIT_dis_2,SIT_dis_3}. 
%is very well studied field. This standard Insulating phase shows the Cooper pair number fixed phase, obtained via adding disorder and/or magnetic field on the Superconductor, but the magnetic field breaks the time-reversal symmetry and disorder breaks the translational symmetry. Recent theoretical and experimental shows the presence of a topological bosonic insulating phase made out of Cooper pairs \cite{SIT1,SIT2,SIT3,SIT4,BoseMetal1,diamantini_jja} between the Superconducting and insulating phase. Hansson et al. \cite{hos} show in their field-theoretic study how coupling to a dynamical gauge field one can effectively restore the symmetry of BCS superconductor to get to topologically ordered phase of Cooper pair. 
While the phenomenological gauge field theories proposed for the CPI offer some insight into its properties, a microscopic approach remains absent. 
%study or Hamiltonian is available. 
Thus, a major finding of our work is the derivation of an effective microscopic Hamiltonian for the CPI that is emergent from a unitary renormalization group (URG) analysis.
\par\noindent
For this, we have worked on a generalized model of a metallic system (i.e., with a repulsive density-density interaction $U$) as well as an attractive pairing interaction. Our URG study of this model offers a phase diagram in terms of a quantum fluctuation scale ($\omega$) and $U$, clearly displaying the existence of a CPI phase of quantum matter at small $\omega$ (i.e., corresponding to energyscales for excitations proximate to the Fermi surface) and for all $U$. 
%formation of Cooper bound state at a low quantum fluctuation and low U value. Beyond a critical value of U no Cooper bound state exists. We get 
The low energy fixed point effective Hamiltonian obtained for the CPI phase is then studied in detail. 
%URG method preserves the translational and time-reversal invariance in this phase. 
As mentioned earlier, the symmetry-preserved CPI phase is found to possess a fixed number of Cooper-pairs but without any global phase coherence among them. Subsequently, we have carried out a detailed analysis of various topological and many-particle entanglement features of this state of quantum matter, establishing thereby the emergence of topological order in the CPI. We have also contrasted the properties of the CPI ground state with its (BCS s-wave) superconducting counterpart, and believe that some of our results provide experimentally testable predictions. Importantly, we have also benchmarked numerically the ground state energy density of the CPI (in the thermodynamic limit) obtained from a finite-size scaling analysis for the RG against a similar finite size scaling analysis of exact diagonalisation calculations. We now end with a discussion of the broader significance of our findings with regards to the subject of topological order.
\par\noindent
%In our way, we show that this symmetry preserved Cooper pair insulating phase shows topological properties, which 
Topological order is proposed to describe the ordering of interacting many-particle quantum system beyond the Ginzburg-Landau-Wilson (GLW) paradigm (see Ref.\cite{wen2017colloquium} for a recent review). The GLW paradigm describes order arising the spontaneous breaking of symmetries, measured in terms of real-space local order parameters and associated with a phase transition whose universality is captured by a set of scaling exponents. On the other hand, a topologically ordered ground state does not arise from breaking any symmetries and thus lacks a local order parameter. Instead, such ground states are invariant under large gauge transformations, can be represented purely in terms of non-local gauge operators (e.g., Wilson loops etc.), and their quantum dynamics can be captured by a topological gauge field theory. When placed on a multiply connected manifold (e.g., a torus), a topologically ordered system displays a non trivial degeneracy of the ground state manifold (protected by a non-zero energy gap), as well as the existence of fractionally charged topological excitations that interpolate between the ground states. While the bulk of such a system is an incompressible insulating state of matter (due to the spectral gap), it can possess gapless current-carrying degrees of freedom at its boundaries. It has also been shown that the ground states of a  topologically ordered system can possess signatures of non-trivial many-particle entanglement, e.g., an entanglement entropy (due to a real-space bipartitioning) proportional to the degeneracy count of the ground state manifold (called the quantum dimension). While all of these properties are widely believed to be the features and diagnostics of a topologically ordered system, an overarching theoretical framework for this subject remains an outstanding challenge. As the pairing instability of the Fermi surface represents a paradigmatic phenomenon for a system of interacting electrons, our insights into the CPI represents an opportunity towards learning the inner workings of emergent topological order in such systems, as well as how it is different from the order captured by ground states belonging to the GLW paradigm (e.g., the BCS ground state).
\par\noindent
The body of results presented for the CPI phase clearly satisfy the diagnostics described above. We have established analytically the topological degeneracy of the ground state manifold using flux insertion arguments, and shown that the zero mode collective effective Hamiltonian for the CPI can be written in terms of Wilson loop operators. This then paves the way for connecting the topological $\theta$ term in the effective theory for the CPI with the 2+1 dimensional topological Chern-Simons gauge field theory proposed for such systems~\cite{diamantini_jja,hos,moroz}. We have shown the origin of the spectral gap that protects the ground state manifold, and shown the spectral flow property of such ground states with a variation in the $\theta$ parameter: ground states form plateaux in $\theta$ labelled by a topological quantum number and with topological quantum phase transitions separating them. Indeed, much of the phenomenology observed by us is common with the properties of topologically ordered fractional quantum Hall ground states~(see Ref.\cite{wen2017colloquium} and references therein). It will be interesting to test these conclusions for systems of interacting electrons in the presence of disorder~\cite{MukherjeeNPB2} or incommensuration~\cite{hatsugai1993,senlal2000}. 
\par\noindent
Our investigations of the entanglement features show clear universal signatures that distinguish the topologically ordered CPI ground states (plateaux) from those found at the transitions between plateaux. The passage to the metallic state upon tuning the effective Aharanov-Bohm flux of the fixed point Hamiltonian is charted at zero as well as finite temperatures, yielding clear signatures once again in the entanglement for the CPI ground states. By carrying out the RG analysis in the presence of a global $U(1)$ symmetry breaking term, a detailed comparison between the CPI and BCS ground states is also offered. This allows us to demonstrate the clear distinctions between these two kinds of ground states in terms of many-particle entanglement and many-body correlations: unlike the BCS state, the CPI ground state is found to possess various measures of entanglement. Further, we show that, as CPI ground states lack phase stiffness, they cannot show the Josephson effect (i.e., upon coupling two such CPI systems through Cooper pair tunneling).
\par\noindent
All of this leads us to conjecture that our results on the CPI offer a broad framework for understanding topological order. Specifically, we believe that various quantum liquid systems displaying the hallmark signatures of topological order described above are likely to be described by effective zero mode collective Hamiltonians described in terms of Wilson loop like non-local degrees of freedom. Using similar flux insertion arguments, it should be possible to show that the ground state manifolds of such Hamiltonians display topological degeneracy on the torus etc. Indeed, similar conclusions have been reached by some of us for the Mott liquid ground states of the 2D Hubbard model discovered recently in Refs.\cite{MukherjeeMott1,MukherjeeMott2}, and the spin liquid ground states of quantum spins coupled through antiferromagnetic exchange on geometrically frustrated lattices~\cite{pal2019,pal2020,pal2019magnetization}. It should be possible, therefore, to chart out in a similar fashion the microscopic origins of various kinds of topologically ordered quantum liquids. This will go a long way in establishing a detailed understanding of the universality of such phenomena.
\par\noindent
We end with a brief discussion on where to search for such CPI ground states. As we have seen here, the CPI state reached from a generic non-nested Fermi surface is strongly susceptible towards the effects of spontaneous symmetry breaking and the emergence of the BCS s-wave superconducting ground state. As mentioned earlier, some hints of the CPI have been found to lie at the superconductor to insulator transition in recent experiments on thin films. Based on our recent study of the 2D Hubbard model~\cite{MukherjeeMott1,MukherjeeMott2}, and its relevance to the physics of the high-temperature superconducting hole doped cuprate Mott insulators, we believe that the CPI ground states may well be observed in those materials too. Specifically, in Refs.\cite{MukherjeeMott2}, we observed the existence at $T=0$ of a pseudogapped CPI state of quantum matter lying above the d-wave superconducting ``dome" obtained upon optimally doping the Mott insulating ground state of the $1/2$-filled 2D Hubbard model with holes. This pseudogapped phase arose from electronic differentiation built into the electronic dispersion of the nested Fermi surface of the 2D tight-binding model at $1/2$-filling. Further, the pseudogap phase showed a clear gapping of the anti-nodal regions in $k$-space that could be described in terms of a state of matter containing condensed Cooper pair bound states but without any global phase coherence. The large superconducting phase fluctuations observed in this pseudogapped phase~\cite{MukherjeeMott2} are a signature of the CPI, and are reminiscent of the findings from Nernst effect measurements on the pseudogap phase of the doped cuprates~\cite{ong-physrevb.73.024510}. We believe, therefore, that the cuprates are excellent candidate systems in which to search for the existence of the CPI phase. Following the suggestion of Ref.\cite{baskaran2015}, pressurised solid H$_{2}$S may be another interesting candidate system in which to search for the CPI.
\acknowledgments
The authors thank A. Mukherjee, S. Pal, R. K. Singh, A. Dasgupta, A. Ghosh, S. Sinha, G. Baskaran, S. Moroz and A. Taraphder for several discussions and feedback. S. P. thanks the CSIR, Govt. of India and IISER Kolkata for funding through a research fellowship. S. L. thanks the DST, Govt. of India for funding through a Ramanujan Fellowship during which a part of this work was carried out.
%Thus we have found a microscopic topological phase of Cooper pairs starting from a general Hamiltonian. This phase is very susceptible to BCS superconductivity as shown by the URG study. The most important question is how to increase this CPI's stability against the symmetry breaking field. Within this CPI phase, is another instability possible leading to a four-electron bound state like the superinsulator, one can study that. We have not studied the CPI to standard insulator transition, one can study this transition by adding a small magnetic field and/or realspace disorder. The larger the gap of this CPI, the larger the corresponding superconductor's gap is. One can study how to increase the gap of this CPI phase to indirectly increase the BCS superconducting gap.
%\acknowledgments
%The authors thank A. Mukherjee, S. Pal, R. K. Singh, A. Dasgupta, A. Ghosh, S. Sinha and A. Taraphder for several discussions and feedback. S. P. thanks the CSIR, Govt. of India and IISER Kolkata for funding through a research fellowship. S. L. thanks the DST, Govt. of India for funding through a Ramanujan Fellowship during which a part of this work was carried out.

\appendix

\begin{widetext}
\appendix

\section{Hamiltonian RG}
\label{section:appendix_A} 
\par\noindent
We first briefly recapitulate the unitary RG method developed in Refs.\cite{MukherjeeMott1,MukherjeeMott2,pal2019,MukherjeeNPB1,
MukherjeeNPB2,MukherjeeTLL}, and then derive
%(cite hubbard and kagome works). \\ We Used Unitary RG method ([kagome][hubbard]) to get 
the RG equations for the generalised pairing Hamiltonian eq.\eqref{rghamiltonian}. The RG method adopted uses a unitary transformation to decouple one single-particle Fock state $|k\sigma\rangle$ from the rest of the states it is interacting with. Very generally, one can write the many-particle Hamiltonian as $\hat{H}=\hat{H}^D + \hat{H}^X_{k\sigma}+ \hat{H}^{\bar{X}}_{k\sigma}$, where $\hat{H}^D$ contains all single-particle and many-particle number diagonal (kinetic energy and interaction) terms. $\hat{H}^X_{k\sigma}$ represents all the off-diagonal interaction terms connected to the single-particle state $|k\sigma\rangle$, while $\hat{H}^{\bar{X}}_{k\sigma}$ represents all off-diagonal interaction terms among all (say, $2^{N-1}$) single-particle states other than $|k\sigma\rangle$. %Let's say we have $N$ single particle fock states.~  
Considering a many-particle eigenstate of the Hamiltonian $|\Psi \rangle$ (a member of the full $2^N$ dimensional Hilbert space), we can write
\begin{eqnarray}
\hat{H} | \Psi\rangle = (\hat{H}^D + \hat{H}^X_{k\sigma}+ \hat{H}^{\bar{X}}_{k\sigma}) |\Psi\rangle = \bar{E} |\Psi \rangle~,
\end{eqnarray}
where $\bar{E}$ is the eigenvalue for $|\Psi\rangle$. One can rewrite the wavefunction $|\Psi\rangle$ in a Schmidt decomposed form as follows
\begin{eqnarray}
|\Psi\rangle &=& a_1 |\Psi_1\rangle \otimes |1_{k\sigma} \rangle + a_0 |\Psi_0\rangle \otimes |0_{k\sigma} \rangle~,
\end{eqnarray}
where $\{|1_{k \sigma}\rangle, |0_{k \sigma}\rangle\}$ live in a $2$-dimensional single-particle Fock space and $\{|\Psi_1\rangle, |\Psi_0\rangle\}$ lives in the remaining $2^{N-1}$ dimensional Hilbert space. We then proceed to remove all quantum fluctuations connected between $|k, \sigma \rangle$ with the other $|k'\neq k, \sigma\rangle$ states. For this, one can define transition operators $\hat{\eta}_{k\sigma}$ and $\hat{\eta}_{k\sigma}^{\dagger}$ as follows
\begin{eqnarray}
a_1 |\Psi_1\rangle \otimes |1_{k\sigma} \rangle = \hat{\eta}_{k\sigma}^{\dagger}~ a_0 |\Psi_0\rangle \otimes |0_{k\sigma} \rangle~,~~~a_0 |\Psi_0\rangle \otimes |0_{k\sigma} \rangle = \hat{\eta}_{k\sigma}~ a_1 |\Psi_1\rangle \otimes |1_{k\sigma} \rangle~,
\label{transition_operation}
\end{eqnarray}
where $|1_{k\sigma} \rangle$ and $|0_{k\sigma} \rangle$ represent the $n_{k\sigma}=1$ and $n_{k\sigma}=0$ states respectively, and 
\begin{eqnarray}
\hat{\eta}_{k\sigma} = \frac{1}{\hat{\omega}-Tr_{k\sigma}(\hat{H}^D(1-\hat{n}_{k\sigma}))(1-\hat{n}_{k\sigma})} Tr_{k\sigma} (c_{k\sigma}^{\dagger} \hat{H}) c_{k\sigma}~.
\end{eqnarray}
Here, $Tr_{k\sigma}(~)$ represents a partial trace in the Fock space over the state $|k, \sigma \rangle$. These transition operators have a fermionic nature 
\begin{eqnarray}
\hat{\eta}_{k\sigma}^{\dagger}\hat{\eta}_{k\sigma}=\hat{n}_{k\sigma}=1-\hat{\eta}_{k\sigma}\hat{\eta}_{k\sigma}^{\dagger},~~\{\hat{\eta}_{k\sigma}^{\dagger},\hat{\eta}_{k \sigma}\}=1,~~[\hat{\eta}_{k\sigma}^{\dagger},\hat{\eta}_{k\sigma}]=2\hat{n}_{k\sigma}-1,~~ \hat{\eta}_{k\sigma}^2=0~.
\end{eqnarray}
\par\noindent
Using the transition operators $(\eta_{k\sigma},\eta_{k\sigma}^{\dagger})$ and the eqs.\eqref{transition_operation}, one can see that
\begin{eqnarray}
|\Psi\rangle &=& a_1 |\Psi_1\rangle \otimes |1_{k\sigma} \rangle + a_0 |\Psi_0\rangle \otimes |0_{k\sigma} \rangle \nonumber\\
&=& a_1 |\Psi_1\rangle \otimes |1_{k\sigma} \rangle +\hat{\eta}_{k\sigma}~ a_1 |\Psi_1\rangle \otimes |1_{k\sigma} \rangle \nonumber\\
&=& a_1 (1+\hat{\eta}_{k\sigma}) |\Psi_1\rangle \otimes |1_{k\sigma} \rangle = a_1 e^{\hat{\eta}_{k\sigma}} |\Psi_1\rangle \otimes |1_{k\sigma} \rangle~.
\end{eqnarray}
Thus,
%one can see that using these $e^{\hat{\eta}_{k\sigma}}$ 
one can construct a unitary operator 
\begin{equation}
U_{k\sigma}=\frac{1}{\sqrt{2}} (1+\eta_{k\sigma}^{\dagger}-\eta_{k\sigma})~,~
\end{equation}
that rotates the many-particle basis in such way that $U_{k\sigma} |\Psi \rangle = \mathcal{N} | \alpha \rangle$,~~${\alpha=0~or~1}$ and $\mathcal{N}$ normalization constant. This unitary rotaion removes all quantum fluctuations between the states $|0_{k\sigma} \rangle$ and $|1_{k\sigma}\rangle$~. Further, using the unitary operator, the Hamiltonian can be written in the rotated basis as
\begin{eqnarray}
U_{k\sigma}\hat{H} U_{k\sigma}^{\dagger} = \frac{1}{2}Tr_{k\sigma}(\hat{H})+\tau_{k\sigma}Tr_{k\sigma}(H \tau_{k\sigma})+ \tau_{k\sigma}\{c_{k\sigma}^{\dagger} Tr_{k\sigma}(\hat{H} c_{k\sigma}),\hat{\eta}_{k\sigma} \}~.
\end{eqnarray}
It is important to note that while $\hat{n}_{k\sigma} \hat{H} (1-\hat{n}_{k\sigma}) \neq 0$ (i.e., there existed non-trivial quantum fluctuations in the occupation of single-particle Fock state given by $n_{\sigma}$) prior to the application of the unitary operator, subsequent to its application we find 
\begin{eqnarray}
%&&\hat{n}_{k\sigma} \hat{H} (1-\hat{n}_{k\sigma}) \neq 0 ~,~~~~~~~\textrm{(in general)} \\
&& \hat{n}_{k\sigma} U_{k\sigma} \hat{H} U_{k\sigma}^{\dagger} (1-\hat{n}_{k\sigma}) ~=~0~~\Rightarrow ~~ [\hat{n}_{k\sigma},U_{k\sigma} \hat{H} U_{k\sigma}^{\dagger}]=0~.
\label{rg_condition}
\end{eqnarray}
The degree of freedom $n_{k\sigma}$ is thus rendered an integral of motion (IOM) of the RG flow. The RG equations can then be obtained from the condition eq.\eqref{rg_condition}.
\par\noindent
Coming to the problem at hand, in the generalised pairing Hamiltonian eq.\eqref{rghamiltonian}, we are working in the subspace given by $n_{k,\sigma}=n_{-k,\sigma}$~\cite{anderson1958random}. Thus, at every step of the RG, we are disentangling two single-particle states $|k,\sigma\rangle,~ |-k$ and $-\sigma\rangle$ simultaneously. We now proceed by rewriting the Hamiltonian in terms of Anderson pseudospins~\cite{anderson1958random}
\begin{eqnarray}
H_{\textrm{pair}}^{q}&=&-\displaystyle\sum_{k}\tilde{\epsilon}_{k,q}(S_{k,q}^z-\frac{1}{2}) -\sum_{k\neq k'} \frac{|W_{kk'}^{q}|}{2}(S_{k,q}^- S_{k',q}^+ +\textrm{h.c.})+U\sum_{k\neq k'} S_{k,q}^z S_{k',q}^z ~,
\label{paireffham} 
\end{eqnarray}
such that the part of the Hamiltonian associated with the $k_N$ pseudo-spin is given by
\begin{eqnarray}
H_{N}^q &=& -\tilde{\epsilon}_{k_N,q} \tau_{k,q}^z  -\sum_{k\neq k_N} \frac{|W_{kk_{N}}^{q}|}{2}(\tau_{k_N,q}^- S_{k,q}^+ + \tau_{k_N,q}^+ S_{k,q}^- )~+\frac{U}{4}~.
\label{pairnodeeffham}
\end{eqnarray}
%In the sector $\tau_{k_{N}}^z=+1/2$, we then obtain the RG equations as
Applying the RG formalism to $H^{q}_{N}$, one obtains from the condition eq.\eqref{rg_condition} the operator level RG equation for the Hamiltonian in the low-energy sector for the quantum fluctuation scale $\omega$ as 
\begin{eqnarray}
\Delta H&=& \bigg(\displaystyle\sum_{k\neq k_N} \frac{|W_{kk_{N}}^{q}|}{2} \tau_{k_N}^+ S_{k}^-\bigg) G_{k_N} \bigg(\displaystyle\sum_{k'\neq k_N} \frac{|W_{k'k_{N}}^{q}|}{2} S_{k'}^+ \tau_{k_N}^-\bigg)~.
\label{rgforeffham}
\end{eqnarray}
From this, we derive the RG equations in the relevant channel ($\tau_{k,q}^z=+\frac{1}{2}$) for Cooper pair condensation as
\begin{eqnarray}
\frac{\Delta \tilde{\epsilon}_{k',q}^{(j)}}{\Delta \log \frac{\Lambda_j}{\Lambda_0}}~= ~\frac{1}{4}\frac{|W_{k_{\Lambda}k'}^{(j)}|^2}{\bigg(\omega-\frac{\tilde{\epsilon}_{k_{\Lambda},q}^{(j)}}{2} - \frac{U}{4}\bigg)}~~~,~~~
\frac{\Delta |W_{k'k''}^{(j)}|}{\Delta \log \frac{\Lambda_j}{\Lambda_0}}~=~- \frac{1}{4}~\frac{|W_{k_{\Lambda}k'}^{(j)}||W_{k_{\Lambda}k''}^{(j)}|}{\bigg(\omega-\frac{\tilde{\epsilon}_{k_{\Lambda},q}^{(j)}}{2} - \frac{U}{4}\bigg)}~.
\label{couplrg}
\end{eqnarray}
\end{widetext}

\begin{widetext}
%\appendix
\section{URG with symmetry breaking field\label{section:appendix_B}}
%\label{section:appendix_B}
We begin by including
% with the pairing Hamiltonian that includes 
a global $U(1)$ symmetry breaking term ($-2|B|\displaystyle\sum_{k} S_k^x$) to the pairing Hamiltonian eq.\eqref{paireffham} (but with the repulsion coupling $U=0$).
%\begin{eqnarray}
%H_{sb}&=& -\displaystyle\sum_{k} \epsilon_k S_{k}^z - \displaystyle\sum_{k\neq k'} \frac{|W_{kk'}|}{2} \bigg( S_{k}^+S_{k'}^- + S_k^- S_{k'}^+ \bigg) -2|B|\displaystyle\sum_{k} S_k^x~.
%\end{eqnarray}
Naturally, the symmetry breaking term now appears in 
%The part of 
the Hamiltonian involving the node $k_N$ (eq.\eqref{pairnodeeffham}), as well as the operator RG equation (eq.\eqref{rgforeffham}).
%is given by
%\begin{eqnarray}
%H^{sb}_N &=& -\epsilon_{k_N} \tau_N -\displaystyle\sum_{k\neq k_N} \frac{|W_{kk'}|}{2} \bigg( \tau_{k_N}^+S_{k}^- + \tau_{k_N}^- S_{k}^+ \bigg) -|B| \bigg(\tau_{k_N}^++\tau_{k_N}^-\bigg)~. 
%\end{eqnarray}
%As above, this leads to the operator RG equation as
%\begin{eqnarray}
%\Delta H_{sb} &=& \tau_{k_N}^+\bigg(\displaystyle\sum_{k\neq k_N} V_{kk_N}  S_{k}^- +|B|\bigg) G_{k_N} \bigg(\displaystyle\sum_{k'\neq k_N} V_{k'k_N} S_{k'}^+ +|B|\bigg)\tau_{k_N}^-~.
%\end{eqnarray}
Subsequently, in the sector $\tau_{k_N}^z=+1/2$, we get the RG equations for $\tilde{\epsilon}_{ k',q}$ and $|W_{k'k''}|$ precisely as in eqs.\eqref{couplrg} (but with $U=0$). Further, we obtain a RG equation for the symmetry breaking field $|B|$
\begin{eqnarray}
%\frac{\Delta \tilde{\epsilon}_{k',q}^{(j)}}{\Delta \log \frac{\Lambda_j}{\Lambda_0}} &=& ~\frac{1}{4}\frac{|W_{k_{\Lambda}k'}^{(j)}|^2}{\bigg(\omega-\frac{\tilde{\epsilon}_{k_{\Lambda},q}^{(j)}}{2} - \frac{U}{4}\bigg)}~,\\
%\frac{\Delta |W_{k'k''}^{(j)}|}{\Delta \log \frac{\Lambda_j}{\Lambda_0}}&=&- \frac{1}{4}~\frac{|W_{k_{\Lambda}k'}^{(j)}||W_{k_{\Lambda}k''}^{(j)}|}{\bigg(\omega-\frac{\tilde{\epsilon}_{k_{\Lambda},q}^{(j)}}{2} - \frac{U}{4}\bigg)}\nonumber\\
\frac{\Delta |B^{(j)}|}{\Delta \log \frac{\Lambda_j}{\Lambda_0}}&=&- \frac{1}{2}~\frac{|B^{(j)}||W_{k_{\Lambda}k''}^{(j)}|}{\bigg(\omega-\frac{\tilde{\epsilon}_{k_{\Lambda},q}^{(j)}}{2}\bigg)}~.
\end{eqnarray}  
\par\noindent 
%\subsection*{II. Gap of symmetry broken phase}
We now compute the spectral gap of the symmetry broken BCS superconducting phase.
%\label{section:appendix_B}
For this, taking $|W_{k_{\Lambda}k''}^{(j)}|\equiv W^{(0)}$ (a constant independent of $k_{\Lambda}$ and $k''$), we note that the solution to the RG equation for $B$ is given by 
\begin{eqnarray}
B^{(j)} = \frac{B^{(0)}}{1+ |W^{(0)}|\displaystyle\sum_{l=0}^{j-1} \frac{1}{2\epsilon^{(j)}- \tilde{\omega} } }~.
\end{eqnarray}
The strong coupling RG fixed point of $B\to\infty$ is reached when the denominator of the above relation for $B$ vanishes:
%\begin{eqnarray}
%1+ |V^{(0)}|\displaystyle\sum_{l=0}^{j-1} \frac{1}{2\epsilon^{(j)}- \tilde{\omega} } &=& 0 \\
%\end{eqnarray}
\begin{eqnarray}
-\frac{1}{|W^{(0)}|}&=&\displaystyle\sum_{l}^{j-1}\frac{1}{2\epsilon^{(l)}-\tilde{\omega}}~~
\approx \displaystyle\int_{E_F+\hbar v_{F}\Lambda^{(0)}}^{E_F+\hbar v_{F}\Lambda^*} \frac{N(E)dE}{2(E-E_{F})}~,
%&=& \frac{N(E_F)}{2} log~\bigg[\frac{E_F+\hbar\omega^*-\frac{\tilde{\omega}}{2}}{E_F+\hbar\omega_D-\frac{\tilde{\omega}}{2}}\bigg]\\
%-\frac{2}{|V_0|N(E_F)}&=&log~\bigg[\frac{E_F+\hbar\omega^*-\frac{\tilde{\omega}}{2}}{E_F+\hbar\omega_D-\frac{\tilde{\omega}}{2}}\bigg]\\
%e^{-\frac{2}{|V_0|N(E_F)}}&=&\frac{\hbar\omega^*}{\hbar \omega_D}~, ~~~~~~E_F=\frac{\tilde{\omega}}{2}\\
\end{eqnarray}
where we have replaced the sum by an integral, $N(E_{F})$ is the electronic density of states (DOS) at the Fermi energy ($E_{F}$), $\epsilon^{(l)}$ by the continuous energy variable $E$ and $\tilde{\omega}$ by $2E_{F}$. $\Lambda^{(0)}$ and $\Lambda^{*}$ correspond to the bare and final $k$-space cutoffs of the RG flow. From here, we obtain the well-known relation for the (exponentially small) BCS gap 
\begin{eqnarray}
\Lambda^*&=&\Lambda^{(0)}~~\exp~\bigg(-\frac{2}{|W_0|N(E_F)}\bigg)~.
\end{eqnarray}
%
%Thus we can see exponentially small gap.
\end{widetext}

%%\subfile{Sections/Appendices/AppendixC}

\bibliography{EPQM_Bibliography}

\end{document}